\documentclass[floatfix,nofootinbib,12pt,onecolumn,prx]{revtex4}
\usepackage{graphicx,amsfonts,amssymb,amsmath,hyperref,hypcap,enumerate}
\usepackage{slashed}
\usepackage{amsmath,amssymb,bm,graphicx,dsfont}
\usepackage[latin9]{inputenc}
\usepackage{amsmath}
\usepackage{slashed}
\usepackage{dsfont}
\usepackage{amstext}
\usepackage{amssymb}
\usepackage{amsbsy}
\usepackage{amsthm}
\usepackage{epsfig}
\usepackage{graphicx}
\usepackage{bbm}
\usepackage{hyperref}
\usepackage{color}
\usepackage{slashed}
\usepackage{slashed}
\newcommand{\be}{\begin{equation}}
\newcommand{\ba}{\begin{align}}
\newcommand{\ee}{\end{equation}}
\newcommand{\bea}{\begin{eqnarray}}
\newcommand{\eea}{\end{eqnarray}}
\newcommand{\beq}{\begin{equation}}
\newcommand{\eeq}{\end{equation}}
\newcommand{\beqn}{\begin{eqnarray}}
\newcommand{\eeqn}{\end{eqnarray}}

\renewcommand{\vec}[1]{{\bf #1}}

\begin{document}

\title{An Adventure in Topological Phase Transitions in $3+1$-D: Non-abelian Deconfined Quantum Criticalities and a Possible Duality}
\author{Zhen Bi,  T. Senthil}
\affiliation{Department of Physics, Massachusetts Institute of Technology, MA, 02139, USA}
\begin{abstract}
Continuous quantum phase transitions that are beyond the conventional paradigm of fluctuations of a symmetry breaking order parameter are challenging for theory. These phase transitions often involve emergent deconfined gauge fields at the critical points\cite{2dDQCP1,2dDQCPprb,2dDQCPjps,2dDQCP2} as demonstrated  in $2+1$-dimensions.  Examples include phase transitions in quantum magnetism as well as those between Symmetry Protected Topological phases. In this paper, we present several examples of Deconfined Quantum Critical Points (DQCP) between Symmetry Protected Topological phases in $3+1$-D for both \emph{bosonic} and \emph{fermionic} systems. Some of the critical theories can be formulated as non-abelian gauge theories either in their Infra-Red free regime, or in the conformal window when they flow to the Banks-Zaks\cite{BZprl,BZnpb} fixed points. We explicitly demonstrate several interesting quantum critical phenomena. We describe situations in which the same phase transition allows for multiple universality classes controlled by distinct fixed points.  We exhibit the possibility - which we dub ``\emph{unnecessary quantum critical points}" - of stable generic continuous phase transitions   \emph{within} the same phase.  We present   examples of interaction driven band-theory-forbidden continuous phase transitions between two distinct band insulators.   The understanding we develop leads us to suggest an interesting possible $3+1$-D field theory duality between $SU(2)$ gauge theory coupled to one massless adjoint Dirac fermion and the theory of a single massless Dirac fermion augmented by  a decoupled topological field theory.

\end{abstract}

\date{\today}

\maketitle
\newpage
\tableofcontents

\section{Introduction}
Ground states of quantum many particle systems can go through phase transitions as the Hamiltonian is tuned. When such a quantum phase transition is continuous, the resulting quantum critical point has many interesting properties which have been explored for many decades\cite{sachdevbook,QPTRMP1} in diverse contexts. Despite this 
 our intuition for what kinds of continuous quantum phase transitions are possible and their theoretical descriptions are very poor. The standard examples involve continuous quantum phase transitions separating a trivial gapped disordered phase from a symmetry breaking phase with a Landau order parameter. In this case the critical phenomena may be described within the framework of a quantum Landau-Ginzburg-Wilson (LGW) theory in terms of a fluctuating order parameter field. 
 
 There are many examples of continuous quantum phase transitions that are beyond the Landau paradigm. First it may happen that one or both phases have non-Landau order (for instance they may have topological order). Then, since  an order parameter based description fails to capture the non-Landau phase, it is not surprising that the critical theory is not within the standard LGW paradigm (see Ref. \onlinecite{QCPcenke} for a review). Perhaps more surprisingly, Landau-forbidden continuous phase transitions may even occur between phases that themselves are Landau-allowed. A classic example is the Neel to Valence Bond Solid transition of spin-$1/2$ quantum magnets on a $2d$ square lattice\cite{2dDQCP1,2dDQCPprb,2dDQCPjps,JQmodelprl1,JQmodelprl2, Kaulprl, ringexchangeprl, twoscales, Wiesejsm, Troyerprl,  Bartoschprb, monopolescaling, Scalapinoprl, Sandvikprb,berryphaseprl}. The theory for this transition is an example of a phenomenon dubbed ``deconfined quantum criticality''.  The critical field theory is conveniently expressed in terms of ``deconfined''  fractionalized degrees of freedom though the phases on either side only have conventional ``confined'' excitations. There are, by now, many other proposed examples of deconfined quantum critical points in 2+1 space-time dimensions\cite{SU(N)dqcp1, SU(N)dqcp2, spintexture, 2dDQCP2, Nahumprx, Nahumprl,classicalmodel1,classicalmodel2, classicalmodel3, senthilfisher, SachdevSO5, Kagomeprl, Xutrangle, HeRG,Honeycombdqcp1,Honeycombdqcp2,Nahum3dloop,MotrunichMC,RibhuSU(N),RibhuCPN, RibhuSO(N)}.  Very similar (sometimes equivalent) theories emerge for critical points between trivial and Symmetry Protected Topological (SPT) phases of bosons in 2+1 space-time dimensions\cite{tarunashvin,LuLee,senthilashvin,Slagleprb,Sp(n)prb, ZYMprx, LuBTPT, 2dDQCP2}. (For a general introduction to SPT phases, see for example \cite{Wengroupcohomology, Wenscience, KitaevClassification, Tenfoldway, senthilashvin,NLSM, Senthilscience, Senthilreview}.) These have been shown\cite{2dDQCP2} to be related by webs of dualities of $2+1$-D conformal field theories discussed in recent years\cite{dualityweb,selfdualcenke,Karch-Tong,Hsin-Seiberg}. 
 
\begin{figure}
  \centering
    \includegraphics[width=0.9\textwidth]{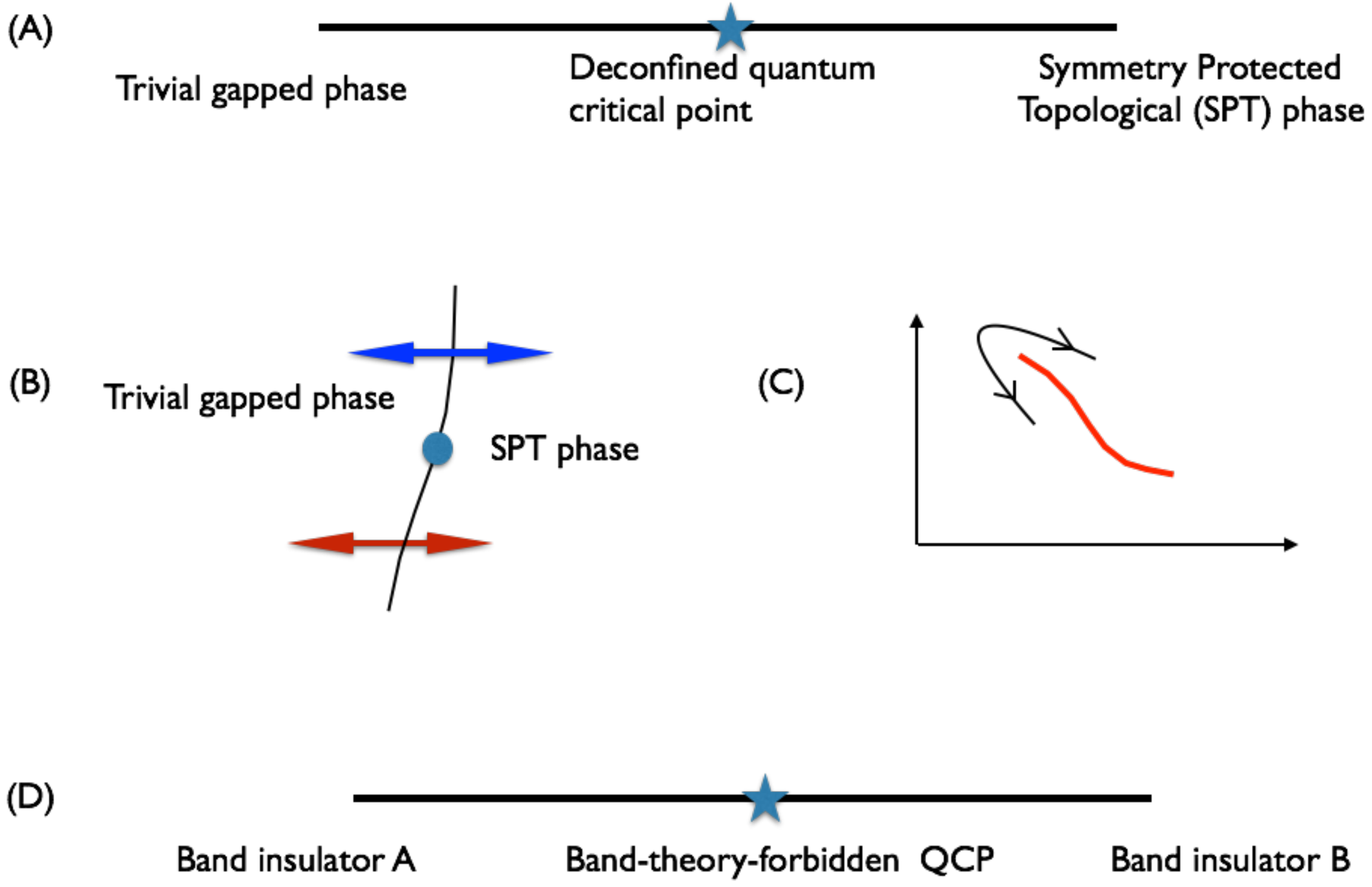}
      \caption{(A) Deconfined quantum criticality at the trivial to SPT phase boundary of systems  of either bosons or of fermions.  (B) Multiple universality classes for the same phase transition. (C) ``Unnecessary quantum critical points" that  live within a single phase of matter
      (D)  Band-theory-forbidden QCP between two band insulators.}
      \label{summary}
\end{figure}

In this paper we will describe a number of surprising quantum critical phenomena for which there are no (or very few) previous examples as far as we know. Fig. \ref{summary}  contains a schematic description of some of our results. We construct examples of deconfined quantum critical points in $3+1$-D of which there are no prior examples.  These examples appear as critical theories separating trivial and SPT phases of either bosons or of  fermions.  We describe situations where the same phase transition admits multiple universality classes depending on where the phase boundary is crossed,   We also introduce and describe the concept of an ``unnecessary continuous phase transition". These are continuous transitions which happen {\em within} the same phase. They are analogous to the liquid-gas transition except that they are continuous. It is common in condensed matter physics when talking about quantum critical points to assume that the most fundamental question is to understand what the distinction is between the phases on either side of the transition. The existence of unnecessary continuous phase transitions shows that quantum critical points may occur that do not at all separate two distinct phases. 
 
In fermionic systems we describe examples of interaction-driven quantum critical points between topological phases  that are not possible within a free fermion description even though the phases themselves can be described by free fermions. These transitions violate band theory rules for which band insulators can be separated by continuous phase transitions. Such band-theory-forbidden continuous phase transition between two band-theory-allowed phases of matter are a  close fermionic analog of the familiar examples of Landau-forbidden continuous transitions between Landau-allowed phases in bosonic systems. We present examples where the critical theory is a deconfined gauge theory.

Many of our results are obtained by considering the phase diagram of non-abelian  gauge theories in space-time dimensions $D = 3+1$. If massless we interpret the corresponding theory as a quantum critical point in the phase diagram and identify the phases obtained by turning on relevant perturbations. As a bonus of the results on  fermionic deconfined quantum critical points, we will discuss a striking possible duality of fermions in $3+1$-D. Specifically we will show that  an $SU(2)$ gauge theory coupled to massless adjoint fermions and massive fundamental bosons may share the same Infra-Red (IR) physics with a theory of a free Dirac fermion supplemented by  a gapped 
 topological field theory. Both theories have the same local operators, and the same global symmetries and anomalies. Further they support the same massive phases.  These checks lend hope that the massless theories may also be infra-red dual. Closely related work on  $SU(2)$ gauge theories with adjoint fermions has recently appeared in Ref. \onlinecite{ThomasSU(2),PoppitzSU(2)}, and we will use some of their results. In $2+1$-D, dualities of Yang-Mills theories with adjoint fermions have been explored in recent work\cite{Gomis2017}. There are many famous examples of dualities of supersymmetric field theories in diverse dimensions\cite{Terning2006}.  Many interesting non-supersymmetric dualities have been found in $2+1$-D (starting from old work\cite{Peskin,bosonvortexdh,bosonvortexfl} on charge-vortex duality in bosonic theories), particulary in recent years\cite{sonphcfl,senthilduality, maxduality, minwalla13, ran13, Aharony2016, dualityweb, selfdualcenke, Karch-Tong, Hsin-Seiberg, Jensen2018, Jensen2017a, Jensen2017b, Karch2017, Tong2016, Karch2018}. However there are no simple dualities of non-supersymmetric theories that are known to us in $3+1$-D.

\section{Preliminaries and summary of results}

\subsection{Free massless Dirac fermion as a quantum critical point}

In this section we review how to interpret free massless Dirac fermion theories in space-time dimensions $D = 3+1$ as quantum critical points. This will enable us to introduce many ideas and methods that will be useful to us later on in a simple setting. 

Consider a free Dirac fermion described by the Lagrangian 
\be
{\cal L} = \bar{\psi} \left(-i \slashed{\partial} + m\right)\psi
\ee
Here $\psi$ is a 4-component Dirac fermion. We will regard this as the low energy theory of electrons with global symmetry $U(1) \times Z_2^T$ (denoted class AIII\cite{KitaevClassification,Tenfoldway} in the condensed matter literature). With this choice the electric charge of the global $U(1)$ symmetry is odd under time reversal $Z_2^T$.  To probe the physics of the system it will be convenient to introduce a background $U(1)$ gauge field $A$ (more precisely a spin$_c$ connection\footnote{Physically this is a device that enables keeping track of the fact that all physical fields with odd charge under $A$ are fermionic. Formally  if we try to formulate this theory on an arbitrary compact oriented space-time manifold, a $Spin_c$ connection is like a $U(1)$ gauge field but with a modified flux quantization condition. Specifically a $Spin_c$ connection satisfies the following condition, $\int\frac{F}{2\pi}=\frac{w_2^{TY_4}}{2}$ mod 1, where $F$ is the field strength for the $U(1)$ gauge bundle, $w_2^{TY_4}$ is the second Stiefel-Whitney class for the tangent bundle\cite{nakaharabook}, and the integral is taken over an arbitrary  oriented 2-cycle.}). We will also allow placing the theory on an arbitrary smooth oriented space-time manifold with metric $g$. Examining the partition function for arbitrary $(A,g)$ will allow us to distinguish phases based on the response to these probes. 

Consider the phase diagram as a function of the mass $m$. So long as $|m| \neq 0$ there is a gap in the spectrum. However the phase with $m > 0$ is {\em distinct } from the one with $m  < 0$. Taking the $m < 0$ phase\footnote{This choice can always be made by suitable UV regulation of the theory.} to be a trivial insulator the $m > 0$ phase will be a symmetry protected topological insulator. Thus the massless Dirac theory can be viewed as sitting at a quantum critical point between a trivial and a topological insulator. 

The topological distinction between the two phases can be understood physically by studying a domain wall in space where the mass $m$ changes sign. It is well known that at this domain wall there is a single massless Dirac fermion. This reveals that the phase for one sign of the mass is topological when the other is trivial. 

It will be extremely useful to us to establish this result in a more formal but powerful way (see Ref. \onlinecite{wittenFP} for a review). Consider the partition function of the free Dirac theory defined by the Euclidean path integral: 
\bea
Z[m; A,g] & =  & \int {\cal D}\psi e^{-\int d^4x \sqrt{g} \bar{\psi} (\slashed{D} + m) \psi }\\
& = & \text{Det}(\slashed{D} + m)
\eea
Here $D$ is a covariant derivative. As $\slashed{D}$ is anti-hermitian it has purely imaginary eigenvalues which we write $i\lambda_i$. Further as 
\be
\{i\slashed{D}, \gamma^5 \} = 0
\ee
for each non-zero eigenvalue $\lambda_i$ there is a partner $-\lambda_i$. Zero modes of the Dirac operator do not have to appear in pairs however. These zero modes can be chosen to have definite helicity, {\em i.e}, have $\gamma^5$ eigenvalues $\pm 1$. Let $N_\pm$ be the number of zero modes with helicity $\pm 1$ respectively.  The index of the Dirac operator is defined to be $J = N_+ - N_-$, and is a topological invariant (it cannot be changed by smooth deformations of $(A,g)$). Then the partition function can be written
\be
Z[m; A,g] = \left(\prod_{\lambda_i > 0} \left(\lambda_i^2 + m^2 \right) \right) (m^{N_+ + N_-})
\ee
Now consider the ratio of the partition functions of the theories with masses $+m$ and $-m$. Clearly 
\be
\label{freeDZrtio}
\frac{Z[m; A,g] }{Z[-m; A,g] }= (-1)^{N_+ + N_-} = (-1)^J=e^{i\pi J}
\ee
Thus the ratio of the partition functions is a topological invariant. Furthermore it is known \cite{nakaharabook} (by the Atiyah-Singer index theorem) that 
\be
J =\frac{1}{2} \int d^4x {F \over 2\pi} \wedge {F \over 2\pi} - {\sigma \over 8}
\ee
where $F = dA$ and $\sigma$ is an integer known as the signature of the space-time manifold. It may be expressed in terms of the Riemann curvature tensor: 
\be
\sigma = -{1 \over 24 \pi^2} \int d^4x\ \text{tr}( R \wedge R)
\label{sigmaRR}
\ee
Eq. \ref{freeDZrtio} thus gives exactly the right $\theta = \pi$ response of a topological insulator for one sign of mass if the other sign is chosen to be trivial. 

We note that the massless Dirac theory has extra symmetries absent in the massive case. For instance, we can write the Dirac fermion as two flavors of Weyl fermions and do a flavor rotation of the two Weyl fermions.  We will regard these symmetries as emergent symmetries of the critical point. These emergent symmetries have 't Hooft anomalies and we will discuss them later as needed. 

We can readily generalize the discussion above to $N$ free Dirac fermions, or equivalently $2N$ Majorana fernions with $SO(2N) \times Z_2^T$ symmetry. Taking the $m < 0$ theory to be trivial, the $m > 0$ theory will describe an SPT phase of fermions with $SO(2N) \times Z_2^T$ symmetry. This is established by calculating the partition function ratio in the presence of a background $SO(2N)$ gauge field $A^{SO(2N)}$ and metric $g$: 
\be
\frac{Z[m; A^{SO(2N)},g] }{Z[-m; A^{SO(2N)},g] }  =  (-1)^{J} 
\label{so2nptnrtio}
\ee
The index $J$ is a topological invariant related by the Atiyah-Singer theorem to $(A^{SO(2N)}, g)$ by
\be
2J =p_1(A^{SO(2N)}) -2N {\sigma \over 8}
\label{so2nindx}
\ee
where $p_1$ is the first Poyntryagin index of the $SO(2N)$ gauge field defined by
\beq
p_1(A^{SO(2N)})=\frac{1}{2}\int_{Y_4} \text{tr}_{SO(2N)} \left(\frac{F^{SO(2N)}}{2\pi}\wedge\frac{F^{SO(2N)}}{2\pi}\right)
\eeq
Therefore, $N$ massless free Dirac fermions can be viewed as the critical theory for the quantum phase transition between the trivial and SPT state of fermions with $SO(2N)\times Z_2^T$ symmetry.

\subsection{Massless $3+1$-D non-abelian gauge theories }
Consider next a generalization to $SU(2)$ gauge theories coupled to $N_f$ flavors of fermionic matter fields. We will study two distinct cases - (i)  matter fields  in the fundamental representation of $SU(2)$ and (ii)  matter fields in the adjoint representation. These two distinct cases correspond ``microscopically" to two very distinct kinds of physical situations.  When the matter fields are in the fundamental representation, all local ({\em i.e}, gauge invariant) operators in the theory (baryons, mesons, .....)
are bosons. We will therefore regard the gauge theory as the low energy theory of a UV theory of these gauge invariant bosons\footnote{This point of view is natural from a condensed matter perspective but may be unfamiliar to some high energy theorists. 
We will find it insightful to view the gauge theory this way. Note in particular that the fermionic matter fields, as well as the gauge fields themselves, are to be regarded as emergent from these UV `local' bosons. }. 
When the matter fields are in the adjoint representation, however, there are local operators that are fermions. We can view the theory as emerging from  a UV system of these fermions (see later for more detail).

The infrared behavior of $3+1$-D quantum chromodynamics with massless matter fields is an extremely important and intensively studied topic in particle physics. The renormalization group (RG) flow equation of the gauge coupling, for $SU(N_c)$ gauge theory with $N_f$ flavors of fermions\footnote{Through out this paper $N_f$ will count the number of Dirac fermions. In the literature however sometimes $N_f$ is used to denote the number of Weyl fermions.}  in the representation $R$, reads
\beq
\beta(g^2)=\frac{dg^2}{dl}=\beta_0(N_c,N_f,R)g^4+\beta_1(N_c,N_f,R)g^6+O(g^8)
\eeq
where $\beta_0$ and $\beta_1$ are functions that depends on $N_c, N_f$ and the representation $R$. For instance, if $R$ is the fundamental representation, the $\beta_0$ and $\beta_1$ are
\beqn
\beta_0&=&\frac{1}{8\pi^2}\frac{1}{3}(11N_c-2N_f)\\
\beta_1&=&\frac{1}{128\pi^2}(\frac{34}{3}N_c^2-\frac{1}{2}N_f(2\frac{N_c^2-1}{N_c}+\frac{20}{3}N_c))
\eeqn

\begin{figure}
  \centering
    \includegraphics[width=0.9\textwidth]{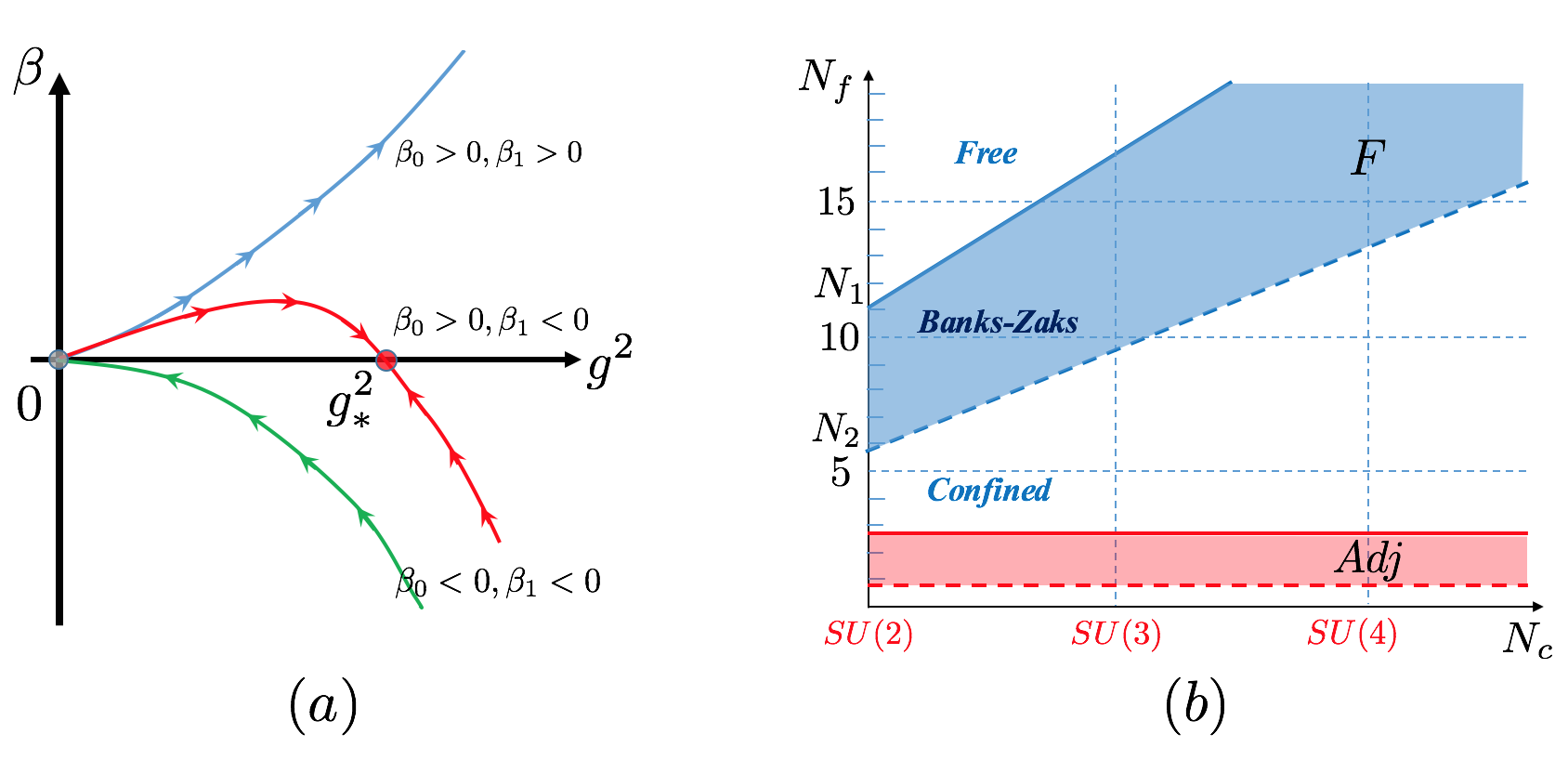}
      \caption{(a) demonstrates the renormalization group flow of the gauge coupling in three different regimes: 1. IR free (green curve); 2. Banks-Zaks fixed point (red curve), conformal; 3. IR confined (blue curve). (b) shows the conformal window for $SU(N_c)$ gauge theories with $N_f$ flavors of fundamental or adjoint fermion fields. The upper edge of the conformal window is sharply defined by the condition $\beta_0(N_c,N_f,R)=0$. The lower edge of the conformal window can only be determined through numerical simulations. Therefore, we should not take the numbers on the dotted line very seriously.}
      \label{BZ}
\end{figure}

Based on the RG equation, the IR phases of the gauge theory can be divided into three classes. Firstly, for $N_f$ bigger than a critical value $N_1(N_c,R)$, the leading term $\beta_0$ is negative ($\beta_1$ is usually also negative for such $N_f$) and gauge coupling $g^2$ flows towards zero under RG, if we start from a weak initial coupling. In the IR, the theory is free, namely decoupled gluons and free fermions. Secondly, for $N_f$ slightly smaller than the critical value $N_1$, $\beta_0$ is a small positive parameter. When we take into account the $g^6$ term in the RG equation, there's a stable fixed point controlled by $\epsilon$ at finite $g^2_*\sim O(\beta_0/|\beta_1|)$ for $\beta_1<0$. This is the famous Bank-Zaks fixed point\cite{BZprl,BZnpb}, which is an example of an interacting conformal field theory in $3+1$-D. As $N_f$ decreases further from $N_1$, in general $|\beta_1|$ decreases and $g_*^2$ becomes larger. Eventually, for $N_f$ approaching certain critical value $N_2(N_c,R)$, $|\beta_1|\rightarrow 0$ and the fixed point goes to infinity, in which case at low energy the gauge theory is belived to be in a confined phases. The RG flows of these three different regimes are summarized in Fig. (\ref{BZ}.a). Naively, the critical $N_2$ can be estimated by solving equation $\beta_1(N_c,N_f=N_2,R)=0$.  However, at that point, perturbative RG is far from a controlled limit. Therefore, the value of $N_2$ is usually determined through numerical calculations. The gauge theory is in the conformal window if $N_f\in (N_2, N_1)$. The conformal windows are confirmed in numerical studies for $SU(2)$ gauge theories with fundamental fermions and adjoint fermions\cite{BZfund,BZadj1,BZadj2}. For fundamental fermions, the conformal window of $SU(2)$ theory is around $8\sim11$. For adjoint fermions, the conformal window is around $1\sim 2$. One can find a plot for the conformal window of $SU(N_c)$ gauge theories in Fig. (\ref{BZ}.b). The IR behavior of $Sp(N_c)$ and $SO(N_c)$ gauge theories are similar to that of $SU(N_c)$ gauge theories. The corresponding conformal windows have also been discussed through various methods\cite{spso, spcft}.

\subsection{Summary of results}

The IR-free gauge theories and the Bank-Zaks fixed points are interesting examples of $3+1$-D conformal field theories. In this paper, we will  show how to interpret them as quantum critical points in the phase diagram of the ``microscopic" degrees of freedom of the system, similar to what we reviewed for the free massless Dirac fermion theories in the previous section.  Remarkably we will find that these theories can be viewed as deconfined quantum critical points for the underlying boson or fermion systems.  The gauge theory description emerges as  a useful one right at the critical point (and its vicinity) though the phases on either side only have conventional excitations ({\em i.e} those that can described simply in terms of the underlying bosons/fermions and their composites.).  In all cases we study, these massless gauge theories provide valuable examples of quantum critical points associated with phase transitions between trivial and Symmetry Protected Topological (SPT) phases of the underlying boson/fermion system.  

\begin{figure}
  \centering
    \includegraphics[width=0.9\textwidth]{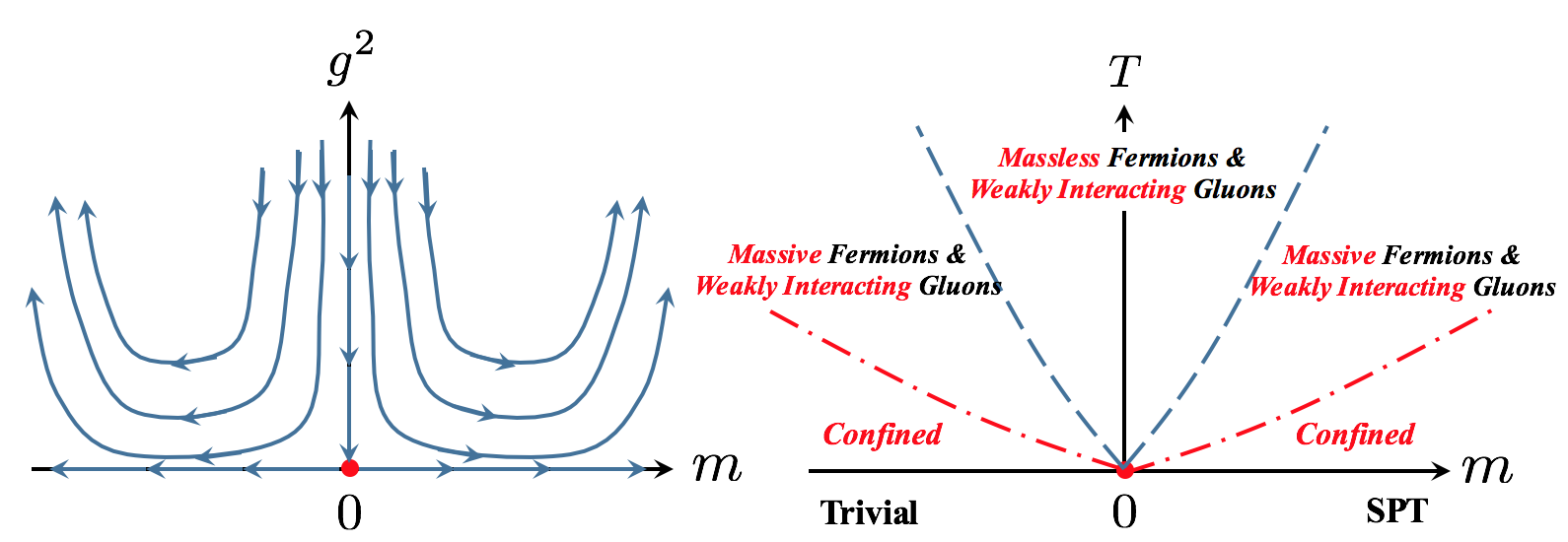}
      \caption{On the left is a schematic demonstration of renormalization flow in $g^2$-$m$ plane for large $N_f$ in the IR free case. The gauge coupling $g^2$ is a dangerous irrelevant operator for the $m=0$ critical point. On the right is the finite temperature phase diagram for the deconfined quantum phase transition. It features two interesting crossover scales. At temperature $T>m$ (or length scale $l<\xi\sim 1/m$), the physics is controlled by the critical point and the system has deconfined massless fermions with weakly interacting gluons. For temperature $m<T<m^y$ (or length scale $1/m<L<1/m^y$) with $y>1$ is a universal exponent, the system has deconfined but massive fermions and weakly interacting gluons. For temperature lower than $\sim m^y$ (or $L >1/m^y$), the gauge theory flows to strong coupling and the system is in a confined phase. }
      \label{dRG}
\end{figure}

Section III describes $3+1$-D deconfined quantum critical points for bosonic systems. In section III.A, we begin with $SU(2)$ gauge theory with $N_f$ fermions in the fundamental representation.  For simplicity we will restrict attention to $N_f$ even in this paper. We will consider the theory in the presence of an arbitrary mass $m$ that preserves the flavor symmetry. When $m \neq  0$ the theory flows, in the IR, to  massive phases. The $m = 0$ point will  correspond to a critical point. For general $m$ the global symmetry of the theory is\footnote{Our notation is $Sp(1)\cong SU(2)$, $PSp(N)=Sp(N)/Z_2$.} 
$PSp(N_f)\times Z_2^T$. We regard this gauge theory as the IR theory of a system of UV (gauge-invariant) bosons with  $PSp(N_f)\times Z_2^T$ global symmetry. 
To begin with consider $N_f$ large enough that the massless point is IR-free. Thus the gauge coupling $g^2$ flows to zero at the IR fixed point when $m = 0$. For any $m \neq 0$ however there is an induced effective action for the gauge field at low energies. The resulting pure $SU(2)$ gauge theory flows to strong coupling and will be confined at long length scales. In Fig. \ref{dRG} we sketch the expected RG flows for this theory in the $g^2, m$ plane for large $N_f$.  For even $N_f$ (the only case we consider)  the confinement results in a trivial vacuum.  Thus the massless IR-free fixed point separates two strongly coupled confined phases with trivial ground states. However we will see that these phases are potentially distinct Symmetry Protected Topological (SPT) phases of the underlying boson system with $PSp(N_f)\times Z_2^T$ global symmetry. Just like in the free Dirac fermion, the massless theory has extra symmetry: we will regard this as an emergent symmetry of the massless fixed point, and not as a fundamental symmetry.

Note that the RG flows show that the Yang-Mills coupling $g^2$ is ``dangerously irrelevant" in the vicinity of the massless fixed point. Naturally there are then two length scales that emerge in the vicinity of the critical point. There is a first length scale $\xi  \sim \frac{1}{m}$  associated with the mass of the gauge charged fermions. At this scale $g^2$ is still small.  Confinement does not set in till a much larger second length scale  $\xi_{conf} \sim \xi^{y}$ where $y > 1$ is a universal exponent\footnote{The precise value of $y$ is readily determined by matching the RG flow for the gauge coupling at the $m = 0$ fixed point with that of the pure gauge theory}. For $SU(N_c)$ gauge theory with $N_f$ fermion flavors in the fundamental  representation $y = \frac{2N_f}{11N_c}$.  Close to the critical point, at length scales smaller than $\xi$, the physics is that of the IR-free massless fixed point of the large-$N_f$ $SU(2)$ gauge theory. For length scales between $\xi$ and $\xi_{conf}$ the physics is that of massive fermions and massless gluons that are weakly interacting. Finally at the longest length scales $\gg \xi_{conf}$ the physics is that of the trivial ground state of the underlying boson system (but potentially in an SPT phase). These critical crossovers are also manifested at non-zero temperature as two distinct temperature scales (see Fig. (\ref{dRG})).

From a condensed matter perspective, consider SPT phases of systems of interacting bosons with $PSp(N_f)\times Z_2^T$ global symmetry with $N_f \in 2\mathbb{Z}$.   As we tune parameters in such a system we can drive phase transitions between the various  SPT phases.  From this point of view,  the   $SU(2)$ gauge theory coupled to $N_f$ flavors of massless Dirac fermions in the   fundamental representation  emerges as  a description of  the quantum critical point between trivial and SPT states of bosons. The $SU(2)$ gauge field  only appears at the critical point. For $N_f<8$, the $SU(2)$ gauge theory is believed to be in a confined phase at low energy. This implies that either the phase transition can be first order or there can exist an intermediate spontaneous symmetry breaking phase separating the two SPT states. For $N_f>10$, the gauge theory provides a description of continuous phase transition between the trivial and SPT state, where the critical point is free $SU(2)$ Yang-Mills theory with decoupled massless Dirac fermions. An interesting situation\footnote{We expect, in this case, that since the fixed point appears at relatively weak coupling, introducing a non-zero bare mass will still drive the system to a confined phase. In other words there is no intermediate phase that appears for small bare mass. This is an {\em assumption} which is reasonable for theories in the  conformal window which are ``close" to the free fixed point.  We will see later when we consider the gauge theory with light adjoint fermions that this assumption fails  for theories far away from the perturbative regime.}  is that, for $N_f=10$ and $8$, the phase transition can be described by the Bank-Zaks fixed point which is an interacting  conformal field theory in $3+1$-D. 

In section III.B, we find generalizations of the above construction. The phase transitions between $PSp(N_f)\times Z_2^T$ bosonic SPT states can also be described by $Sp(N_c)$ gauge theories coupled to $N_f$ fundamental massless Dirac fermion for any $N_c=4\mathbb{Z}+1$. The transition is continuous provided that $N_f$ is inside or above the conformal window of $Sp(N_c)$ gauge theories. These theories are weakly dual to the $SU(2)$ gauge theory described above  in the sense that they are distinct low energy descriptions of the same underlying UV physical system (in our case bosons with global $PSp(N_f) \times Z_2^T$ symmetry). Furthermore they describe the same phases and phase transition of this system. However, clearly the theories with fixed $N_f$ and different $N_c$ are truly distinct conformal field theories.  First they clearly have different numbers of low energy massless fields - this may be formalized by computing their $a$-coefficients which are clearly different for these different theories. Further the emergent symmetries (and their 't Hooft anomalies) of these theories at the massless point are different. Thus these theories provide  valuable examples where the same continuous phase transition admits multiple distinct universality classes, controlled by distinct fixed points.  The IR theories are not dual in a strong sense and are distinct conformal field theories.

In section III.C, we discuss an interesting phenomenon which we call unnecessary phase transitions. We define unnecessary phase transition as {\em generic continuous} phase transitions {\em within} the same phase. We provide several explicit examples for this phenomenon. The first example is a bosonic system with $PSp(N_f)\times Z_2^T$ symmetry at $N_f=4\mathbb{Z}$.  We show that there can be a generic continuous phase transition {\em inside} the topologically trivial phase of this bosonic system. The critical theory is an emergent $Sp(N_c)$ gauge theory at $N_c=4\mathbb{Z}$ with $N_f=4\mathbb{Z}$ massless fundamental fermions. As the phases on the two sides of this critical point are identical, the transition can be bypassed by some symmetric path in the whole parameter space. However, the transition is locally stable. We give another example which do not involve emergent gauge fields. We consider 16 copies of topological superconductor in DIII class with an additional $SO(2)\times SO(7)$ global symmetry. In the topologically trivial phase of this system, there can exist a generic second order transition characterized by 16 gapless free Majorana fermions in $3+1$-D. The transition can be circumvented by adding strong interaction. In condensed matter physics, it is common that two phases separated by a discontinuous ({\em i.e.} first order, as for the liquid-gas transition) phase transition can actually be the same phase. The examples in this section teach us that even a generic continuous phase transition does not necessarily change the nature of the state. 

Sections \ref{fdqcp} and \ref{nfaone} contain  examples of deconfined quantum critical points in fermionic systems for which there are very few previous examples. We study $3+1$-D fermionic deconfined quantum critical points that can be formulated as an $SU(2)$ gauge theory coupled to $N_f^A$ flavors of adjoint Dirac fermions. This theory has local fermion operators (baryons) and we will therefore regard it as a low energy theory of a microscopic system of these local fermions. However, to enable this point of view we need to  augment the theory by including a massive spin-$\frac{1}{2}$ (under the $SU(2)$ gauge transformation) scalar particle in our spectrum. Otherwise the theory has physical loop degrees of freedom corresponding to `electric' field lines in the spin-$1/2$ representation\footnote{A formal but very useful description is to say that the $SU(2)$ gauge theory with adjoint matter but no fundamental matter has a global $Z_2$ $1$-form symmetry (denoted $(Z_2)_1$). Of course a microscopic condensed matter system of fermions has no such $1$-form symmetry. Therefore we allow for an explicit breaking of the $(Z_2)_1$ symmetry by including the massive spin-$1/2$ scalar.}. We call this massive spin-$\frac{1}{2}$ scalar the spectator field. To complete the theory, we need to   specify its symmetry quantum numbers under the global symmetry, especially its time reversal properties\footnote{From a formal point of view this corresponds to how to define the theory on non-orientable manifolds.}.   The adjoint $SU(2)$ theory can actually describe different quantum phase transitions depending on the time reversal symmetry properties of the spectator field.

For $N_f^A>2$, the massless theory is free in the infrared limit. This theory, by tuning the fermion mass $m$, describes a quantum phase transition between a trivial and SPT state protected by the global symmetry which is $SO(2N_f^A)\times Z_2^T$. We first discuss the fermion SPT classification for this symmetry. For example for $N_f^A \in 2\mathbb{Z} + 1$, we show that it is $\mathbb{Z}_8 \times \mathbb{Z}_2$, generalizing the known results\cite{senthilDIII, YouXuintclass} for $SO(2) \times Z_2^T$ symmetry (known in the condensed matter literature as a class AIII topological superconductor.)  This means that such systems form distinct SPT states labelled by a pair of integers $(n, \eta)$ where $n = 0,1.,....,7$ mod 8, and $\eta = 0,1$ mod 2.  Phases with $\eta = 0$ are accessible within free fermion band theory. The IR-free massless gauge theory with $N_f^A > 2$  sits at the critical point between two such SPT phases. A subtlety arises with the time reversal properties of the theory. The precise SPT phase is changed depending on the symmetry properties of the massive spectator field. With one choice of spectator field, it describes the phase transition between the $n=0$ (trivial) state and the $(n=3,\eta=0)$ SPT state. This is a quantum phase transition that is not generically second order in the free fermion system where $n$ can only jump by $1$.  Thus this is an example of an interaction-driven band-theory-forbidden quantum critical point between two band insulators.  With a different choice of the spectator field, the adjoint $SU(2)$ theory can describe the phase transition between the trivial state and $(n=-1, \eta = 0 )$ SPT state. This transition can also occur within band theory where it is described by a free Dirac theory of physical fermions. The gauge theory however yields a {\em distinct} fixed point for the same transition. This is yet another example of multiple universality classes for the same phase transition. For $N_f^A\in2\mathbb{Z}$, the $m>0$ phase does not depend on the choice of spectator field. 

If we banish the fundamental scalars from the spectrum, at the IR-free massless point, the $1$-form $(Z_2)_1$ symmetry is spontaneously broken. Turning on a small mass to the fermions confines the symmetry and restores the $(Z_2)_1$ symmetry. In other words electric loops in the spin-$1/2$ representation are tension-full in the massive phase. These loops are decoupled from the physical excitations of this phase (which are the local fermions). Now if we re-introduce the fundamental scalars, they will have no effect on the low energy properties at the critical point. However in the massive phase the scalars allow the loops to break. At the same time they also affect the SPT characterization of the phase.  

In Sec. \ref{nfaone} we consider the interesting case $N_f^A = 1$ (augmented as above with a spectator fundamental scalar). This describes the familiar system of fermions with $SO(2) \times Z_2^T$ symmetry (the class AIII topological superconductor).  This is an asymptotically free theory and there is some numerical evidence that it flows to a CFT in the IR\cite{BZadj1}. We will therefore first consider the fate of this theory in the presence of a large mass (of either sign) when trivial confined phases will indeed result.  The precise  SPT identification of these massive phases depends on the symmetry realization on the spectator boson in exactly the same way as for general $N_f^A \in 2\mathbb{Z} + 1$.   In contrast to the previous examples, here the gauge theory description of the massless point is strongly coupled. In Sec. \ref{4ddrcdual} we explore the possibility that the low energy theory consists of a free Dirac fermion together with a decoupled topological field theory.  This may be viewed as a duality of the $SU(2)$ gauge theory with $N_f^A = 1$ adjoint Dirac fermions and the theory of a free Dirac fermion augmented with a decoupled topological field theory\footnote{A somewhat similar duality in $2+1$-D was proposed\cite{Gomis2017} recently for $SU(2)$ gauge theory with $N_f = 1/2$ adjoint fermions ({\em i.e} with a single Majorana fermion in the adjoint representation. The IR theory was argued to consist of a free massless Majorana fermion augmented with a decoupled topological theory.}. The latter is needed to be able to match all the anomalies of the theory (in the absence of the spectator field) identified recently in Ref. \onlinecite{ThomasSU(2)}.  We discuss physical properties of this topological order. We will show that the free massless Dirac + topological theory has the same local operators and  the same global symmetries (both exact and emergent), and further enable matching all 't Hooft anomalies of the emergent symmetries.  While these checks are necessary to claim a duality they are not sufficient as a proof.  A small mass in the gauge theory will map to a small mass of the physical Dirac fermions of the IR theory but will not  destroy  the extra topological order. This leads to a situation where between the two large mass insulators there is an intermediate phase which has an additional topologically ordered sector.

Several details are in the Appendices. In particular we present some simple models - not involving emergent gauge fields - for some of the phenomena depicted in Fig. \ref{summary}.  We also briefly discuss the fate of $SU(2)$ gauge theory coupled to arbitrary $N_f^A$ flavors of adjoint Dirac fermions. 

\section{Bosonic Deconfined Critical Points in $3+1$-D}
\label{bdqcp}

In this section, we study quantum phase transitions between trivial and SPT phases in $3+1$-D systems of interacting bosons. The critical theories we construct   for such transitions  resemble the features of deconfined quantum phase transitions in $2+1$-D\cite{2dDQCP1,2dDQCP2}.  In particular,  the critical point has emergent non-abelian deconfined gauge fields and associated `fractionalized' matter fields.  To get an understanding of certain phase transition, it is often helpful to firstly identify the nature of the two nearby phases, which provide crucial information about the critical fluctuations at the phase transition. Here however we pursue a reversed logic by asking the following question: given some deconfined gauge theory in $3+1$-D, what phase transition can this theory describe? To complete the phase diagram, we will start from the deconfined gauge theory and then identify its nearby gapped phases by perturbing the theory with a relevant perturbation. 

\subsection{$SU(2)$ gauge theory with $N_f\in2\mathbb{Z}$ fundamental fermions}
\label{su2nf}

Consider $SU(2)$ gauge theory with $N_f$ Dirac fermions in the fundamental representation. We will label it $SU(2)+N_f^F$ theory. A key observation is that in this theory all local ({\em i.e}, $SU(2)$ gauge invariant)  operators are bosonic because they are composed of even number of fundamental fermions\footnote{From a formal point of view, despite the presence of fermionic matter fields,  this theory can be defined on non-spin manifolds by choosing gauge bundles  in $\frac{SU(2) _g\times Spin(4)}{Z_2}$ .  On a non-spin manifold we require $w_2(SO(3)_g) = w_2(TY_4)$ mod 2, where the left side is the second Stiefel-Whitney class of the $SO(3)$ gauge bundle and the right side is the second Stiefel-Whitney class of the tangent bundle $TY_4$ of the 4-manifold $Y_4$. That the theory can be so defined on a non-spin manifold without imposing any conditions on bundles for background gauge field is  an alternate way of seeing that the  theory  describes a physical system of bosons.}.Therefore, the theory describes a phase transition in a purely bosonic system. 

A relevant perturbation that can drive the massless theory away from the critical point is the Dirac mass term that is uniform for all flavors. 
\beq
\mathcal{L}_{QCD}=\sum_{i=1}^{N_f}i\bar{\psi}_i\gamma_\mu D_\mu\psi_i-m\bar{\psi}_i\psi_i
\label{qcddirac}
\eeq
We first show that both $m<0$ and $m>0$ phases (at least for large $|m|$) are trivial  gapped phases if $N_f\in 2\mathbb{Z}$. Let us assume that in the $m<0$ phase integrating out the massive fermions generates a trivial $\Theta$-term for the $SU(2)$ gauge theory. This is always possible by certain UV regularization. Then on the $m>0$ side, the massive fermions contribute a $\Theta$-term for the $SU(2)$ gauge field at $\Theta=\pi N_f$. With the condition $N_f\in2\mathbb{Z}$, both phases have trivial $SU(2)$ $\Theta$-terms because of the $2\pi$ periodicity of the $\Theta$-angle. Therefore, the $SU(2)$ gauge theory enters a trivial confined phase at low energy and the system has gapped spectrum in both cases. Importantly it is believed that  when pure $SU(2)$ gauge theory confines the resulting ground state is also topologically trivial: there is a unique ground state on all spatial manifolds. In condensed matter parlance, we expect a ``Short Ranged Entangled" (SRE) ground state\cite{Wengroupcohomology,Wenscience}. In contrast, if $N_f\in 2\mathbb{Z}+1$, we have an $SU(2)$ gauge theory with $\Theta=\pi$ for the $m>0$ phase.  The dynamics of this gauge theory  is nontrivial at low energy\cite{zohar2017}, and the ground state likely has long range entanglement. To keep things simple in this paper we will henceforth focus on the case $N_f \in 2\mathbb{Z}$. 

With $N_f\in 2\mathbb{Z}$, by tuning the uniform Dirac mass from negative to positive, the system goes between two gapped phases through a quantum phase transition, which is described by the massless $SU(2)+N_f^F$ theory. For large enough $N_f$, the IR physics of the $SU(2)+N_f^F$ theory is either free or controlled by the Bank-Zaks fixed point. Therefore, it describes a continuous phase transition. In the following, we will explain that the $SU(2)+N_f^F$ theory with uniform Dirac mass has $PSp(N_f)\times Z_2^T$ symmetry. With this global symmetry, the uniform Dirac mass is the only symmetry allowed relevant perturbation at the critical point. The $m<0$ and $m>0$ phases are the trivial and the symmetry protected topological phases of this global symmetry, respectively.

In order to illustrate the global symmetry in an explicit way, let us construct the $SU(2)+N_f^F$ theory in a more systematic way. First, we consider $4N_f$ flavors of Majorana fermions in $3+1$-D
\beq
\mathcal{L}_0=\sum_{j=1}^{4N_f}i\bar{\chi}_j\gamma_{\mu}\partial_\mu\chi_j-m\bar{\chi}_j\chi_j,
\label{so4n}
\eeq
with $\{\gamma_0,\gamma_1,\gamma_2,\gamma_3\}=\{\sigma^{12},i\sigma^{03},-i\sigma^{22},-i\sigma^{01}\}$ and $\bar{\chi}=\chi^T\gamma_0$. $\gamma_5=i\gamma_0\gamma_1\gamma_2\gamma_3=\sigma^{32}$. ($\sigma^{ij}$ is the short notation of $\sigma^i\otimes\sigma^j$.) At this stage, the system has an $SO(4N_f)$ flavor symmetry and time reversal symmetry $Z_2^T$, whose actions on the Majorana fields are as the following.
\beqn
SO(4N_f)&:& \chi_i\rightarrow O_{ij}\chi_j \\
\mathcal{T}&:&\chi_i(\vec{x},t) \rightarrow \gamma_0\gamma_5\ \chi_i(\vec{x},-t)=-i\sigma^{20}\chi_i(\vec{x},-t),\ \ i\rightarrow -i
\eeqn
It is  easy to check that the $SO(4N_f)$ and $Z_2^T$ symmetry\footnote{This is usually denoted as $\mathcal{CT}$ symmetry in the literature because in the Dirac fermion representation the $Z_2^T$ symmetry also flip the charge. The theory also has the usual time reversal symmetry $\mathcal{T}$ and parity symmetry $\mathcal{P}$. However, they are not relevant to our constructions.} commute with each other and $\mathcal{T}^2=(-1)^F$.  Next, we will gauge a diagonal $SU(2)$ subgroup of the flavor symmetry. To specify the $SU(2)$ subgroup, we reorganize the fermion fields into a matrix form\cite{2dDQCP2}. Let us  split the Majorana flavor index into two indices, namely labeling the Majorana fields as $\chi_{v,j}$ with $v=1,2,..,N_f$ and $j=0,1,2,3$. The matrix fermion fields are defined as follows,  
\beq
X_{v,\alpha;\beta}=\frac{1}{\sqrt{2}}(\chi_{v,0}\mathbb{I}_{\alpha\beta}+i\sum_{\mu=1}^{3}\chi_{v,\mu}\sigma^{\mu}_{\alpha\beta})
\eeq
where $\sigma^\mu$'s are pauli matrices and $\alpha,\beta=1,2$. This step can be viewed as combining four real fields into one quaternion field. The theory written in terms of $X$ is manifestly invariant under right $SU(2)$ rotation and a left unitary rotation. 
\beq
X\rightarrow LXR_{SU(2)}
\eeq
The left rotation $L$ must satisfy the reality condition of Majorana fermions. As a result, $L$ actually belongs to $Sp(N_f)$ group. It turns out the $Sp(N_f)$ group is the maximal symmetry group that commutes with the $SU(2)$. They share the same center symmetry, namely $SO(4N_f)\supset\frac{SU(2)\times Sp(N_f)}{Z_2}$. We now gauge the $SU(2)$ symmetry and get our $SU(2)+N_f^F$ theory
\beq
\mathcal{L}_{QCD}=\text{tr}(i\bar{X}\gamma_\mu D_\mu X-m\bar{X}X),
\eeq
where $\bar{X}=X^\dagger\gamma_0$. We can map this formulation back to the complex Dirac fermions in Eq. (\ref{qcddirac}) by $\psi_{\alpha,i}=i\sigma^y_{\alpha,\beta}X_{1,i;\beta}$, where $\alpha$ is the $SU(2)$ spin index, $i$ represents the flavor index.  

The global symmetry after gauging the $SU(2)$ subgroup is manifestly $\mathcal{G}=PSp(N_f)\times Z_2^T$. One can check that with this global symmetry the uniform Dirac mass is the only allowed mass term. For example, the $im\bar{X}\gamma_5X$ mass is not time reversal invariant. Any mass term of the form $\bar{\chi}_iS_{ij}\chi_j$ or $i\bar{\chi}_i\gamma_5S_{ij}\chi_j$, with $S_{ij}=S_{ji}$, is not invariant under $PSp(N_f)$ rotation. 

In the two gapped phases, on any closed spatial manifold, the system has non-degenerate ground state and no spontaneous symmetry breaking. The distinction of the two phases can only come from their topological properties. They can be different Symmetry Protected Topological phases of the global symmetry $\mathcal{G}$. Let us assume $m<0$ phase is the trivial disordered phase under this symmetry. Now we want to understand the nature of the $m>0$ phase. The strategy is to couple background gauge fields of the global symmetry $PSp(N_f)$ to the system and identify its topological response that is a signature of the SPT state. To achieve this, we will firstly turn on a background gauge field for the whole $SO(4N_f)$ flavor group and find its topological response. Then we will reduce the response theory down to its $SU(2)$ and $PSp(N_f)$ subgroups. 

Let us start from Eq. (\ref{so4n}) and turn on a background $SO(4N_f)$ gauge field $A^{SO(4N_f)}$. We consider the response to the $SO(4N_f)$ gauge field after integrating out the massive fermions. To cancel dynamical contributions to the partition function, we calculate the ratio between the Euclidean partition functions with $m<0$ and $m>0$. From Eqns. \ref{so2nptnrtio} and \ref{so2nindx}, we get a purely topological response. 
\beq
\mathcal{S}_{topo}=Log\left\{\frac{\mathcal{Z}[m<0,A,g]}{\mathcal{Z}[m>0,A,g]}\right\}=i\frac{\pi}{2}\left(p_1(A^{SO(4N_f)})-\frac{N_f}{2}\sigma\right)
\eeq
The topological action contains the $\Theta$-terms of $SO(4N_f)$ gauge field in terms of the first Pontryagin class $p_1$ and the gravitational $\Theta$-term (written in terms of the manifold signature $\sigma$ - see Eqn. \ref{sigmaRR}).
\beq
p_1(A^{SO(4N_f)})=\frac{1}{2}\int_{Y_4} \text{tr}_{SO(4N_f)} \left(\frac{F^{SO(4N_f)}}{2\pi}\wedge\frac{F^{SO(4N_f)}}{2\pi}\right)=2l_{SO(4N_f)}
\eeq
The Pontryagin class is equal to twice of the instanton number of $SO(4N_f)$ gauge field. More details about the definition for the Pontryagin class and instanton number are given in the Appendix \ref{appinstanton}. 

We restrict the $SO(4N_f)$ to particular configurations which have seperate $Sp(N_f)$ and $SU(2)$ gauge fields.
\beqn
p_1(A^{SO(4N_f)})&=&2p_1(A^{Sp(N_f)})+2N_fp_1(a^{SU(2)_g})\\
&=&2l_{Sp(N_f)}+2N_f l_{SU(2)}\\
&=&2l_{PSp(N_f)}+2N_f l_{SO(3)}\\
&=&\mathcal{P}(w_2^{PSp(N_f)})+2w_4^{PSp(N_f)}+\frac{N_f}{2}\mathcal{P}(w_2^{SO(3)}) \ \ \text{mod}\ 4
\label{topoact}
\eeqn
where $l$ represents the instanton number for the gauge bundle, $\mathcal{P}(a)$ is the Pontryagin square operator (for a definition see Ref. \onlinecite{akpsq,2dDQCP2} and references therein), $w_2$ and $w_4$ are the second and fourth Stiefel-Whitney classes\cite{nakaharabook}. Here we have used the following relations between the instanton numbers and characteristic classes for the vector bundles\cite{seiberg2013,2dDQCP2}.
\beqn
2l_{PSp(N_f)}&=&\mathcal{P}(w_2^{PSp(N_f)})+2w_4^{PSp(N_f)}\ \text{mod}\ 4\ \text{for}\ N_f\in2\mathbb{Z} \\
4l_{SO(3)}&=&\mathcal{P}(w_2^{SO(3)})\ \text{mod}\ 4
\eeqn

Since our fermion transforms projectively under the $SO(3)$ and $PSp(N_f)$ gauge bundle, in order for the theory to be consistently defined on any manifold with or without spin structure, we should impose the following constraint on the gauge bundles. 
\beq
w_2^{SO(3)}+w_2^{PSp(N_f)}+w_2^{TY_4}=0\  \text{mod}\ 2
\label{w2relation}
\eeq
This is the obstruction free condition to lift a $SO(3)\times PSp(N_f)\times SO(4)^{TY_4}$ bundle to $(SU(2)\times Sp(N_f)\times Spin(4)^{TY_4})/(\mathbb{Z}_2\times \mathbb{Z}_2)$. Based on this relation and the following few useful identities (for references, see Wang et al.\cite{2dDQCP2}), 
\beqn
\mathcal{P}(a+b)&=&\mathcal{P}(a)+\mathcal{P}(b)+2a\cup b\  \text{mod}\ 4\label{relation1} \\
\mathcal{P}(a)&=&a\cup a\ \text{mod}\ 2 \\
a\cup w_2^{TY_4}&=&a\cup a \text{  for any $a\in H^2(Z_2)$} \\
\int_{Y_4}\mathcal{P}(w_2^{TY_4})&=&\sigma \ \text{mod} \ 4 
\label{relation4}
\eeqn
we can simplify the response theory in Eq. (\ref{topoact}). There are four types of response theories depending on $N_f/2=k$ mod $4$. 

\begin{description} 

\item[$\bullet$] $k=0$ mod $4$, $e.g.$ $N_f=8, 16, 24,...$
\beq
S_{topo}^{k=1}=i\pi l_{PSp(N_f)}
\eeq
This is the usual $\Theta-$term for the $PSp(N_f)$ gauge field, and the value of $\Theta=\pi$ is protected by $Z_2^T$ symmetry.
\item[$\bullet$] $k=1$ mod $4$, $e.g.$ $N_f=2, 10, 18,...$
\beq
S_{topo}^{k=1}=i\pi w_4^{PSp(N_f)}
\eeq
This topological term is robust against $Z_2^T$ breaking because $w_4$ is a $\mathbb{Z}_2$ class. However, if $Z_2^T$ is broken, the $\bar{\psi}i\gamma_5\psi$ mass is also allowed at the critical point. Therefore, the $Z_2^T$ symmetry must be preserved in order to have a generic second order transition. 

\item[$\bullet$] $k=2$ mod $4$, $e.g.$ $N_f=4, 12, 20,...$
\beq
S_{topo}^{k=2}=i\pi l_{PSp(N_f)}+i\pi w_2^{PSp(N_f)}\cup w_2^{PSp(N_f)}
\eeq 
The first term is the $\Theta$-term for the $PSp(N_f)$ gauge fields, which requires $Z_2^T$ symmetry to be stable. The second term is an independent topological term that can be non-trivial on a non-spin manifold. The second term is a $\mathbb{Z}_2$ class and hence is stable against $Z_2^T$ breaking. 

\item[$\bullet$] $k=3$ mod $4$, $e.g.$ $N_f=6, 14, 22,...$
\beq
S_{topo}^{k=3}=i\pi w_4^{PSp(N_f)}+i\pi w_2^{PSp(N_f)}\cup w_2^{PSp(N_f)}
\eeq 

Both terms are stable against $Z_2^T$ symmetry breaking.

\end{description}

Numerically, the conformal window for $SU(2)+N_f^F$ theory is $N_f\sim 6-11$. For $N_f>11$, the theory is free. Therefore, we have many examples of $3+1$-D deconfined quantum phase transitions, which are described by free $SU(2)+N_f^F$ theory, between the trivial and the $PSp(N_f)\times Z_2^T$ SPT state, for even $N_f>11$. Assuming further that for $N_f = 8, 10$, a small mass drives the Banks-Zaks  theories to the large mass fixed points, we have  two explicit examples of $3+1$-D DQCP, which are described by strongly interacting CFTs. They separate  trivial and the $PSp(N_f)\times Z_2^T$ bosonic SPT states.

\subsection{Multiple universality classes}
\label{multiuni} 
\begin{figure}
  \centering
    \includegraphics[width=0.7\textwidth]{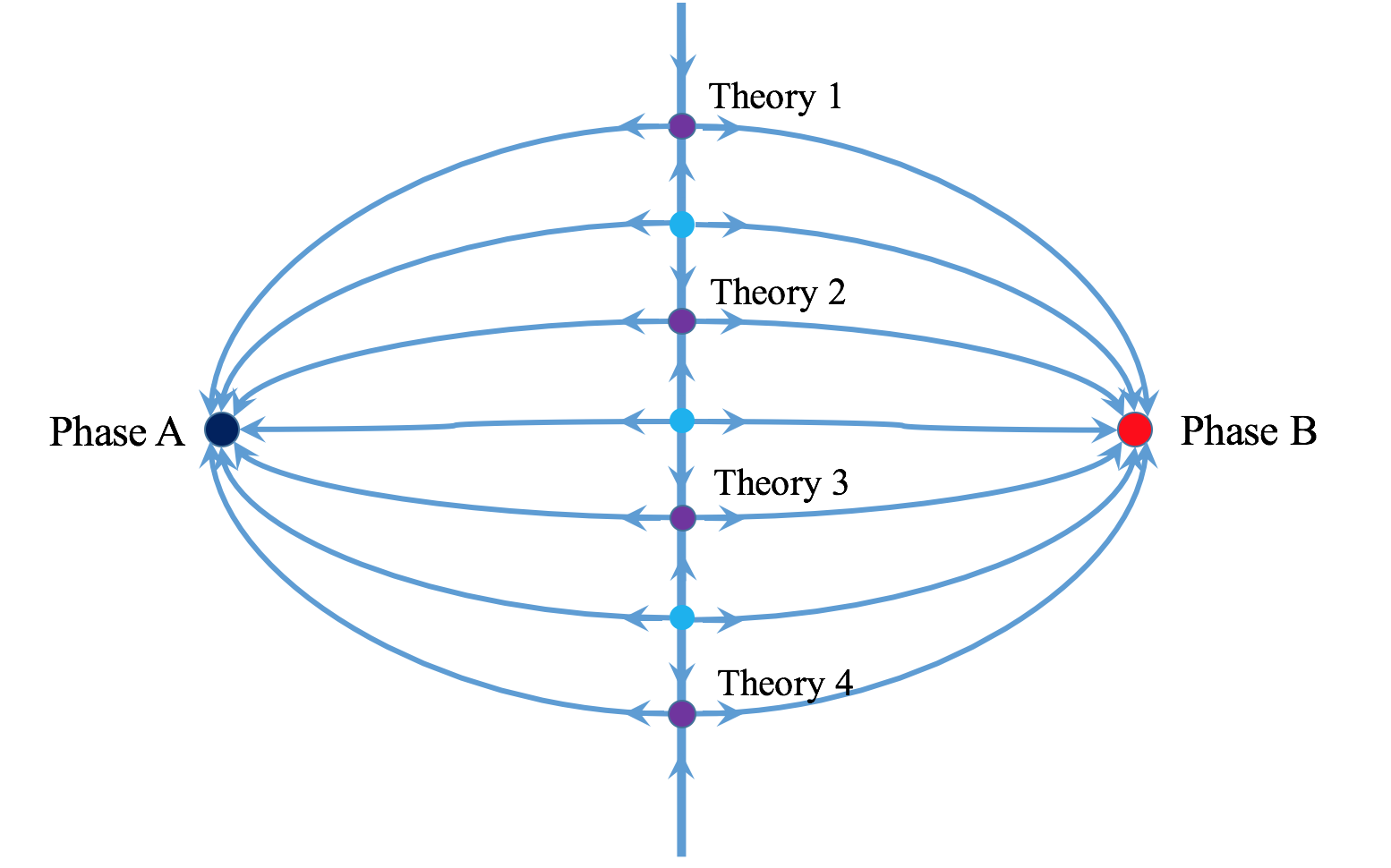}
      \caption{A schematic renormalization flow diagram for degenerate quantum critical points. }
      \label{degenerate}
\end{figure}

In this section\footnote{We thank Nathan Seiberg for a crucial discussion that directed us to the results of this section.}, we demonstrate that the phase transition between the trivial and SPT state protected by $PSp(N_f)\times Z_2^T$ symmetry can have many descriptions which are distinct from each other. A schematic renormalization flow diagram is shown in Fig. (\ref{degenerate}). In practice, such situation, although not forbidden by any physical principle, is not commonly observed in critical phenomena. It is interesting that here we can show such an example explicitly in a controlled way. 

To introduce these different transition theories, we consider a generalization of our previous construction of $3+1$-D bosonic DQCP. We start with $4N_cN_f$ flavors of Majorana fermion in $3+1$-D. The total flavor symmetry is $SO(4N_cN_f)$. 
\beq
\mathcal{L}_0=\sum_{j=1}^{4N_cN_f}\bar{\chi}_j(i\gamma_{\mu}\partial_\mu-m)\chi_j 
\eeq
There is a well known group decomposition for the $SO(4N_cN_f)$ group.
\beq
SO(4N_cN_f)\supset \frac{Sp(N_c)\times Sp(N_f)}{Z_2}
\eeq
We can understand the general group decomposition intuitively as follows. First, we use 4 real fermions to form a quaternion fermion. Then we arrange the $N_fN_c$ quaternion fermions into a $N_f\times N_c$ quaternion matrix fermion field $\mathcal{X}$. The $Sp(N_f)$ transformation can be packed into a $N_f\times N_f$ quaternion matrix $\mathcal{L}$ and it has a natural action on a $N_f$ dimensional quaternion vector. So the $Sp(N_f)$ action on the $\mathcal{X}$ field is the left multiplication on the $\mathcal{X}$ matrix, namely $\mathcal{X}\rightarrow \mathcal{L}\mathcal{X}$. Similarly, $Sp(N_c)$ action is the right multiplication on $\mathcal{X}$ by a $N_c\times N_c$ quaternion matrix $\mathcal{R}$, namely $\mathcal{X}\rightarrow \mathcal{X}\mathcal{R}$.\footnote{For more details, see for example the appendix in \cite{MengSpn}.}  The group decomposition we used in the previous section is a special case with $N_c=1$ and $N_f=2$.

Let us gauge the $Sp(N_c)$ part of the flavor symmetry. The result is an $Sp(N_c)$ gauge theory with $N_f$ fundamental fermions, which we label as $Sp(N_c)+N_f^F$ theory. 
\beq
\mathcal{L}_{QCD}=\text{tr}(i\bar{\mathcal{X}}\gamma_\mu D_\mu \mathcal{X}-m\bar{\mathcal{X}}\mathcal{X}),
\eeq
The global symmetry of this theory is again $PSp(N_f)\times Z_2^T$. Notice the global symmetry only depends on $N_f$ but not on $N_c$. Next, we need to identify the nature of $m<0$ and $m>0$ phases by their topological response to the background field of the global $PSp(N_f)$ symmetry. After integrating out fermions, we get the following topological action for $m>0$ phase.
\beqn
\mathcal{S}_{topo}&=&i\frac{\pi}{2}\left(2l_{SO(4N_cN_f)}-\frac{4N_cN_f}{8}\sigma\right) \\
&=&i\frac{\pi}{2}\left(2N_cl_{PSp(N_f)}+2N_fl_{PSp(N_c)}-\frac{N_cN_f}{2}\sigma\right)
\eeqn
The instanton numbers have the following algebraic relations with the Stiefel-Whitney classes\cite{seiberg2013}. For $n\in \mathbb{Z}$,
\beqn
4l_{PSp(2n+1)}&=&\mathcal{P}(w_2^{PSp(2n+1)})\ \  \text{mod}\ \  4 \\
2l_{PSp(2n)}&=&\mathcal{P}(w_2^{PSp(2n)})+2w_4^{PSp(2n)}\ \  \text{mod}\ \  4
\eeqn
Let's consider a case in which $N_f=2p, p\in \mathbb{Z}$, $N_c=4q+1, q\in\mathbb{Z}$. With the above relations, we can simplify the topological action.
\beqn
\mathcal{S}_{topo}&=&i\frac{\pi}{2}\left\{2N_cl_{PSp(N_f)}+2N_fl_{PSp(N_c)}-\frac{N_cN_f}{2}\sigma\right\} \\
&=&i\frac{\pi}{2}\left\{(4q+1)(\mathcal{P}(w_2^{PSp(N_f)})+2w_4^{PSp(N_f)})+p\mathcal{P}(w_2^{PSp(N_c)})-p(4q+1)\sigma\right\} \\
&=&i\frac{\pi}{2}\left\{\mathcal{P}(w_2^{PSp(N_f)})+2w_4^{PSp(N_f)}+p\mathcal{P}(w_2^{PSp(N_c)})-p\sigma\right\} 
\eeqn
If $p\in4\mathbb{Z}+1$, namely $N_f\in8\mathbb{Z}+2$, we get 
\beqn
\mathcal{S}_{topo}
&=&i\frac{\pi}{2}\left\{\mathcal{P}(w_2^{PSp(N_f)})+2w_4^{PSp(N_f)}+\mathcal{P}(w_2^{PSp(N_c)})-\sigma\right\} \\
&=&i\pi w_4^{PSp(N_f)}+i\frac{\pi}{2}\left\{\mathcal{P}(w_2^{PSp(N_f)})+\mathcal{P}(w_2^{PSp(N_c)})-\sigma\right\}\label{topoaction}
\eeqn
There is the following consistency relation for the gauge and tangent bundles, which is the analog of Eq. (\ref{w2relation}). 
\beq
w_2^{PSp(N_f)}+w_2^{PSp(N_c)}+w_2^{TY}=0\ \ \text{mod} \ \ 2
\eeq
We can prove the second term in Eq. (\ref{topoaction}) vanishes mod 4. 
\beqn
\mathcal{P}(w_2^{PSp(N_f)})+\mathcal{P}(w_2^{PSp(N_c)})-\sigma=2\mathcal{P}(w_2^{PSp(N_c)})+2w_2^{PSp(N_c)}\cup w_2^{TY}+\mathcal{P}(w_2^{TY})-\sigma=0 \ \text{mod}\ 4
\eeqn
In the derivation, we have again used relations in Eq. (\ref{relation1}-\ref{relation4}) to simplify the result. In the end, the topological response for the background $PSp(N_f)$ gauge field is quite simple and farmilar. The topological action reads
\beq
\mathcal{S}_{topo}=i\pi w_4^{PSp(N_f)}\ \ \text{for}\ \  N_c\in 4\mathbb{Z}+1, N_f\in 8\mathbb{Z}+2.
\eeq

\begin{figure}
  \centering
    \includegraphics[width=0.6\textwidth]{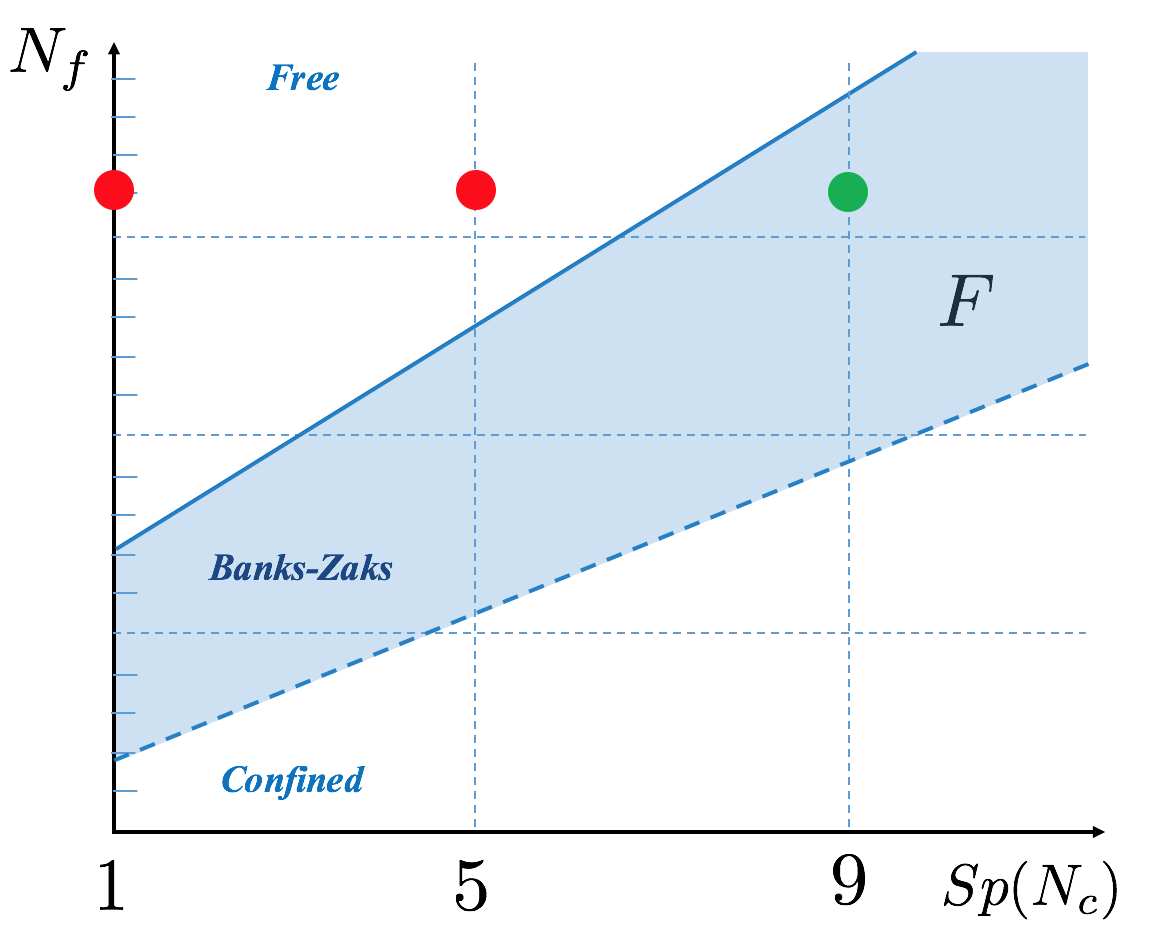}
      \caption{A sketch for the conformal window of $Sp(N_c)$ gauge theories (numbers on the $N_f$ axis are not precies). The red and green dots are different gauge theories. The red ones are free and the green one is strongly interacting. However, they all describe the topological phase transition from the trivial state to the same $PSp(N_f)\times Z_2^T$ bosonic symmetry protected topological phase.  }
      \label{spbz}
\end{figure}

One interesting observation is that the topological action does \emph{not} depend on $N_c$ as long as $N_c\in 4\mathbb{Z}+1$. For a fixed but very large $N_f\in8\mathbb{Z}+2$ and small enough $N_c\in4\mathbb{Z}+1$, the $Sp(N_c)+N_f^F$ theory is free in the infrared limit. By increasing $N_c\in4\mathbb{Z}+1$ before it hits some critical value, we have different free $Sp(N_c)$ gauge theories (They are labeled by the red dots in Fig. (\ref{spbz})). Most importantly these theories all describe a phase transition between the trivial state and the same SPT state protected by $PSp(N_f)\times Z_2^T$ symmetry. 

These free theories are truly distinct conformal field theories. For instance they have different numbers of emergent low energy degrees of freedom. This may be formalized based on the $a$-theorem. The quantity $a$ is a universal number used to characterize 4D CFT. It is the 4D analogy of the central charge of 2D conformal field theories. The trace of the stress energy tensor of a 4D CFT can be expressed as the following, 
\beq
\langle T_{\mu}^\mu\rangle=-aE_4+cW^2
\label{stress}
\eeq
where $E_4$ is the Euler density and $W^2$ is the square of Weyl tensor. The $a$-value is a universal signature for  a 4D CFT, similar to the central charge of 2D CFTs. It was conjectured and subsequently proven to be a monotonic function under RG flow\cite{cardyatheorem}, so-called $a$-theorem. Since these $Sp(N_c)+N_f^F$ theories are IR-free theories, we know the answer for the $a$-values\cite{cardyatheorem}, 
\beq
a(N_c,N_f)=22N_cN_f+62N_c(2N_c+1)
\label{avalue}
\eeq
For fixed $N_f$, different $N_c$'s give different $a$-values indicting that they are distinct 4D CFTs. 

Furthermore, if $N_c$ is in an appropriate range, the $Sp(N_c)+N_f^F$ theory can possibly fall into the conformal window of $Sp(N_c)$ gauge theory (labeled by the green dots in Fig.(\ref{spbz})), which is described by the Bank-Zaks fixed point. It is a strongly interacting deconfined gauge theory, which is clearly distinct from free theories. For instance, it has different scaling dimensions for the gauge invariant operators from those of the free theories\cite{BZfund}.

The $Sp(N_c)$ generalization provides an explicit example for the phenomenon that there can exist multiple distinct critical theories that describe the transition between the same two nearby phases. In this controlled example, we are certain that these critical points are \emph{not} dual to each other. We call them Multiversality classes. In later sections we will provide more examples of such phenomena for fermionic deconfined critical points. 

\subsection{Unnecessary continuous phase transitions}
\label{uncpt}

In this section, we introduce a phenomenon which we name unnecessary phase transition. Unnecessary phase transitions are generic stable continuous phase transitions between two identical phases. We will show examples of such a phenomenon within the $Sp(N_c)+N_f^F$ theory. We will also discusss examples that do not involve gauge fields. 

\paragraph*{Example 1: $Sp(N_c)+N_f^F$ theory with $N_c\in 4\mathbb{Z}$ and $N_f\in4\mathbb{Z}$. }

The first example we consider is the $Sp(N_c)+N_f^F$ theory with different $N_c$ and $N_f$ from previous sections. An interesting situation is $N_c=4q\in4\mathbb{Z}$ and $N_f=4p\in4\mathbb{Z}$. With such condition, the two phases with $m<0$ and $m>0$ are actually the same phase. We can show the topological response for $m>0$ phase is $\mathcal{S}_{topo}=i2\pi \mathbb{Z}$,
\beqn
\mathcal{S}_{topo}&=&i\frac{\pi}{2}(2N_cl_{PSp(N_f)}+2N_fl_{PSp(N_c)}-\frac{4N_cN_f}{8}\sigma)\\
&=&i\frac{\pi}{2}\left\{4q(\mathcal{P}(w_2^{PSp(N_f)})+2w_4^{PSp(N_f)})+4p(\mathcal{P}(w_2^{PSp(N_c)})+2w_4^{PSp(N_c)})-\frac{16qp}{2}\sigma\right\} \\
&\sim&2\pi i\mathbb{Z},
\eeqn
Namely, $m<0$ and $m>0$ have identical partition functions for any gauge configuration. This means the two phases are identical. 
 
\begin{figure}
  \centering
    \includegraphics[width=0.5\textwidth]{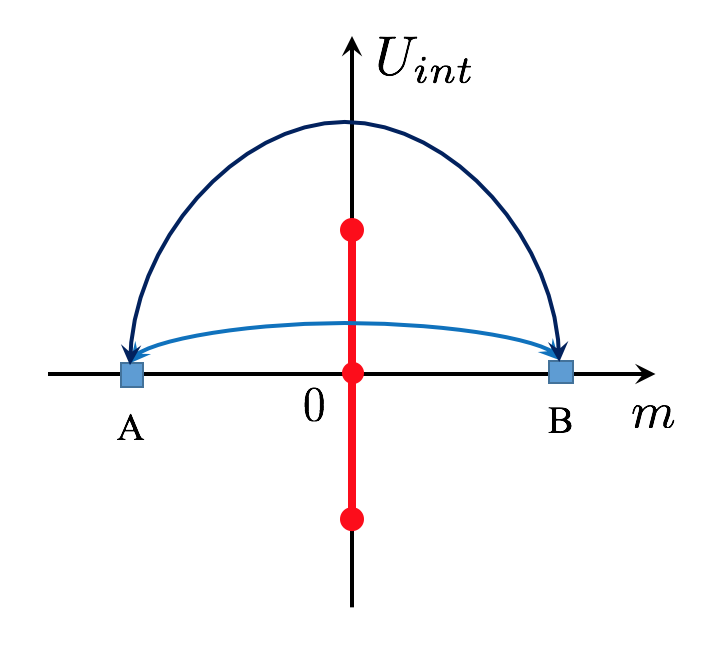}
      \caption{A schematic phase diagram for unnecessary phase transitions. }
      \label{upt}
\end{figure}

Nevertheless, there is a generic continuous phase transition in the phase diagram by tuning $m$ from negative to positive. In the large $N_f$ limit, the $m=0$ theory is IR free. The uniform mass $m$, as the driving parameter, is the only relevant symmetric perturbation at the critical point. Other generic interactions which respect the symmetry are perturbatively irrelevant. In other words, the IR free gauge theory controls the whole  critical line which exists within a single phase of matter. A schematic phase diagram is shown in Fig. (\ref{upt}).\footnote{The end point of the critical line may also be an interesting critical point, which may relate to the phenomenon of symmetric mass generation in $2+1$-D\cite{smg2017}. }

\paragraph*{Example 2: 16 copies of $^3$He-B phases. }
We now describe an example for an unnecessary continuous phase transition in $3+1$-D fermionic system without gauge fields. Let us consider a $3+1$-D topological superconductor (TSC) in DIII class, namely TSC protected by time reversal symmetry and $\mathcal{T}^2=-1$. The interacting fermionic SPT has a $\mathbb{Z}_{16}$ classification\cite{kitaevz16,fcav2013,senthilDIII} labeled by an integer $\nu$. The low energy theory for the $\nu=1$ DIII TSC state can be represented by a massive Majorana fermion in $3+1$-D. Reversing the Majorana mass can tune a trivial superconductor to topological superconductor phase transition. Now we take 16 copies of the $\nu=1$ DIII TSC states. We can consider the phase transition from the trivial state to the topological state for all copies of the system. Certainly, 16 copies of $^3$He-B is adiabatically connected to a trivial phase because the surface has no time-reversal anomaly\cite{senthilDIII}. However, the transition is not guaranteed to be a single generic transition. Different copies of the system can go through phase transition successively. In order to have a single transition, there must be some flavor rotation symmetry. The most naive one is an $SO(16)$ symmetry, which rotates the 16 copies of TSC's. This symmetry together with $Z_2^T$ symmetry will allow only one Majorana mass term. The low energy theory is 
\beq
\mathcal{L}^{\times 16}_{TSC}=\sum_{i=1}^{16}\bar{\chi}_i(i\gamma_\mu\partial_\mu-m)\chi_i+...
\eeq
Therefore, there is a generic continuous phase transition when we tune the mass from negative to positive. However, there is a problem in this situation. The two sides of the phase transition are different topological phases protected by the $SO(16)$ symmetry. In particular, on one side, $m<0$, we can regularize the system to be in the trivial phase, where we have a trivial response theory for the $SO(16)$ background gauge field. On the other side, $m>0$, the response theory of background $SO(16)$ gauge field has a $\Theta$-term with $\Theta=\pi$, which indicates that the system is an SPT state protected by the $SO(16)$ symmetry. Since the two sides are distinct topological phases of the $SO(16)$ symmetry, there will always be a phase transition separating them. This seems to be a disappointing case. However, a slight modification of the symmetry gives us an example of another unnecessary continuous phase transition.

Consider breaking the flavor symmetry from $SO(16)$ to $SO(2)\times SO(7)$   symmetry. The symmetry action on the fermions can be understood in the following way. Let us pack the 16 Majorana fields into a $2\times 8$ matrix. The $SO(2)$ and $SO(7)$ symmetry are implemented by the left and right multiplication by orthogonal matrices. Here, the right multiplications are in the 8-dimensional spinor representation of the $SO(7)$ group. This symmetry only allows the $\bar{\chi}\chi$ mass. To see this, we can write down the general form of Lorentz and time reversal symmetric mass term $\bar{\chi}_iS_{ij}\chi$, where $S_{ij}$ is a real symmetric matrix in flavor space. $S$ can in general be decomposed into two classes $S\sim S_2\otimes S_8$ or $S\sim A_2\otimes A_8$, where $S_4$ and $S_8$ denote real symmetric matrices of dimension $2$ and $8$, $A_2$ and $A_8$ are antisymmetric matrices of dimension $2$ and $8$. The $SO(2)$ generators are $A_2\otimes \mathcal{I}_8$, and $SO(7)\in \mathcal{I}_2\otimes A_8$. We can explicitly check that the only matrix that commutes with all the $SO(2)$ and $SO(7)$ generators is $\mathcal{I}_2\otimes \mathcal{I}_8$, which is the identity. Therefore, the $\bar{\chi}\chi$ term is the only allowed mass term. This means, with $SO(2)\times SO(7)$ symmetry, there is still a generic phase transition as we tune the mass from negative to positive. Since the phase transition is described by free fermions, it is stable against small interactions.  

Next we show that in the $SO(2)\times SO(7)$ case the $m<0$ and $m>0$ phases are actually the same phase. We argue this through the surface state of the system. At free fermion level, the natural $2+1$-D surface state of the $m>0$ phase has 16 gapless Majorana fermions. We will argue that the surface state can become symmetrically gapped out by interaction, which indicates the bulk state is in the same class of the trivial state. First, let us organize the 16 Majorana fermions into 8 Dirac fermions. 
\beq
H=\sum_{i=1}^8\psi^\dagger_i(p_x\sigma_x+p_y\sigma_z)\psi_i
\eeq
The $SO(2)$ or the $U(1)$ now is the phase rotation of the $\psi$ fermion. The time reversal action is $\mathcal{T}:\psi\rightarrow i\sigma^y\psi^\dagger$. The $\psi_i$'s also form the spinor representation of the $SO(7)$ symmetry. Then we introduce a spin singlet pairing in the theory which completely gaps out the surface state. 
\beq
H_{pairing}=\sum_{i=1}^8\Delta \psi_i^Ti\sigma^y\psi_i+h.c.
\label{pairing}
\eeq
This pairing obviously preserves the $SO(7)$ symmetry. It breaks both $U(1)$ and the time reversal $\mathcal{T}$. However, it preserves another anti-unitary symmetry $\tilde{\mathcal{T}}=\mathcal{T}U(\pi/2)$\cite{senthilDIII}. The next step is to quantum fluctuate the pairing order parameter to restore the symmetries. This can be done by condensing the $2\pi$ vortices of the pairing order parameter. There are two key requirements for getting a symmetric gapped surface state after the condensation. First of all, in order to restore both $U(1)$ and $\mathcal{T}$, the condensation has to preserve the $\tilde{\mathcal{T}}$ symmetry. Secondly, the vortices must carry a gapped spectrum. These conditions need special care because the vortex core of the system carries majorana zero modes\cite{FuKaneSC}. For our system, in a $2\pi$ vortex ($\pi$-flux for fermions), there will be 8 majorana zero modes, $\chi_i$, $i=1,...,8$. Their $\tilde{\mathcal{T}}$ transformation is $\tilde{\mathcal{T}}: \chi_i\rightarrow \chi_i$ because the $\tilde{\mathcal{T}}$ does not change the vortex background. This time reversal symmetry will forbid us to gap out the zero modes by fermion bilinear term. However, it is well known that an $SO(7)$ invariant four fermion interaction term, which is the so-called Fidkowski-Kitaev interaction\cite{FKinteraction}, can give rise to a gapped spectrum for 8 majorana modes. With this interaction, we can condense the $2\pi$ vortices and get a symmetric gapped surface state. This indicates that the bulk state is topologically trivial. 

The phase diagram of the system is demonstrated in Fig. (\ref{upt}). The $m$ term precisely corresponds to the free fermion mass and $U_{int}$ to the Fidkowski-Kitaev interaction. The free fermion phase transition in $3+1$-D is stable against small interaction. In the limit of large interaction, we can first diagonalize the interaction and treat the kinetic term as a perturbation. In the large interaction limit, the system is essentially a trivial insulator with tensor product wavefunction. Therefore, the phase transition can be avoided by going through the strongly interacting part of the phase diagram. 

\section{Fermionic Deconfined Critical Points in $3+1$-D}
\label{fdqcp}

In this section and following sections we study quantum critical points that can be formulated as $3+1$-D $SU(2)$ gauge theory coupled to $N_f$ flavors of massless adjoint Dirac fermions, denoted as $SU(2)+N_f^A$, and some generalizations of it. Based on the perturbative calculation, for $N_f^A>2$, the theory is free in the infrared limit. Numerically, the $N_f^A=2$ theory is inside the conformal window\cite{BZadj2}. There are also numerical indications that the $N_f^A=1$ theory is conformal in the IR\cite{BZadj1}. In this section, we study in details the IR-free $SU(2)$ gauge theories with $N_f^A =  3$ massless adjoint Dirac fermions and interpret them as a quantum critical points between fermionic SPT states. Since the gauge theory is free in the IR, we can make a lot of precise statements.

\subsection{$SU(2)$ gauge theory with $N_f=3$ adjoint fermions}

We consider a quantum critical point that can be described by $3+1$-D $SU(2)$ gauge theory with 3 flavors of adjoint Dirac fermions. The story will be very similar for all the odd $N_f^A>3$. We label the fermions by $\psi^a_i$, where $a=1,2,3$ is the $SU(2)$ index, $i=1,2,3$ is the flavor index. A key difference from the fundamental fermion case is that there are  gauge singlet fermion operators (the baryons) such as $\epsilon_{abc}\psi^a\psi^b\psi^c$ and $\epsilon_{abc}\psi^{a\dagger}\psi^b\psi^c$. Indeed all local operators of the theory carry quantum numbers that can  be built up as composites of these baryons. Therefore, the $SU(2)$ gauge theory with adjoint fermion fields describes a critical theory in intrinsic fermionic systems. 

The Lagrangian for the $N_f^A=3$ theory reads 
\beq
\mathcal{L}^{N_f^A=3}_{Adj}=\sum_{j=1}^3i\bar{\psi}_j\gamma_\mu(\partial_\mu-a_{\mu}^iT^i)\psi_j-m\bar{\psi}_j\psi_j+...,
\label{su2adj}
\eeq
where $\{\gamma_0,\gamma_1,\gamma_2,\gamma_3\}=\{\sigma^{12},i\sigma^{03},-i\sigma^{22},-i\sigma^{01}\}$, $\gamma_5=i\gamma_0\gamma_1\gamma_2\gamma_3=\sigma^{32}$, and $T^i$, $i=1,2,3$, are the $SU(2)$ generators in spin-1 representation. The theory has a $Z_2^T$ symmetry\footnote{This is usually denoted as $\mathcal{CT}$ symmetry in the literature because it also involves a particle-hole transformation.} whose transformation on the fermion fields are as following. 
\beq
Z_2^T:\psi(\vec{x},t)\rightarrow \gamma_0\gamma_5\ \psi^\dagger(\vec{x},-t)=-i\sigma^{20}\psi^\dagger(\vec{x},-t),\ \  i\rightarrow -i
\label{AIIIT}
\eeq
Following the method in previous sections, we can construct the adjoint $SU(2)$ theory from 18 Majorana fermions, and then gauge the diagonal $SO(3)$ part of the total $SO(18)$ flavor symmetry. Since $SO(18)\supset SO(3)\times SO(6)$, the global symmetry after gauging is $\mathcal{G}=SO(6)\times Z_2^T$.\footnote{As written in Eq. (\ref{su2adj}), the theory also has the usual time reversal symmetry $\mathcal{T}$, which does not flip the $U(1)$ charge, as well as the parity symmetry $\mathcal{P}$. In this construction, the $\mathcal{T}$ and $\mathcal{P}$ symmetries are not important.} 
The Dirac mass in Eq. (\ref{su2adj}) is the only mass term allowed by the global symmetry. 

As written the theory  in Eq. (\ref{su2adj}) also has a global 1-form $Z_2$ center symmetry\cite{1form}, because we did not include any matter field in the $SU(2)$ fundamental representation. The physical manifestation of the 1-form symmetry is that the Hilbert space of the system contains unbreakable spin-$\frac{1}{2}$ electric flux loops.  However if we are to view the gauge theory as emerging from a UV system of gauge invariant fermions (defined perhaps on a lattice), the 1-form symmetry can only be an infrared emergent symmetry. Therefore, we should allow for explicit breaking of the 1-form symmetry in the UV. To that end we will introduce a \emph{massive} spin-$\frac{1}{2}$ particle into our theory, which we call the \emph{spectator} field. The spectator field   allows the spin-$1/2$ electric flux loops to break.  We emphasis that, from the point of view adopted in this paper, the theory in Eq. (\ref{su2adj}) is not complete yet because we did not specify the properties of the spin-$\frac{1}{2}$ spectator fields under the global symmetry $\mathcal{G}$. To have a complete theory, we need to specify the symmetry charges of the spectator field under the 0-form global symmetry $\mathcal{G}$. (This is in some sense equivalent to defining the symmetry properties of the spin-$\frac{1}{2}$ electric flux lines.) Perhaps surprisingly, the symmetry charges of the massive spectator field crucially determine the nature of the $m\neq0$ phases of this theory, although it does not participate in the low energy physics at all. We will explain this phenomenon in detail later. For now, let us restrict our attention only to the 0-form global symmetry of the system, which is $\mathcal{G}=SO(6)\times Z_2^T$. 

The theory in Eqn. \ref{su2adj} at the massless point is a free theory in the infrared. The fermion mass is a relevant perturbation which will drive the system to the infinite mass fixed point. Thus the massless theory describes a continuous quantum phase transition between $m<0$ and $m>0$ phases. The schematic renormalization group flow of the fermion mass and gauge coupling is in Fig. (\ref{dRG}). Let us identify the phases with \emph{large} negative or positive fermion masses. For large fermion mass, we can integrate out the fermions first. We choose a UV regularization such that in the $m<0$ phase the $SU(2)$ $\Theta$-term generated by integrating out the massive fermions is zero. The $SU(2)$ gauge theory confines at low energy and the resulting state is a trivial gapped state. For large $m>0$ phase, one can show that the $\Theta$-angle is $12\pi$ for the $SU(2)$ gauge fields\footnote{Integrating out the fermion will generate an $SO(3)$ $\Theta$ angle at $6\pi$ and the $\Theta$ angle is $12\pi$ once we restrict to $SU(2)$ gauge bundle }. This is also trivial because of the $2\pi$ periodicity of the $\Theta$ angle, and the $SU(2)$ gauge theory is again in a confined phase. In particular both confined phases are believed to be in a Short-Ranged-Entangled ground state.  For both signs of the mass, in the large mass limit we expect a gapped and non-degenerate ground state with no symmetry breaking. They must fall into the classification of the fermionic SPT states with $SO(6)\times Z_2^T$ symmetry. Since this symmetry class is not usually considered in the literatures, let us first discuss the interacting classification of such SPT phases. 

The classification of fermion SPTs for this symmetry in $3+1$-D is $\mathbb{Z}_8\times \mathbb{Z}_2$ which can be labeled by two indices $n\in\mathbb{Z}_8$ and $\eta\in\mathbb{Z}_2$. The $\mathbb{Z}_2$ part comes from the pure $Z_2^T$ SPT labeled\cite{WS2013} by $efmf$. The $\mathbb{Z}_8$ part is the reduced classification from the free fermion SPT with the same symmetry. Note that at the free fermion level SPTs with this symmetry have a $\mathbb{Z}$ classification which we will label by the same index $n$. The root $n = 1$ state of the free fermion SPT with $SO(6)\times Z_2^T$ symmetry can be viewed as 6 copies of topological superconductor with $Z_2^T$ symmetry, namely the DIII class. The 6 copies form a vector representation under $SO(6)$. At the surface, the $n=1$ state has (within free fermion theory) 6 massless Majorana fermions. For general $n$ there will correspondingly be $6n$ massless Majorana fermions at the surface.  With interactions, we need to consider whether for some special $n$ the surface is anomaly free. The anomaly on the surface has two parts: 1). pure time reversal anomaly; 2). mixed anomaly between the $SO(6)$ and $Z_2^T$ which is sometimes called (generalized) parity anomaly in the literature. The pure time reversal anomaly is $\mathbb{Z}_{16}$-fold. Physically this means 16 copies of Majorana in $2+1$-D is time reversal anomaly free. Therefore, at least 8 copies of the root states are needed to cancel the surface time reversal anomaly. The mixed anomaly between $SO(6)$ and $Z_2^T$ or the parity anomaly is 4-fold\cite{EWnonorientable}. The physical diagnosis for this mixed anomaly is the quantum number of the background $SO(6)$ monopole. One can show that for 4 copies of the root state the monopole of the background $SO(6)$ gauge field is a trivial boson. Therefore, the surface of the $n = 8$ state is totally anomaly free. Hence with interactions the free fermion SPT classification collapses to $\mathbb{Z}_8$. In addition, the $n=4$ state corresponds to the $eTmT$ state\cite{senthilashvin,WS2013}. For $n = 4$ there is no parity anomaly involving $SO(6)$ and $Z_2^T$. The surface anomaly purely comes from the time reversal anomaly. For $n=4$, the surface theory has $4\times6=24$ Majorana fermions. Since the time reversal anomaly is $\mathbb{Z}_{16}$ periodic, the surface corresponds to the surface of $\nu=24\sim 8$ state in the DIII class, which is precisely equivalent to the $eTmT$ anomalous surface. 

Let us always assume the $m<0$ phase is the trivial state $(n=0,\eta=0)$ which can be achieved by certain UV regularization. The question is which ($n,\eta$) the $m>0$ phase falls into. To answer this, we derive the topological response to a background $SO(6)$ gauge field through the same method used before, namely gauging the total $SO(18)$ group and restricting the gauge configurations to its subgroups. The topological action for the $m>0$ phase (on an arbitrary closed oriented spacetime manifold) is
\beqn
\mathcal{S}_{topo}&=&i\frac{\pi}{2}\left(p_1(A^{SO(18)})-\frac{9}{4}\sigma\right)\\ \nonumber
&=&i\frac{\pi}{2}\left(3p_1(A^{SO(6)})+6p_1(a^{SO(3)})-\frac{9}{4}\sigma\right)\\ \nonumber
&=&i3\pi\left(\frac{1}{2}p_1(A^{SO(6)})-\frac{3}{8}\sigma\right)+i\pi p_1(a^{SO(3)})\\
&=&i3\pi\left(S_{\theta}^{SO(6)}-\frac{3}{8}\sigma\right)+i\pi \mathcal{P}(w_2^{SO(3)}).
\eeqn
where $S_\theta^{SO(6)}$ is the usual $\Theta$-term for the $SO(6)$ background gauge field, and the combination $(S_\theta^{SO(6)}-3\sigma/8)$ is always an integer. The response theory until now indicates that the $m>0$ phase is a non-trivial topological state. However, it is not enough to exactly determine the topological index of the state. In particular, we cannot tell whether the system belongs to $n=3$ state or $n=7\sim-1$ state. As the difference between the two is the $n=4$ state or the $eTmT$ state, whose partition function is always trivial on an orientable manifold. It turns out that to settle this we have to consider the symmetry properties of spectator field. We shall see that different symmetry properties of the spectator field lead to different topological phases on the $m>0$ side. 

To demonstrate the importance of the spectator field, we consider the following two different choices of spectators. There are other ways to choose spectator fields. We will leave them to future studies. From the discussions below, we shall see that the symmetry properties of the spectator field crucially determine the nature of the $m>0$ phase. 

\subsection{Band-theory-forbidden phase transition between band-theory-allowed insulators}

The simplest choice of the spectator is a bosonic particle which is neutral under all global symmetries, namely an $SU(2)$ spin-$\frac{1}{2}$ boson which is a scalar under $SO(6)$ and has $\mathcal{T}^2=1$. We will see that this choice of spectator field leads to an interesting type of band-theory-forbidden phase transition between two band theory allowed states.

To consistently define this spectator field, we have a constraint on the gauge connections
\beq
\label{w2=0}
w_2^{SO(3)}=0\  \text{mod}\  2.
\eeq
This relation must be satisfied on any base manifold $Y_4$. Then the topological response can be simplified as
\beq
\label{n=3}
\mathcal{S}^A_{topo}=i3\pi(S_\theta^{SO(6)}-\frac{3}{8}\sigma),
\eeq
which suggests $n=3$ in the $\mathbb{Z}_8$ classification.

To confirm the nature of the topological phase, let us investigate the surface state of the system. The natural surface state of the system is a QCD$_3$ theory with an $SU(2)$ gauge field coupled to 3 flavors of massless adjoint Dirac fermion\cite{2dDQCP2}. The action for the $2+1$-D surface theory can be written as follows. 
\beq
\label{surfaceQCD1}
\mathcal{L}_{surf}=\sum_{j=1}^3i\bar{\psi}_j\gamma_\mu(\partial_\mu-ia^i_\mu T^i)\psi_j+|(\partial_\mu-ia^i_\mu \frac{\sigma^i}{2})z|^2+m|z|^2+...
\eeq
where $\{\gamma_0,\gamma_1,\gamma_2\}=\{\sigma_y,-i\sigma_z,i\sigma_x\}$. Here we explicitly include the massive spectator field labeled by $z$. The time reversal symmetry and gauge transformations are 
\beqn
Z_2^T&:&\psi\rightarrow i\gamma_0\ \psi^\dagger \\
\label{spectator1}
Z_2^T&:&z\rightarrow z^*\\
SU(2)&:&\psi^a\rightarrow (e^{i\theta^aT^a})_{ab} \psi^b \\
SU(2)&:&z\rightarrow e^{i\theta^a\frac{\sigma^a}{2}}z
\eeqn

The surface theory in Eq. (\ref{surfaceQCD1}) is not very illuminating to us because it involves gauge fields. We want to deform the surface theory in a symmetry preserving manner to a more familiar surface state. Notice that the spectator boson $z$ is only charged under the $SU(2)$ gauge group. Let us condense the $z$ boson, i.e. go into a ``Higgs'' phase with $\langle z\rangle\neq 0$. This condensate preserves the $SO(6)\times Z_2^T$ global symmetry. Further the condensate completely Higgses the $SU(2)$ gauge fields, because $z$ carries fundamental representation of the $SU(2)$ gauge group. There are no residual gauge fields left on the surface. As a result, the $\psi$ fermions becomes physical fermions. The surface state now consists of \emph{18 physical massless Majorana fermions} with $SO(6)\times Z_2^T$ symmetry. This is precisely the surface of $n=3$ state in the $SO(6)\times Z_2^T$ fermionic SPT classification.
\begin{figure}
  \centering
    \includegraphics[width=0.5\textwidth]{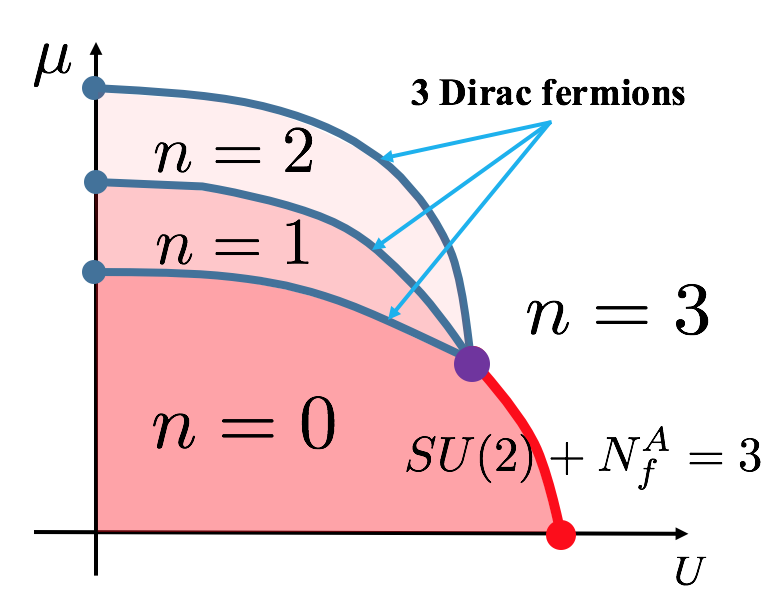}
      \caption{A schematic phase diagram. The $\mu$ axis represents some parameter which can tune through the transition from $n=0$ to $n=3$ SPT states with $SO(6)\times Z_2^T$ symmetry in the free fermion limit. Generically, the free fermion transition will split into three transitions each of which is described by 3 massless Dirac fermions (or 6 Majorana fermions) in the bulk. The three transitions in principle can be merged into a single transition in the strongly interacting region. The critical theory for the single transition is the strongly coupled $SU(2)+N_f^A=3$ gauge theory. The multi-critical point in the middle may be described by $SU(2)+N_f^A=3$ QCD$_4$-Higgs transition.}
      \label{n=3}
\end{figure}

This theory implies a very interesting schematic phase diagram shown in Fig. (\ref{n=3}). The phase transition between the trivial state and $n=3$ state in the $SO(6)\times Z_2^T$ class can happen in two different routes. In the weakly interacting limit, a trivial superconductor can only become $n=3$ TSC by \emph{three} successive topological phase transitions. At each step, the topological index can only jump by 1 and the low energy theory is described by 6 massless Majorana fermions with $SO(6)$ symmetry. However, the $SU(2)+N_f^A=3$ formulation suggests another very striking possibility that, in the strong interaction region, it is possible to go between the trivial topological state and $n=3$ state through a \emph{single} generic second order transition. It is a quantum phase transition between two band insulators which is \emph{forbidden by band theory}. In Appendix \ref{bbtqcp2d} we give a very simple example of this phenomenon that does not involve emergent gauge fields.

These two possibilities for the phase transition may merge at a multi-critical point somewhere in the phase diagram. One possible theory for the multi-critical point is the Higgs transition of the bosonic spin-$\frac{1}{2}$ spectator in the $3+1$-D bulk. Once the spectator is condensed in the bulk, the $SU(2)$ gauge fields are completely Higgsed out and \emph{each} flavor of the adjoint fermions becomes three physical fermions with topological band structure. Let us label the physical fermions by $c^a_j$, $a,j=1,2,3$. They can be expressed by the gauged fermion $\psi$ and spectator field $z$ in the following form.
More explicitly, the resultant physical fermions have the following form. 
\beqn
\label{fermions}
\nonumber
c^1_j&\sim& \vec{\psi}_j\cdot (z^\dagger \vec{\sigma}z)\\ 
c^2_j&\sim& \vec{\psi}_j\cdot \text{Re}(z^t i\sigma^y\vec{\sigma}z)\\ \nonumber
c^3_j&\sim& \vec{\psi}_j\cdot \text{Im}(z^t i\sigma^y\vec{\sigma}z)
\eeqn
The $c$ fermions are gauge invariant operators. It can be easily checked that the $c$ fermions share the same symmetry transformation as the $\psi$ fermions. The three successive phase transitions can be viewed as the mass inversion transition for each flavor of the $c$ fermion. 

It is interesting to ask what happens if we first take the mass of the fundamental spectator scalar to infinity. Then the gauge theory has the $Z_2$ $1$-form symmetry associated with the spin-$1/2$ electric flux loops. This symmetry is spontaneously broken in the free theory that emerges at the massless point. Upon perturbing with a fermion mass the gauge theory enters a confined phase. Then the electric flux loops acquire a line tension, and the $(Z_2)_1$ symmetry is restored. The spin-$1/2$ electric loops are however decoupled from other excitations. If now we re-introduce the fundamental scalars to explicitly break the $(Z_2)_1$ symmetry, in either phase the loops  can break but the sole effect on the phase is to determine the SPT character. At the massless critical point the explicit breaking of the $(Z_2)_1$ has no effect on low energy critical properties of the fermions. The spectator scalars will be deconfined at the critical point and gapped away from it. 

\subsection{Multiple universality classes in fermionic phase transitions}

Another choice of the massive matter content is a spin-$\frac{1}{2}$ bosonic particle which is an $SO(6)$ scalar but a Kramers doublet under time reversal, namely a spin-$\frac{1}{2}$ boson with $Q_{SO(6)}=0$ and $\mathcal{T}^2=-1$. This choice of spectator field implies the following constraint on the gauge connection. 
\beq
\label{w2=w1^2}
w_2^{SO(3)}=w_1^{TY}\cup w_1^{TY}\ \text{mod}\ 2
\eeq
Eq. (\ref{w2=0}) and Eq. (\ref{w2=w1^2}) are the fundamental difference between the two theories. The difference only arises when we consider the gauge bundle on non-orientable\footnote{$w_1^{TY}=1$ mod 2 only on non-orientable manifold. } manifold such as $\mathbb{RP}^4$. This relation implies that on an orientable manifold, $w_2^{SO(3)}=0$ mod $2$, meaning that the $SO(3)$ connection can always be lifted to an $SU(2)$ connection. In this case, we get the same topological response theory Eq.(\ref{n=3}) on an orientable manifold.
\beq
\label{n=-1}
\mathcal{S}^B_{topo}=i3\pi(S_\theta^{SO(6)}-\frac{3}{8}\sigma),
\eeq
It appears that this state also corresponds to the $n=3$ state. However, it is known that the $eTmT$ state, which corresponds to $n=4$ state\cite{senthilDIII,bridge}, is only visible in the partition function on a non-orientable manifold\cite{EWnonorientable, Maxnonorientable, Kapustin1,Kapustin2}. Therefore, the topological response on an orientable manifold cannot tell us precisely what topological phase the $m>0$ state belongs to. In the following, we will instead use physical surface arguments to determine the topological index of this system. 

To determine the nature of the $m>0$ phase, we again look at the boundary state. The surface theory has the same form as the QCD$_3$ theory written in Eq. (\ref{surfaceQCD1}), while the only difference is the time reversal symmetry transformation on $z$\footnote{This $Z_2^T$ transformation commutes with the $SU(2)$ gauge rotation.}
\beq
Z_2^T: z\rightarrow i\sigma_yz, \ \ \ i\rightarrow -i.
\label{TRforz}
\eeq
In this situation, it appears that condensing the bosonic spectator field may break the time reversal symmetry. But the condensate actually preserves the physical time reversal symmetry. The reason is that the time reversal transformation on $z$ can always be combined with an $SU(2)$ gauge rotation. Physical time reversal symmetry is preserved so long as such a combination of the time reversal action in Eqn. \ref{TRforz} and $SU(2)$ gauge rotation is preserved.  To be explicit about this,  we consider a gauge equivalent time reversal transformation $\tilde{Z}_2^T$: \footnote{This $\tilde{Z}_2^T$ transformation does not commute with the $SU(2)$ gauge transformation. It however commutes with the $SO(6)$ global symmetry.}
\beq
\label{newTRz}
\tilde{Z}_2^T: z\rightarrow i\sigma^ye^{-i\frac{\pi\sigma^y}{2}}\ z=z, \ \ \ i\rightarrow -i.
\eeq
The boson is a Kramer's singlet for this time reversal symmetry transformation.\footnote{One would think that because of the gauge transformation, the $\mathcal{T}^2$ of the spectator is actually meaningless. This is true if we only have spin-$\frac{1}{2}$ boson in our theory. However, we also have adjoint fermion matter with fixed time reversal transformation. The $\mathcal{T}^2$ for the spectator has physical implication in this case.}  Notice this time reversal transformation also has a different action on the gauged fermion fields by an additional gauge rotation, (here we suppress the flavor index because all the operations are identical for the three flavors.)
\beq
\label{newTRf}
\tilde{Z}_2^T: \psi^a\rightarrow i\gamma_0(e^{-i\pi T^y})_{ab}\psi^{b\dagger}, \ \ \ i\rightarrow -i.
\eeq
In component form, 
\beqn
\nonumber
\tilde{Z}_2^T: &&\psi^{1,3}\rightarrow -i\gamma_0\psi^{1,3\dagger},\\
 &&\psi^{2}\rightarrow i\gamma_0\psi^{2\dagger}.
\label{newT}
\eeqn
Now let us condense the spectator boson with $\langle \text{Im}(z)\rangle=0$ and $\langle \text{Re}(z)\rangle\neq0$. This condensate completely Higgses the $SU(2)$ gauge theory while preserving the $SO(6)\times \tilde{Z}_2^T$ symmetry. The three adjoint fermions become 9 physical Dirac fermions. But we need to be careful about their time reversal transformation in order to determine the topological index. In particular, the relative sign of the time reversal transformation of the surface Dirac fermions plays an important role here. In our convention, the Dirac fermion with the ``+" transformation, namely $\psi\rightarrow +i\gamma_0\psi^\dagger$, contribute $n=+1$ to the topological index for the bulk. Correspondingly, the ``$-$" transformation will contribute $n=-1$\cite{wittenFP}. Based on the transformation in Eq. (\ref{newT}), the surface state corresponds to the $n=-1+1-1=-1$ state in the $\mathbb{Z}_8$ classification.

From the above physical arguments, we see that the spectator field plays an important role in defining the global structure of the gauge fields and determining the nearby topological phase, although it is massive and never appears at low energy near the critical point. To our knowledge, this is not a widely appreciated phenomenon. However, it is not uncommon. We include an example of this phenomenon in $2+1$-D bosonic Mott insulator to a time reversal symmetry enriched $Z_2$ spin liquid transition in Appendix \ref{2dZ2SL} 

This provides a clear example of multiple universality classes in fermionic systems. The transition between the $n=0$ state and $n=-1$ state can happen within band theory. The critical theory is described by 3 massless Dirac fermions in the bulk with the $SO(6)\times Z_2^T$ symmetry. The $SU(2)+N_f^A=3$ theory gives another phase transition theory between $n=0$ and $n=-1$ states. We know in the IR this theory contains just free $SU(2)$ gluons and 9 Dirac fermions which is clearly different from the critical theory in free fermion limit. These two theories not only differ by their matter contents and but also by the emergent symmetries at the critical point. In particular, the gauge theory has an emergent $Z_2$ 1-form symmetry which is spontaneously broken in the IR. 

The theory discussed in this section is readily generalizable to all odd $N_f^A>3$. With general $N_f^A$, the global symmetry of the system is $SO(2N_f^A)\times Z_2^T$. The interacting fermionic SPT classification for this class is again $\mathbb{Z}_8\times\mathbb{Z}_2$. With a Kramers singlet bosonic spectator field ($SU(2)$ gauge spin-$\frac{1}{2}$ and $SO(2N_f^A)$ scalar), the massless $SU(2)+N_f^A$ theory describes a phase transition between $n=0$ and $n=3$ SPT states in this symmetry class. This provides new examples of band-theory-forbidden continuous phase transition between band theory allowed states. For a Kramers doublet bosonic spectator, the massless $SU(2)+N_f^A$ theory is a theory of continuous phase transition between $n=0$ and $n=-1$ SPT states. Since $N_f^A$ is large enough, this theory contains only free gluons and free fermions in the IR. Another route for this phase transition is a free fermion phase transition which is characterized by free massless Dirac fermions. These two critical theories are obviously distinct from each other. This is another example of multiple universality class for phase transitions in fermionic systems.

The $SU(2)$ gauge theories with even $N_f^A$ flavors of massless adjoint Dirac fermions can also be interpreted as quantum critical points between fermionic SPT states. The phenomenology of the even series is slightly different from the odd series. In particular, the topological phase on the $m>0$ side of the phase diagram does not depend on the choice of the spectator field. We present the example of $N_f^A=2$ in App. \ref{nfeven}. It is straightforward to generalize the theory to larger even $N_f^A$. We also provide generalizations to the $SU(2)+N_f^A$ theories with half integer $N_f^A$, as well as a generalization to $SU(4)$ gauge theory with $N_f^A=1$ flavor of adjoint fermion in the App. \ref{nfhalf} and \ref{su4}

\section{The $SU(2)$ gauge theory with one flavor adjoint Dirac fermion}
\label{nfaone}

The $SU(2)+N_f^A=1$ theory is a special case for the odd $N_f^A$ series. The global symmetry in this case is $SO(2)\times Z_2^T\sim U(1)\times Z_2^T$ which is the symmetry of the topological superconductor in the AIII class. Since this theory is an strongly interacting gauge theory in the IR, its low energy fate is more subtle than previous examples. We will discuss this theory in detail in the following sections. There is some numerical evidence that this theory is conformal in the IR\cite{BZadj1}. We will explore its interpretation as a quantum critical point.

Note that the fermion mass is a relevant perturbation for the massless $SU(2)+N_f^A=1$ theory\cite{BZadj1}. However, the massless $SU(2)+N_f^A=1$ theory is strongly coupled in the gauge theory description. A priori, we do not know whether an infinitesimal mass $m$ perturbation will flow to the infinite mass fixed point. If small mass does lead to a flow to the infinite mass limit, we will have a direct second order phase transition between the two gapped phases. If this is not the case, there may be an intermediate phase in the small mass limit. In this section, we only discuss the properties of the system with large fermion mass, and determine the distinct gapped phases. Inspired by this understanding, in section \ref{4ddrcdual}, we describe a possible IR theory of the massless $SU(2)+N_f^A=1$ theory. We will see that within this proposed IR theory there are indeed intermediate phases for small $m$ which differ from the large $m$ phases by the presence of an extra topological ordered state.

\subsection{Global symmetry and topological response}

As mentioned in the previous section, the $SU(2)$ gauge theory with adjoint fermion fields describes a critical theory in intrinsically fermionic systems. The Lagrangian for the $N_f^A=1$ theory reads 
\beq
\mathcal{L}^{N_f^A=1}=i\bar{\psi}\gamma_\mu(\partial_\mu-a_{\mu}^iT^i)\psi-m\bar{\psi}\psi+....
\label{nf=1}
\eeq
The theory has $U(1)\times Z_2^T$ global symmetry whose transformations on the fermion fields are as following,
\beqn
U(1)&:& \psi\rightarrow e^{i\theta}\psi. 
\label{AIIIU1}\\
Z_2^T&:&\psi(\vec{x},t)\rightarrow \gamma_0\gamma_5\ \psi^\dagger(\vec{x},-t),\ \  i\rightarrow -i.
\label{AIIIT}
\eeqn
$U(1)\times Z_2^T$ is the symmetry for topological superconductor in AIII class in condensed matter language. The Dirac mass in Eq. (\ref{nf=1}) is the only mass term allowed by the symmetry. As written the theory in Eq. (\ref{nf=1}) also has a global 1-form $Z_2$ center symmetry, because of the absence of the matter fields in the $SU(2)$ fundamental representation. However as we emphasized before, this gauge theory is to be viewed as an emergent theory from a UV system of gauge invariant fermions where there is no 1-form symmetry. Therefore, we will impose explicit breaking of the 1-form symmetry in the UV by introducing a \emph{massive} spin-$\frac{1}{2}$ spectator field into our theory. In this section, we will only consider the 0-form global symmetry of the system, which is $\mathcal{G}=U(1)\times Z_2^T$. 

We want to explore the theory in the large fermion mass limit.  We can then analyze the theory by integrating out the fermions first. We choose a UV regularization such that in the $m<0$ phase the $SU(2)$ $\Theta$-term is zero. The $SU(2)$ gauge theory is confined at low energy and the resulting state is a trivial gapped state. For large $m>0$ phase, one can show that the $\Theta$-angle is $4\pi$ for the $SU(2)$ gauge fields. This is also trivial because of the $2\pi$ periodicity, and the $SU(2)$ gauge theory is again in a confined phase. In particular both confined phases are believed to be in a Short-Ranged-Entangled ground state. 

For both signs of the mass, in the large mass limit we expect a gapped and non-degenerate ground state with no symmetry breaking. They must fall into the classification of the AIII topological superconductor (TSC) in $3+1$-D, which as we mentioned before is  $\mathbb{Z}_8\times \mathbb{Z}_2$ once we include interaction effects\cite{senthilDIII}. We can denote different AIII TI states by two labels $n\in\mathbb{Z}_8$ and $\eta\in\mathbb{Z}_2$. The $n\neq0$ states are descendent of the free fermion AIII TSC. The typical $2+1$-D surface state is $n$ flavors of massless Dirac fermions. The $n=4$ state is in the same phase of a bosonic SPT protected by $Z_2^T$ symmetry, which is usually signified by its surface $Z_2$ topological order, the so-called $eTmT$ state\cite{senthilashvin,WS2013, NLSM}. The $\eta=1$ state is another $Z_2^T$ bosonic SPT state, whose surface $Z_2$ topological order is the so-called $efmf$ state\cite{senthilashvin}. Let us always assume the $m<0$ phase is the trivial state $(n=0,\eta=0)$. We want to determine which ($n,\eta$) the $m>0$ phase falls into. 

We derive the topological response to a background $U(1)$ gauge field through the same method used before. The topological action for the $m>0$ phase (on an arbitrary closed oriented spacetime manifold) is
\beqn
\mathcal{S}_{topo}&=&i\frac{\pi}{2}\left(p_1(A^{SO(6)})-\frac{3}{4}\sigma\right)\\ \nonumber
&=&i\frac{\pi}{2}\left(3p_1(A^{SO(2)})+2p_1(a^{SO(3)})-\frac{3}{4}\sigma\right)\\ \nonumber
&=&i3\pi\left(\frac{1}{2}p_1(A^{SO(2)})-\frac{1}{8}\sigma\right)+i\pi p_1(a^{SO(3)})\\
&=&i3\pi\left(S_{\theta}^{U(1)}-\frac{1}{8}\sigma\right)+i\pi \mathcal{P}(w_2^{SO(3)}).
\eeqn
The response theory implies that the $m>0$ phase is a non-trivial topological state. However, as before we cannot tell precisely which class the system belongs. There may be a $n=4$ state or the $eTmT$ state attached to the system, whose partition function is always trivial on an orientable manifold. This can be settled by considering the symmetry properties of the spectator field. Just like in the previous section we will demonstrate that different symmetry properties of the spectator field lead to different topological phases on the $m>0$ side. 

\subsection{An alternate argument to identify the massive phases}

It is straightforward to use the argument in the previous section to justify that 1). with $\mathcal{T}^2=1$ charge neutral spin-$\frac{1}{2}$ spectator boson, the $m>0$ phase is the $n=3$ state in AIII class; 2) with $\mathcal{T}^2=-1$ spectator boson, the $m>0$ phase is the $n=-1$ state. We will not repeat this argument again. However, in this section we provide a different argument to support this result. 

We can justify the nature of the gapped phases from another point of view. Let us first consider the structure of the massive phases in the infinitely massive spectator limit.  Later we will reinstate the finite mass of the spectator. We will particularly be interested in understanding the anomaly of the surface theory as a window into which SPT phase the bulk system is in. The way to identify the anomaly of the surface state is through the method of anomaly inflow. 

Deep in the confined phases, all the $SU(2)$ electric flux lines have line tension. In the infinitely massive spectator limit, the spin-$\frac{1}{2}$ electric flux lines cannot end in the bulk. In other words the system has an exact 1-form $Z_2$ symmetry. The physical difference between the two spectator choices in this case lies in the properties of the spin-$\frac{1}{2}$ electric flux lines. While for the $\mathcal{T}^2=1$ case the spin-$\frac{1}{2}$ line has nothing special associated with it, the $\mathcal{T}^2=-1$ case physically corresponds to the situation that each spin-$\frac{1}{2}$ line is decorated with a Haldane chain protected by the time reversal symmetry\cite{senthilashvin,decoDW}. 

\begin{figure}
  \centering
    \includegraphics[width=0.7\textwidth]{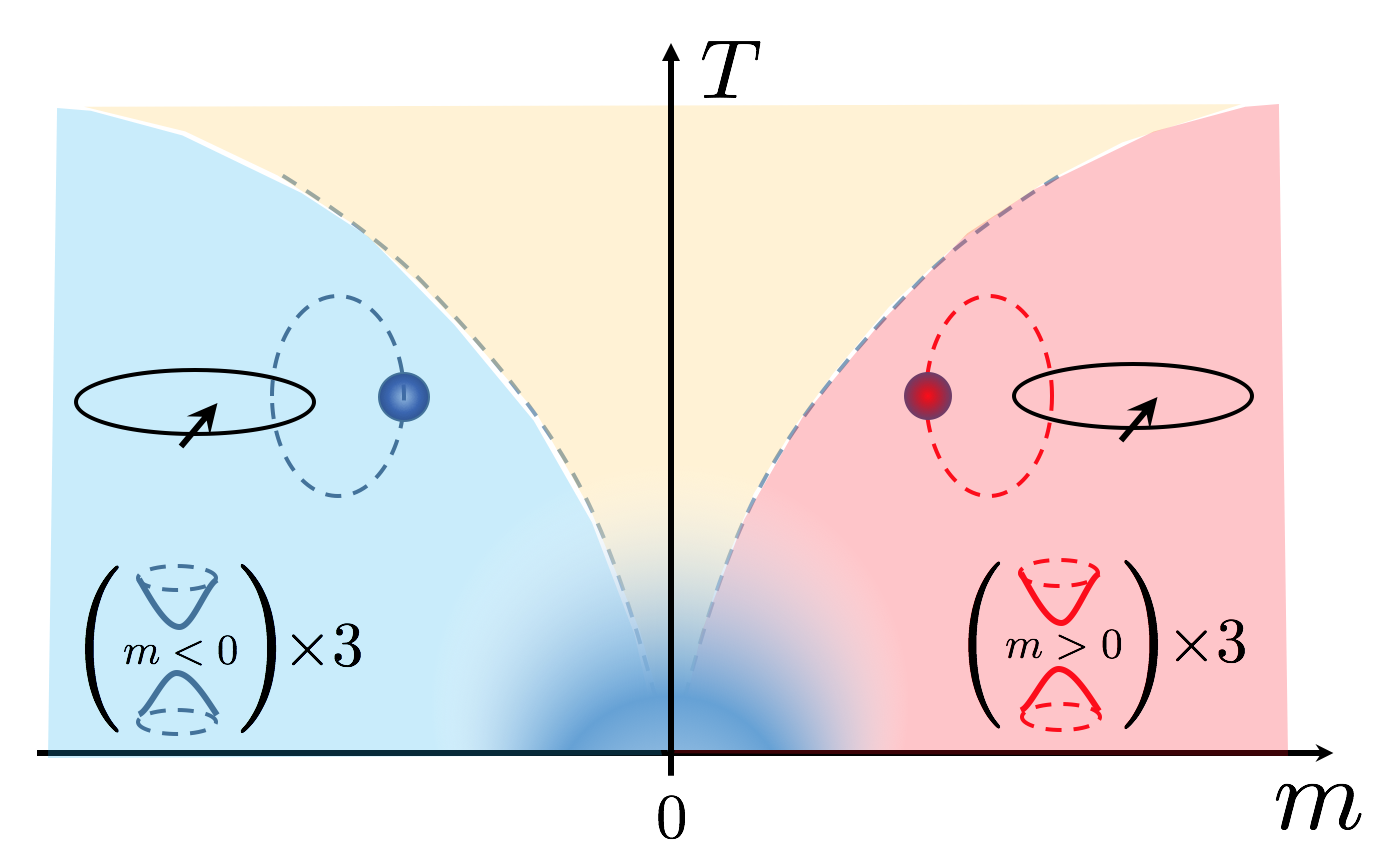}
      \caption{{\em Deep in the large $m<0$ and $m>0$ phases} (we will discuss the regime with small fermion mass later), in the bulk the $SU(2)$ gauge theory is in a confined phase where all electric flux lines have line tension. In the infinitely massive spectator limit, the spin-$\frac{1}{2}$ electric flux loops cannot break. The system has a global 1-form $Z_2$ symmetry. We can couple the system to a background 2-form $Z_2$ gauge field. The $Z_2$ flux of the 2-form gauge field physically corresponds to the $SO(3)$ magnetic monopole. It has a nontrivial $\pi$ mutual braiding phase with the spin-$\frac{1}{2}$ electric flux loops. The statistical and symmetry properties of $SO(3)$ monopole are different between $m<0$ phase and $m>0$ phase because of the topological band structure of the adjoint fermions. }
      \label{SO(3)}
\end{figure}

For our system, the surface anomaly contributions come from both the bulk massive adjoint fermions and the unbreakable spin-$\frac{1}{2}$ electric flux loop sector. Here we want to do a comparison between the $\mathcal{T}^2=1$ and $\mathcal{T}^2=-1$ spectator cases. Notice that the only physical difference between the two cases is whether we decorate the spin-$\frac{1}{2}$ loops with a Haldane chain protected by $Z_2^T$. Since the adjoint fermions are topologically decoupled from the spin-$\frac{1}{2}$ loops, changing the symmetry properties of these loops should not change the surface anomaly contributed by the bulk adjoint fermions. Therefore, we will be focusing on the differences in the surface anomalies contributed by the loop sector for the $\mathcal{T}^2=1$ and $\mathcal{T}^2=-1$ spectator cases. 

A useful formal approach to identify the surface Hilbert space and anomalies is to couple the system with background gauge field of the global symmetry. We can study the statistical and symmetry properties of the background symmetry fluxes in the bulk and then use the anomaly inflow argument to identify the surface excitations\cite{maxBTI}. Since the anomaly on the surface is a renormalization group invariant property of the SPT phase, we can consider the anomalies in the weak coupling or UV limit in which we can do reliable calculations. 

The symmetry we are interested here is the 1-form $Z_2$ symmetry. Let us coupled the system to a background 2-form gauge field for the 1-form $Z_2$ symmetry and consider a background $Z_2$ flux. This corresponds to an $SO(3)$ magnetic monopole configuration. The time reversal symmetry quantum number and statistics of this background monopole will be different between $m<0$ and $m>0$ phases because the adjoint fermions respond to the $SO(3)$ monopoles. In particular, the fermions in topological nontrivial band will contribute zero modes to the monopole configurations and potentially change the time reversal quantum number and the statistics of the monopole. We will show that in the $m>0$ phase the \emph{gauge neutral and global $U(1)$ charge neutral} $SO(3)$ monopole is a \emph{Kramers singlet fermion}. This statement is independent of the properties of the spin-$\frac{1}{2}$ Wilson lines. It is an analog of the statistical Witten effect in bosonic topological insulator with $U(1)\rtimes Z_2^T$ symmetry\cite{maxBTI}.

To demonstrate this, let us put the system on a large sphere and consider a configuration of $2\pi$ magnetic flux of the $SO(3)$ gauge field coming out of the bulk. For $m<0$, we know that there is no gapless surface state and the $SO(3)$ monopole carries trivial time reversal quantum number and bosonic statistics. However for $m>0$, the surface theory is a $QCD_3$ with massless adjoint fermions. It is sufficient to calculate the symmetry and statistical properties of the $SO(3)$ monopole in the weak coupling limit - the answers will be unmodified in the strong coupling limit. Let us write down the surface action with a background $SO(3)$ gauge flux along the $z$ direction in color space. 
\beq
\label{surfaceQED}
\mathcal{L}_{surf}=i\bar{\psi}\gamma_\mu(\partial_\mu-ia_\mu^z T^z)\psi,
\eeq
where $T^z$ is the $SO(3)$ generators along $z$ direction. We can diagonalize the $T^z$ matrix by unitary rotations of the fermions and it has eigenvalues $\pm1$ and $0$. Let us label the three flavors of fermions as $\psi_+,\psi_-,\psi_0$ ($\psi_+\sim \psi^x+i\psi^y, \psi_-\sim \psi^x-i\psi^y, \psi_0\sim \psi^z$). Only $\psi_+$ and $\psi_-$ are coupled to $a_\mu^z$ with charge $+1$ and $-1$ respectively. Hence, $\psi_+$ feels $2\pi$ flux and $\psi_-$ feels $-2\pi$ flux. With rotational symmetry in the color space,  every monopole can always be viewed this way. The gauge fluxes in our case are time reversal invariant since time reversal symmetry flips the gauge charges instead of the fluxes. From the surface theory in Eq.(\ref{surfaceQED}), we know that $2\pi$-flux of $a^z$ will trap two complex fermion zero modes guaranteed by the index theorem. One zero mode is associated with the $\psi_+$ fermion, which we label as $f_+$. The other one is associated with the $\psi_-$ fermion, which we label as $f_-$. Let us denote the flux background with both zero modes empty by $|0\rangle$. There are in total four states which are labeled by $|0\rangle, f_+^\dagger|0\rangle, f_-^\dagger|0\rangle, f_+^\dagger f_-^\dagger |0\rangle$. $f_+$ and $f_-$ carry opposite gauge charges but the same global $U(1)$ charge. The gauge neutral states from the four states are $|0\rangle$ and $f_+^\dagger f_-^\dagger|0\rangle$. But they carry opposite global $U(1)$ charge of $\pm 1$. We can attach $\psi_0$ fermion to the monopole state to compensate the $U(1)$ charge. However, this makes the monopole a fermionic object. Let us label the two states as $|M_1\rangle\sim \psi_0^\dagger|0\rangle$ and $|M_2\rangle\sim  \psi_0 f_+^\dagger f_-^\dagger|0\rangle$. Under time reversal, $Z_2^T: |0\rangle \rightarrow f_+^\dagger f_-^\dagger|0\rangle, \psi_0\rightarrow \gamma_0\gamma_5\psi_0^\dagger$. The time reversal transformations on $f_+$ and $f_-$ is a bit subtle. After carefully solving the zero mode wavefunction in Appendix \ref{zeromodes}, we find $f_+^\dagger\rightarrow -if_+, f_-^\dagger\rightarrow if_-$, where the relative minus sign is because the fluxes are opposite. With these, we can work out the time reversal transformation on the flux as follows. 
\beqn
\nonumber
&&|M_1\rangle\sim  \psi_0^\dagger|0\rangle \rightarrow \gamma_0\gamma_5\psi_0 f_+^\dagger f_-^\dagger|0\rangle \sim \gamma_0\gamma_5|M_2\rangle\\
&&|M_2\rangle\sim  \psi_0f_+^\dagger f_-^\dagger|0\rangle \rightarrow \gamma_0\gamma_5\psi_0^\dagger(-if_+if_-)f_+^\dagger f_-^\dagger|0\rangle \sim -\gamma_0\gamma_5|M_1\rangle
\eeqn
Since $(\gamma_0\gamma_5)^2=-1$, the $SO(3)$ monopole is a Kramer's singlet fermion\cite{ZouU(1)}. Note that this result cannot be altered if we redefine the $Z_2^T$ transformation by combining with $U(1)$ phase rotation, because the two states are gauge and global charge neutral. 

Let consider an interface between the vacuum and our system. Now imagine a process in which we take a background $SO(3)$ monopole in the vacuum and drag it into our system. This process can be viewed as an instanton event for the $2+1$-D interface, where the background $SO(3)$ flux changes from 0 to $2\pi$. The $SO(3)$ monopole is a neutral boson in the vacuum; however it becomes a neutral fermion in the bulk system. As a result, the instanton event, besides creating a $2\pi$ background flux on the surface, must also nucleate a neutral fermion excitation, labeled by $f$, in order to conserve the fermion parity of the whole system. Therefore the surface must have a neutral fermion excitation.   

Now let us introduce a finite mass spin-$\frac{1}{2}$ spectator boson on the surface which can be viewed as the end point of the spin-$\frac{1}{2}$ electric flux line on the boundary. We label it by $e$. In the weak coupling limit, they are deconfined particles on the surface. We need to determine the braiding statistics between $e$ and $f$. The instanton event we described above is a local process on the surface. The locality implies that, if we adiabatically drag the spectator boson $e$ around the location of the instanton event, there should be no difference in the accumulated Berry's phase before and after the instanton event. As a result, the braiding phase between the spectator and the neutral fermion $f$ must cancel that between the spectator and the $2\pi$ background flux. Since the spin-$\frac{1}{2}$ spectator can be viewed as the half charge under $SO(3)$, the braiding phase between the spectator and the $2\pi$ flux is $\pi$. Therefore, $e$ and $f$ have a mutual $\pi$ braiding phase and they form a $Z_2$ topological order on the surface. Now let us consider the time reversal properties of the $Z_2$ topological order. For the first case with $\mathcal{T}^2=1$ spectator, we have a vanilla $Z_2$ topological order which is not anomalous. For the other case with $\mathcal{T}^2=-1$ spectator, since $f$ is a Kramers singlet, the bound state $m\sim ef$ is also a Kramers doublet boson. The $Z_2$ topological order is the so-called $eTmT$ state which carries time reversal anomaly of the $n=4$ state in the AIII class. 

We can also include the spin-$\frac{1}{2}$ matter and break the 1-form $Z_2$ symmetry in the bulk. Dynamically the $SU(2)$ gauge theory will be in a confined phase for large fermion mass, which means all electric flux lines have finite line tension. With 1-form $Z_2$ symmetry, in the confined phase the system has unbreakable tension-full spin-$\frac{1}{2}$ electric flux loops. With finite mass spectators, these loops 
will break dynamically in the bulk and the system is in an ordinary confined phase. However since the time reversal anomaly on the surface does not involve the 1-form symmetry, it will survive even with a finite spectator mass.

\section{A possible $3+1$-D duality}
\label{4ddrcdual}

\begin{figure}
  \centering
    \includegraphics[width=0.6\textwidth]{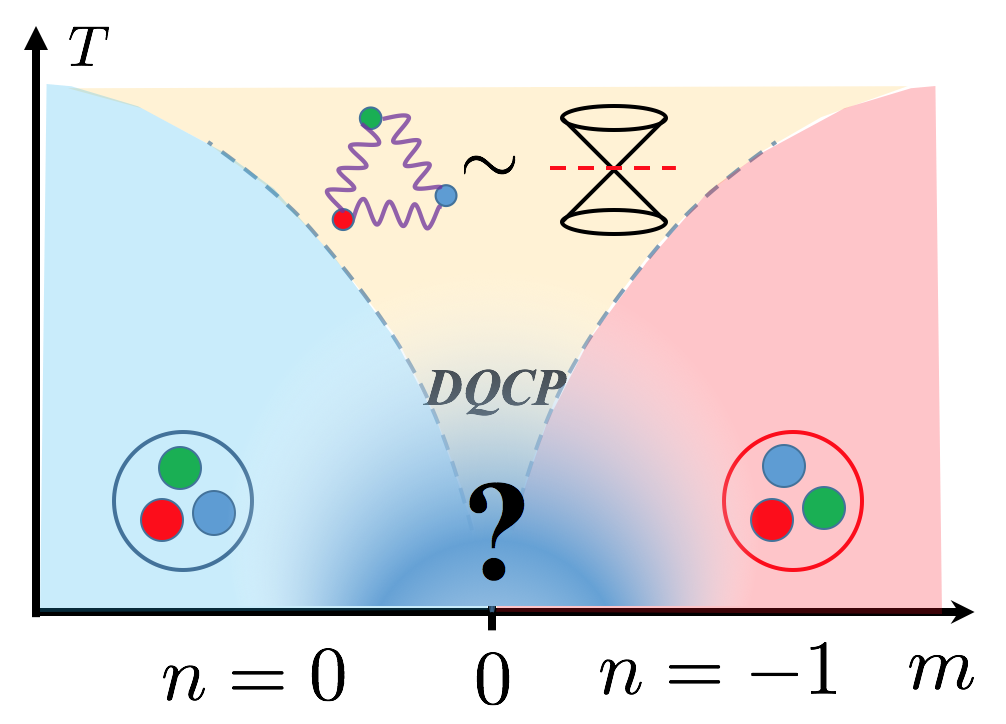}
      \caption{The transition between $n=0$ state and $n=-1$ state can happen in two ways. One way is a free fermion transition with a single gapless Dirac fermion as the critical theory. The other way is through a strongly coupled non-abelian gauge theory, which we labeled $SU(2)^*+N_f^A=1$. A very exciting possibility is that the two $3+1$-D conformal field theories are dual to each other in the infrared limit. Unfortunately, this is not likely the case. We will argue that the a possible IR theory of the theory is a single Dirac fermion plus a topological field theory. }
      \label{n=-1}
\end{figure}

The $SU(2)+N_f^A=1$ theory with $\mathcal{T}^2=-1$ spectator field potentially provides a continuous phase transition theory between $n=0$ and $n=-1$ states in the AIII class. The same phase transition can also happen in a free fermion setting which is described by a free massless Dirac fermion. There are several possible scenarios about the relationship between the strongly coupled gauge theory and the single Dirac fermion theory. For examples, we can imagine 1). a simple possibility is the low energy theory of $SU(2)^*+N_f^A=1$\footnote{We add a star as a reminder that the theory has a specific choice of the spectator field.} theory is a completely different critical theory from the single Dirac fermion.\footnote{An example of such low energy candidate theories, a $CP^1$ state with confinement and chiral symmetry breaking, is discussed in \cite{ThomasSU(2)}. };  2). Perhaps the most exciting scenario is that the $SU(2)^*+N_f^A=1$ theory in the IR is strictly dual to a single Dirac fermion. Unfortunately, we will argue that this scenario is very unlikely. Instead a candidate low energy theory of the $SU(2)^*+N_f^A=1$ theory can be very close to a single Dirac fermion. In particular, we will suggest a possible IR theory which contains a single free Dirac fermion plus an decoupled gapped topological sector. For energy lower than the gap of the topological order, the theory is described purely by a free Dirac fermion.

An important consistency check on any proposed IR theory is anomaly matching with the UV theory.  Our UV theory in the $m=0$ limit has emergent global symmetries which are anomalous. Matching the emergent symmetries and their anomalies between IR and UV provides nontrivial constraints. In particular, our theory in the infinite spectator mass limit is closely related to the celebrated Seiberg-Witten theory\cite{SW,ThomasSU(2)} whose global symmetry and anomaly structure are well understood in the high energy literature.  Exploiting this, Ref. \onlinecite{ThomasSU(2)} recently provided a  very nice discussion of the various anomalies of the $SU(2)$ gauge theory with a single massless adjoint Dirac fermion. The exact 1-form $(Z_2)_1$ symmetry of this theory was shown to have mixed anomalies with the emergent global symmetry and geometry\cite{ThomasSU(2)}, which put more constraints on the possible low energy theories. Therefore we will start our discussion from the infinitely massive spectator limit and later reinstate the finite mass to the spectator field. We first  identify the emergent 0-form global symmetry and their anomalies in the $SU(2)+N_f^A=1$ theory. We will see that the 0-form emergent symmetry and anomaly can indeed be matched by a single Dirac fermion theory. However, the single Dirac fermion does not have the $Z_2$ 1-form symmetry and hence cannot match the UV anomalies associated with it. This indicates that the low energy theory must contain additional either gapless or gapped topological degrees of freedom which could compensate the anomalies associated with the 1-form symmetry. Ref. \onlinecite{ThomasSU(2)} obtains such a candidate IR theory consisting of a single Dirac fermion plus a decoupled $U(1)$ gauge theory in the Coulomb phase through supersymmetry breaking deformations from the Seiberg-Witten theory. We will propose a different candidate theory which has a single Dirac fermion plus a decoupled topological order.\footnote{An easy consistency check is the $a$-theorem. As we introduced in Eq. (\ref{stress}) and (\ref{avalue}), the quantity $a$ is a universal property of every 4D CFT. It is known that $a$ is a monotonic decreasing function under renormalization group flow, namely $a_{UV}>a_{IR}$\cite{cardyatheorem}. The UV theory for the $SU(2)+N_f^A=1$ theory is free $SU(2)$ Yang-Mills theory with decoupled three free Dirac fermions. For free theories, we know the simple formula for the $a$ value. Therefore, the UV value of $a$ for the adjoint $SU(2)$ theory is $a_{UV}=3\times11+62\times(2^2-1)=219$, which is indeed larger than the $a$ value of a single Dirac fermion, $a_{Dirac}=11$. Hence, our proposed IR theory is consistent with the $a$-theorem conjecture. } The possibility of a topologically ordered state was also mentioned in Ref. \onlinecite{ThomasSU(2)}.

\subsection{The IR Dirac fermion}

Let us label the proposed gauge invariant Dirac fermion in the IR theory by $\Psi$. (The notation for the UV degrees of freedom in the $SU(2)$ gauge theory is defined in Eq. (\ref{su2adj}). ) The massless $\Psi$ theory describes a phase transition from $n=0$ to $n=-1$ state in AIII class. Therefore, the $\Psi$ fermion should carry the following quantum number under the global symmetry $U(1)\times Z_2^T$.
\beqn
U(1)&:& \Psi\rightarrow e^{i\theta}\Psi \\
Z_2^T&:& \Psi\rightarrow -\gamma_0\gamma_5\Psi^\dagger
\label{psiTR}
\eeqn
The ``$-$'' sign in the $Z_2^T$ transformation has physical consequence\cite{wittenFP}. (Notice that no linear transformation of the fermion field can change this sign.) The convention is that a gapless Dirac fermion with the ``$+$" $Z_2^T$ transformation describe a phase transition from the $n=0$ to the $n=1$ state in AIII class. Correspondingly, a Dirac fermion with the ``$-$" $Z_2^T$ transformation describe a transition from the $n=0$ to the $n=-1$ state.

By matching symmetry quantum numbers, the IR Dirac fermion operator $\Psi$ in terms of the UV degrees of freedom is
\beq
\Psi\sim \epsilon^{abc}(i\bar{\psi}_a\psi_b) (\gamma_5\psi_c)-\epsilon^{abc}(\bar{\psi}_ai\gamma_5 \psi_b) \psi_c.
\label{cmap}
\eeq
The right hand side is an $SU(2)$ gauge singlet operator. The global $U(1)$ quantum number obviously matches. The $\bar{\psi}_a\psi_b$ is a Lorentz scalar and $\bar{\psi}_ai\gamma_5\psi_b$ is a Lorentz pseudo-scalar. The reason for the choice of this specific combination of scalar and pseudo-scalar in the mapping is two-fold. Firstly, it is chosen to match the time reversal transformation of the $\Psi$ fermion. Secondly, as we discuss later, with such a combination, the single Dirac fermion theory matches the 't Hooft anomalies of the emergent symmetries in the $SU(2)$ gauge theory. Let us see how the time reversal symmetry works out first. We can check explicitly that the $\Psi$ in Eq. (\ref{cmap}) satisfies the transformation in Eq. (\ref{psiTR}). First of all, let us write down $\Psi^\dagger$.
\beq
\Psi^\dagger\sim \epsilon^{abc}(-i\bar{\psi}_b\psi_a)(\gamma_5\psi_c^\dagger)- \epsilon^{abc}(i\bar{\psi}_b\gamma_5\psi_a)\psi_c^\dagger
\label{dict}
\eeq
Recall that the time reversal action on the $\psi$ fermions is $\psi\rightarrow \gamma_0\gamma_5\psi^\dagger$. Also notice that the scalar $\bar{\psi}\psi$ is invariant under time reversal while the pseudo-scalar $\bar{\psi}i\gamma_5\psi$ is odd under time reversal. Therefore, the transformation of $\Psi$ is 
\beqn\nonumber
Z_2^T: \Psi&\rightarrow& \epsilon^{abc}(-i\bar{\psi}_b\psi_a)\gamma_5\gamma_0\gamma_5\psi_c^\dagger-\epsilon^{abc}(-\bar{\psi}_bi\gamma_5\psi_a)\gamma_0\gamma_5\psi_c^\dagger \\
&=&-\gamma_0\gamma_5\left(\epsilon^{abc}(-i\bar{\psi}_b\psi_a)\gamma_5\psi_c^\dagger-\epsilon^{abc}(i\bar{\psi}_b\gamma_5\psi_a)\psi_c^\dagger\right)=-\gamma_0\gamma_5\Psi^\dagger,
\eeqn
which is indeed what we want. We list partially the gauge invariant Lorentz scalar and spinor operators in the Appendix \ref{operators}

Since the operator $\bar{\Psi}\Psi$ and $\sum_{a=1}^3\bar{\psi}^a\psi^a$ share the same quantum numbers under all the global symmetries, they will have finite overlap in the IR. The conjecture is that $\Psi$ is free in the IR. Therefore, the anomalous dimension for the $\sum_{a=1}^3\bar{\psi}^a\psi^a$ operator should be zero. This could be checked in future numerical calculations.

\subsection{The emergent symmetries and anomalies}

For both the $SU(2)^*+N_f^A=1$ theory and the Dirac theory, the global $\mathcal{G}=U(1)\times Z_2^T$ symmetry is a non-anomalous symmetry of the system for all value of the mass $m$. When the system is tuned to the critical point at $m=0$, it has enlarged global symmetries. These emergent symmetries usually have 't Hooft anomalies. Coupling these emergent symmetries to background gauge fields will lead to an inconsistency in the theory which can be cured\footnote{From a formal point of view the we extend the background gauge fields  but not the dynamical degrees of freedom to the higher dimensional bulk. The difference between two different such extensions is described by a topological action in terms of these background gauge fields.  The boundary theory by itself is not gauge invariant but its combination with the bulk action is gauge invariant.} by regarding the theory as living at the boundary of a higher dimenisonal SPT phase. In this section, we compare the emergent symmetries and their anomalies of the two theories at their critical points.

For the massless $SU(2)^*+N_f^A=1$ theory in the UV, the emergent symmetry is $\mathcal{G}=\frac{SU(2)_f\times Z_8^A}{Z_2}$. The $SU(2)_f$ is a flavor rotation symmetry and $Z_8^A$ is a discrete axial rotation. The meaning of these symmetries will be clear in a moment. To understand these symmetries, let us look at the theory in Eq. (\ref{su2adj}) without the gauge field $a_\mu$. We can write down the Dirac fermions in the Weyl basis (we use a different set of $\gamma$ matrices than we were using previously), in which a single Dirac fermion can be written as two Weyl fermions with different chiralities,
\beq
\psi=\left(
\begin{array}{c}
\xi_{1}\\ i\sigma^y\xi_{2}^\dagger
\end{array}
\right).
\eeq
Here $\xi_1$ and $\xi_2$ are both two component left-handed Weyl fermions. The $i\sigma^y\xi_2^\dagger$ is particle-hole transformation of $\xi_2$ and has the opposite chirality. We can decompose our 3 Dirac fermions in Eq. (\ref{su2adj}) into 6 left-handed Weyl fermions (after a particle-hole transformation). The Lagrangian can be written as 
\beq
\mathcal{L}=i\sum_{a=1}^3\sum_{\alpha=1}^2\xi_{a,\alpha}^\dagger\bar{\sigma}_\mu\partial_\mu\xi_{a,\alpha},\ \ \bar{\sigma}_\mu=\{1,-\vec{\sigma}\},
\eeq
The largest unitary symmetry on the system is $U(6)$. Next, we want to gauge the diagonal $SU(2)$ subgroup of the $U(6)$ symmetry. Since the fermions are in the spin-1 representation, the gauge rotations on the Weyl fermions are $SO(3)$ rotations, 
\beq
SO(3)_g: \xi_{i,\alpha}\rightarrow O_{ij}\xi_{j,\alpha}, \ \text{with}\ \ O\in SO(3). 
\eeq
For convenience, we will use $SO(3)_g$ to denote the gauge group in the following. (But keep in mind that eventually this is an $SU(2)$ gauge field because of the spin-$\frac{1}{2}$ spectator field.) The $U(6)$ symmetry can be decomposed as $U(6)\supset \frac{SU(3)\times SU(2)\times U(1)}{Z_3\times Z_2}\supset \frac{SO(3)_g\times SU(2)\times U(1)}{Z_2}$. Therefore, the global symmetry left after gauging is naively $\frac{SU(2)\times U(1)}{Z_2}$. The $SU(2)$ is a flavor rotation, therefore we denote it as $SU(2)_f$. Its action on the Weyl fermions is:
\beq
SU(2)_f: \xi_{i,\alpha}\rightarrow U_{\alpha\beta}\xi_{i,\beta},\  \text{with}\ \ U\in SU(2).
\eeq
The 6 Weyl fermions form three fundamental representations of the $SU(2)_f$. The action of the $U(1)$ symmetry is 
\beq
U(1)_{A}: \xi_{i,\alpha}\rightarrow e^{i\phi}\xi_{i,\alpha}.
\eeq
Because of the particle hole transformation on the $\xi_{i,2}$ fields, this $U(1)$ rotation is the $\gamma_5$ rotation of the original Dirac fermion, which is usually called the axial rotation. We label it as $U(1)_A$. The familiar charge $U(1)$ rotation of the Dirac fermion is now the $S_z$ rotation of the $SU(2)_f$. 

The $U(1)_A$ suffers from chiral anomalies. It is explicitly broken down to $Z_8$ after considering the mixed anomalies with the $SO(3)_g$. This is seen from the following anomaly equation,
\beqn
\partial_\mu j^{\mu5}&=&\text{Tr}_{SO(3)}\left(\frac{F^{SO(3)}}{2\pi}\wedge\frac{F^{SO(3)}}{2\pi}\right)=4\text{Tr}_{SU(2)}\left(\frac{F^{SU(2)}}{2\pi}\wedge\frac{F^{SU(2)}}{2\pi}\right)\label{axialanomaly1} \\
\int_{Y^4}\partial_\mu j^{\mu5}&=&2p_1(SO(3))=8p_1(SU(2))\in 8\mathbb{Z} 
\label{axialanomaly2}
\eeqn
The first part of the equation is the standard Fujikawa's calculation for abelian anomalies\cite{nakaharabook}. In the second part, we use the relation between the Pontryagin classes between $SO(3)$ and $SU(2)$ groups. The Pontryagin class of $SU(2)$ counts the instanton number of the $SU(2)$ gauge field and takes value in integers. The equation means the axial charge will change by 8 if we insert an $SU(2)$ instanton configuration with winding number 1. Therefore, the axial charge is only well defined up to 8. The $U(1)_A$ is broken down to $Z_8^A$. 

Note that  there is no mixed anomaly between the $SU(2)_f$ and $SO(3)_g$. The divergence of the $SU(2)_f$ current is 
\beq
\partial_\mu j^{\mu}_{\alpha}=\frac{1}{24\pi^2}\text{Tr}\left[\sigma_\alpha\partial_\kappa\epsilon^{\kappa\lambda\mu\nu}(A_{\lambda}\partial_\mu A_{\nu}+\frac{1}{2}A_\lambda A_\mu A_\nu)\right]=0
\label{mixed}
\eeq
where $A_\mu=\sum_{a=1}^{3}A_\mu^aT_a$, $T_a$'s are $SO(3)$ generators and $\sigma_\alpha$'s are Pauli matrices. The anomaly equations are determined by calculating certain triangle loop diagrams\cite{nakaharabook,nairbook}. The essential part of the right hand side of the equation involves the trace of three symmetry generators. In this case, it is clearly zero because the $SO(3)$ generator and $SU(2)$ generator acting on different spaces. In the flavor space the trace of an $SU(2)$ generator is zero. This tells us that the $SU(2)_f$ is still a symmetry after gauging the $SO(3)_g$. Thus we see that the global symmetry for the critical $SU(2)^*+N_f^A=1$ theory is $\mathcal{G}=\frac{SU(2)_f\times Z_8^A}{Z_2}$.

In the infrared limit, it is possible that the $Z_8$ symmetry is dynamically enhanced to $U(1)$.  There are many examples of such phenomenon in $2+1$-D deconfined quantum critical points\cite{2dDQCP1,2dDQCP2,dualitynumerics}. Though we can not be  sure that  this  enlargement actually happens in our case, we are encouraged by the matching of anomalies with the free Dirac theory at its massless point (which has emergent $U(2) = \frac{SU(2) \times U(1)}{Z_2}$ symmetry)  discussed below\footnote{However we will also need to postulate an additional decoupled gapped sector in which there is no such dynamical enhancement. Nevertheless, as the free Dirac sector is decoupled, we can ask about realization of the $Z_8$ on this gapless sector. The more correct assumption then is that the $Z_8$ is dynamically enhanced to $U(1)$ in this decoupled sector. } . 
Henceforth in talking about the free Dirac theory we will simply   treat the $Z_8$ axial symmetry of the gauge theory as though it is a $U(1)$ symmetry. A proper discussion of the anomalies involving the $Z_8$ without this simplification is in Ref.\onlinecite{ThomasSU(2)}.

Now let us consider the anomaly structure for the $\mathcal{G}$. Firstly, we discuss the t' Hooft anomaly of $SU(2)_f$. The $SU(2)_f$ itself has no perturbative anomaly but has the global Witten anomaly. The Witten anomaly is a $Z_2$ anomaly \cite{wittenanomaly} which depends only on the parity of the number of $SU(2)_f$ fundamental Weyl fermions. Here we have three $SU(2)_f$ fundamental Weyl fermions. Therefore, they carry the $SU(2)$ Witten anomaly. Dynamically gauging the $SU(2)_f$ symmetry will lead to vanishing partition function. 

The $Z_8$ symmetry has self 't Hooft anomaly and mixed anomalies with $SU(2)_f$ and gravity. The anomaly is summarized in the following equation 
\beqn
\partial_\mu j^{\mu}_A&=&3\left(\frac{f_A}{2\pi}\wedge\frac{f_A}{2\pi}\right)+\frac{3}{2}\text{Tr}_{SU(2)}\left(\frac{F_{SU(2)_f}}{2\pi}\wedge\frac{F_{SU(2)_f}}{2\pi}\right)-\frac{6}{8}\sigma.
\label{axial}
\eeqn

Next we look at the IR Dirac fermion $\Psi$ at its massless point. In the Weyl basis, the Dirac theory reads, 
\beq
\mathcal{L}=i\sum_{\alpha=1}^2\eta_\alpha^\dagger\bar{\sigma}_\mu\partial_\mu\eta_\alpha\ ,\ \ \bar{\sigma}_\mu=\{1,-\vec{\sigma}\},
\eeq
where $\eta_1$ and $\eta_2$ are both left-handed Weyl fermions. According to our dictionary in Eq. (\ref{dict}), the $\eta$ fermions can be written as composite operators from $\xi$ fermions in the $SU(2)$ gauge theory. 
\beqn
\nonumber
\eta_1&\sim& \epsilon^{abc}i(\xi_{a,2}i\sigma^y\xi_{b,1})\xi_{c,1}\\
\eta_2&\sim& \epsilon^{abc}i(\xi_{a,1}i\sigma^y\xi_{b,2})\xi_{c,2}
\label{weyldict}
\eeqn
The theory manifestly has $\mathcal{G}^A=\frac{SU(2)_f\times U(1)_A}{Z_2}$ symmetry. These symmetries are in one to one correspondence with the emergent symmetries in the $SU(2)^*+N_f^A=1$ theory if as we assumed the $Z_8$ symmetry of the latter theory is enhanced to $U(1)$ in the gapless sector of the proposed IR theory. The $SU(2)_f$ transformation is
\beq
SU(2)_f: \eta_\alpha\rightarrow U_{\alpha\beta}\eta_\beta,\  \text{with}\ \ U\in SU(2).
\eeq
This transformation is consistent with the dictionary in Eq. (\ref{weyldict}). From the dictionary, the $\eta$ fermions carry charge 3 under the axial $U(1)_A$ symmetry in the $SU(2)$ gauge theory. 
\beq
U(1)_A: \eta_\alpha \rightarrow e^{i3\phi}\eta_\alpha
\eeq
This property is crucial for matching the anomalies with the UV theory.

Now we study the 't Hooft anomalies of the emergent symmetry. First the $SU(2)_f$ symmetry has the same global Witten anomaly\cite{wittenanomaly} because we have a single $SU(2)_f$ fundamental Weyl fermion. The anomalies associated with $U(1)_A$ are summarized in the following anomaly equation,
\beqn
\partial_\mu j^{\mu}_A&=&27\left(\frac{f_A}{2\pi}\wedge\frac{f_A}{2\pi}\right)+\frac{3}{2}\text{Tr}_f\left(\frac{F_{SU(2)_f}}{2\pi}\wedge\frac{F_{SU(2)_f}}{2\pi}\right)-\frac{6}{8}\sigma,
\eeqn
where the coefficient 27 and 3 precisely comes from the fact that the $\Psi$ fermions carry charge-3 under the axial $U(1)_A$ symmetry. This will match the anomalies in Eq. (\ref{axial}) if we consider the discrete $Z_8$ axial symmetry instead of the $U(1)_A$. This indicates that the low energy theory cannot be a simple Dirac fermion but needs some additional sector which remembers that the $U(1)_A$ is broken down to $Z_8$.

\subsection{The 1-form symmetry anomalies and the additional $Z_2$ topological order }

Thus far we argued that the IR Dirac fermion $\Psi$ matches almost all the 0-form symmetry and anomalies in the UV theory. Now we  focus on the 1-form $(Z_2)_1$ symmetry of the system in the infinitely massive spectator limit. The IR Dirac fermion does not have the 1-form symmetry. As shown in Ref. \onlinecite{ThomasSU(2)}, this $(Z_2)_1$ symmetry has mixed anomalies with both the $Z_8$ and with gravity. Therefore there must be other degrees of freedom in the IR which carry the 1-form symmetry and its anomalies.
 
The anomalies involving the $Z_2$ 1-form symmetry have two pieces according to Ref. (\cite{ThomasSU(2)}). The first part is a mixed anomaly between the $Z_2$ 1-form symmetry and the $Z_8^A$ discrete axial symmetry. Let us call this type I anomaly. The mixed anomaly means that dynamically gauging the $Z_2$ 1-form symmetry will break the $Z_8$ down to $Z_4$ on spin manifold, $Z_2$ on non-spin manifold. Formally, we can couple the system to a background 2-form $Z_2$ gauge field $B$. By definition, a symmetry operation on a quantum system should preserve the partition function. However, in this system in the presence of  the 2-form background gauge field $B$, the partition function is no longer invariant under $Z_8$ axial rotation. The $k^{th}$ element of $Z_8$ axial rotation will shift the partition function by a phase $\exp[i\frac{\pi k}{2}\int_{Y^4}\mathcal{P}(B)]$, where $\mathcal{P}(B)$ is the Pontryagin square of $B$. On a spin manifold, $\int_{Y^4}\mathcal{P}(B)$ is quantized as an even number\cite{seiberg2013}. Therefore the $Z_8$ is broken down to $Z_4$ by the mixed anomaly. On a non-spin manifold, $\int_{Y^4}\mathcal{P}(B)$ is an arbitary integer\cite{seiberg2013} and the axial symmetry is then broken down to $Z_2$. The second anomaly is more abstract. It is a mixed anomaly between the $Z_2$ 1-form symmetry and geometry. We will call this type II anomaly. This anomaly has the following formal interpretation. We again couple the system to a 2-form $Z_2$ gauge field $B$. The 2-form gauge field is a $Z_2$ gauge field which means a redefinition of the 2-form gauge field, $B\rightarrow B+2x$ with $x$ another 2-form $Z_2$ gauge field, should not change the partition function of the system. However, in this theory, such a redefinition change the partition function by a factor $\exp[i\pi x\cup w_2^{TY}]$, which can be $-1$ on a non-spin manifold. 

It is useful for us to have a  more concrete physical picture for both types of anomalies. The type I anomaly in the UV has the following physical interpretation. Let us remind ourself from Eq. (\ref{axialanomaly2}) that the change of axial charge is equal to 8 times the instanton number of the $SU(2)$ gauge field. Coupling the $SU(2)$ gauge theory to the $Z_2$ 2-form gauge field $B$ is effectively turning the $SU(2)$ bundle to an $SO(3)$ bundle which has magnetic monopole excitations. The instanton number for the $SU(2)$ bundle is quantized to be integer. However when we extend the $SU(2)$  bundle to the $SO(3)$ bundle, we have new field configurations involving the $SO(3)$ monopoles, and the quantization of the instanton number is changed. On spin manifolds, the $SO(3)$ instanton number is quantized as half integer. On non-spin manifolds, the smallest $SO(3)$ instanton number can be a quarter. 

The 1/2 instanton event for the $SO(3)$ bundle have the following physical picture. We take two $2\pi$ magnetic flux loops\footnote{The normalization is that the magnetic flux coming out of a single $SO(3)$ monopole is $2\pi$.} initially separated in space and then move them cross each other to form a link\cite{GuQiTI}.\footnote{Notice this event is not allowed in the pure $SU(2)$ bundle because the minimal flux unit is twice as that of the $SO(3)$ bundle.} This spacetime process produce the 1/2 instanton.\footnote{In practice, we can take the first part of Eq. (\ref{axialanomaly1}) involving the $SO(3)$ gauge field and then restrict it to a $U(1)$ subgroup. Inserting a spacetime event as described here, the result of the integral will be 4 instead of 8.} 

We can now give a physical description of the mixed anomaly between $Z_8$ and $(Z_2)_1$. We assign an axial charge of $4$ to two $2\pi$ $SO(3)$ flux loops that have linking number $1$.  The instanton event changes this linking number and hence breaks the $Z_8$ to $Z_4$.  On a non-spin manifold, for example $CP^2$, there is an even smaller instanton event. It can be roughly thought as creating a self-linking of the $2\pi$ $SO(3)$ magnetic flux. 

The type II anomaly involves the second Stiefel-Whitney class of the tangent bundle which detects the spin structure of the base manifold. This anomaly tells us that there is an ambiguity on the quantum statistics of the $2\pi$ $SO(3)$ monopole. Below we will build on these physical characterizations to augment the free Dirac theory with a gapped sector that enables matching the $1$-form anomalies.

Note that the extra anomalies discussed in this section are of the discrete unitary symmetry $Z_8 \times (Z_2)_1$.  For ordinary $0$-form discrete unitary symmetries (at or above $2+1$-D) it is known that their anomalies can always be satisfied by a symmetry preserving gapped topological order. Inspired by this we ask if there can be some symmetry preserving\footnote{Preserving the $1$-form symmetry means the ``physical" loops are tension-full.} $3+1$-D topological order that captures the anomalies of $Z_8 \times (Z_2)_1$. Further note that with anomalous 0-form symmetry, the charge particles will be fractionalized into partons that carry projective representations of the symmetry. Here the anomalous $Z_2$ 1-form symmetry acts on loops. Thus we are lead to search for a topological ordered state of matter that has ``fractionalized" loop excitations. A short introduction and example of such a fractionalized loop phase is given in Appendix \ref{fractionalloops}. Now we describe a postulated topological order that can match the anomalies associated with the 1-form symmetry. It has the following properties. 
\begin{enumerate} 
\item This is a ``loop fractionalized" topological phase that preserves the $(Z_2)_1 \times Z_8$ symmetry. 
\item The specific theory is a $Z_2$ gauge theory where the ``microscopic" loops (we can call them $2\pi$-flux loops) have fractionalized into two $\pi$-flux loops. The physical manifestation of the $(Z_2)_1$ is that the $2\pi$ flux loops are unbreakable. 
\item The $Z_2$ gauge charge carries fermionic statistics. 
\item Two linked electric loops of the $Z_2$ gauge theory carries axial charge 16. These loops can unlink dynamically as there are sources for the electric loops. The linked loops are therefore mixed in with the unlinked loops by the Hamiltonian. The ground state wavefunction contains all electric loop configurations (linked or unlinked), hence the state has global $Z_{16}$ symmetry. 
\item Each electric loop should be thought of as a ribbon. A self-linked loop is assigned axial charge of $8$. Events in the theory that create a single such self-linked loop will break the axial symmetry to $Z_8$.
\end{enumerate} 

Now let us explain why this topological order can match the $Z_8 \times (Z_2)_1$ anomalies. The fermi statistics of electric charge 1 objects ensures that the $(Z_2)_1$ symmetry has the right mixed anomaly with gravity. Gauging the $(Z_2)_1$ symmetry introduces electric charge 1/2 particles. Since the fusion result of two charge 1/2 particles must be the charge 1 particle which is a fermion, these charge 1/2 particles have indefinite statistics. In contrast in a strictly $3+1$-D system it should be possible to assign definite statistics to these particles. This is the manifestation of the mixed anomaly between $(Z_2)_1$ and geometry. 

Introducing electric charge 1/2 particles into the theory implies that the system must also allow strength 1/2 electric loops. These $1/2$ strength electric loops can form links. A link of two 1/2 electric loops will carry axial charge 4. However as there are sources for these loops the linking number can change dynamically.  An event in which two linked strength-$1/2$ electric loops is created changes the axial charge by $4$. This breaks the axial symmetry down to $Z_4$. We also need to consider single strength-$1/2$ loop  that is self-linked. As a self-linked strength-$1$ loop is assigned axial charge $8$, a self-linked charge-$1/2$ loop should be assigned axial charge $2$. Dynamically again the self-linking number can change as there are sources for the loops. It follows that an event where a self-linked strength $1/2$ electric loop created changes the axial charge by $2$. Therefore the axial symmetry is broken down to $Z_2$. These precisely match the mixed anomaly between $(Z_2)_1$ and $Z_8$ axial symmetry.

To recap, the proposed low energy theory is a free massless Dirac fermion augmented with the topologically ordered state just described. What we have argued is that this theory has the same global symmetries, the same local operators, and the same anomalies as the $SU(2)$ gauge theory with an $N_f^A  = 1$ adjoint Dirac fermion (and no spectator fundamental scalar). We do not of course know if the gauge theory really flows to the free Dirac + topological theory but are encouraged by these checks. Alternate possibilities have been discussed in Ref. \onlinecite{ThomasSU(2)}. 

Let us now introduce a finite mass for the spin-$\frac{1}{2}$ spectator fields in our UV theory. With a finite mass spectator, the $Z_2$ 1-form symmetry is explicitly broken. Physically this means that the $2\pi$ flux loops can be broken dynamically. The question is whether the $Z_2$ topological order we described is immediately destroyed dynamically by a finite but large spectator mass. In our case, since the topological order is in a ``fractionalized" loop phase, the $\pi$ flux loops still cannot break and remain as non-trivial excitation in our system. Therefore with a large but finite spectator mass, the $Z_2$ topological order is still stable. 

\begin{figure}
  \centering
    \includegraphics[width=0.8\textwidth]{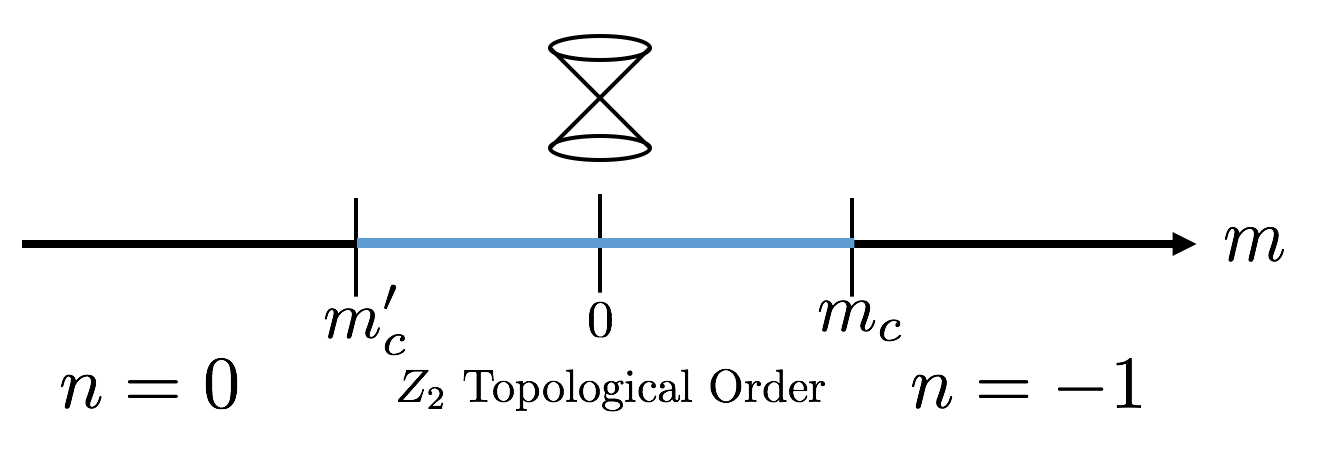}
      \caption{If the $SU(2)+N_f^A=1$ theory is dual to a single Dirac fermion supplemented by a $Z_2$ topological order, the adjoint fermion mass will not drive the system to the large mass confined phase immediately as the $Z_2$ topological order is stable against small perturbation. Increasing the fermion mass the $Z_2$ topological order should go through another phase transition to the confined phases. } 
      \label{PDZ2}
\end{figure}

If the low energy theory of massless $SU(2)+N_f^A=1$ theory with finite spectator mass is indeed a free Dirac fermion plus a decoupled $Z_2$ topological order, then the phase diagram of the theory will be as shown in Fig. (\ref{PDZ2}). Since the $Z_2$ topological order is stable against small perturbation, it will survive until a critical fermion mass $m_c$. The phase transition at $m =0$ occurs entirely in the gapless free Dirac sector, and describes the topological phase transition between the $n = 0$ and either $n = -1$ or $n = 3$ Class AIII topological superconductors (depending on the time reversal properties of the spectators).

\section{Discussion}
From a condensed matter perspective, the main results in this paper are the numerous examples of unusual quantum critical phenomena. Here we briefly describe some general lessons we can learn. 

\begin{enumerate} 
\item
The possibility of multiple universality classes for the same phase transition  (of which we found many examples) arises in many different contexts. As far as we are aware previous examples of this phenomenon are known only in systems with quenched randomness (for instance the $\pm J$ spin glass).  An important context is at heavy electron quantum critical points between a Fermi liquid and an antiferromagnetic metal. The standard Moriya-Hertz-Millis `spin density wave' theory for the onset of antiferromagnetism in a metal has difficulties with the phenomenology observed in some systems. Alternate theories invoke the idea of Kondo breakdown and posit a distinct universality class. However it has never been very clear whether the resulting antiferromagnetic phase is {\em necessarily} sharply distinct from the one obtained through the spin density wave route. It is interesting therefore to contemplate that the heavy fermi liquid to antiferromagnetic metal transition may admit  (at least) two distinct universality classes between the same two phases.  

\item
The existence of ``unnecessary quantum critical points" should be kept in mind when the question of what two phases are separated by a quantum critical point is posed in some condensed matter system. The two phases may actually not be sharply distinct. 

\item
We found a number of examples of band-theory-forbidden quantum critical points between band insulators. This raises the general question of what the criteria are - beyond free fermion theory - for which transitions between free fermion topological phases are allowed to be continuous. Consider for instance integer quantum Hall transitions of  electrons. At the free fermion level it is well known that the quantized values of both electrical and thermal Hall conductivities will generically jump by 1 at a continuous transition. Is this still true in the presence of  interactions? 

\item
Previous examples of deconfined quantum critical points have been in $2+1$-D systems (as far as we are aware). It is encouraging that we have been able to find a number of examples of this phenomenon in $3+1$-D, and which furthermore have emergent non-abelian gauge fields at the critical  point.  All of our  examples describe transitions between gapped short range entangled phases (possibly distinguished as SPT phases). 
It will be interesting to search for other $3+1$-D examples - like in the Neel-VBS transition of $2+1$-D magnets - where a Landau forbidden transition occurs between two symmetry broken Landau-allowed phases. 

\item
 Continuous phase transitions between SPT phases in $3+1$-D have not been much explored (beyond free fermion theory). The examples we have found and the resulting novel phenomena should give impetus to a systematic study of such transitions.

\end{enumerate}

From a high energy perspective, one of our results is to provide an interpretation of some massless gauge theories as quantum critical points. We saw that even when the gauge theory is IR-free it has an interesting place in the phase diagram as a deconfined quantum critical point. Perhaps the most interesting aspect (for quantum field theorists) is our discussion of the possible duality of the $SU(2)$ gauge theory with a massless $N_f = 1$ adjoint (Dirac) fermion, and a massive fundamental boson to a free massless Dirac fermion with an additional decoupled topological field theory. It will be interesting to scrutinize this possibility through numerical studies of  the gauge theory.

\section*{Acknowledgement}
We thank Max Metlitski for innumerable discussions and constant encouragement. We also thank Nati Seiberg for a crucial conversation that 
pointed us toward the examples of multiple universality classes. ZB thanks Liujun Zou, Yuji Tachikawa and Chao-ming Jian for numerous stimulating discussions. We thank the participants of the Aspen Winter Conference on ``Field Theory Dualities and Strongly Correlated Matter" for many useful discussions and feedback. This work was supported by NSF grant
DMR-1608505, and partially through a Simons Investigator
Award from the Simons Foundation to Senthil Todadri. Part of this work was  performed at the Aspen Center for Physics, which is supported by National Science Foundation grant PHY-1607611. ZB acknowledges support from the Pappalardo fellowship at MIT.

\appendix

\section{Instanton number}\label{appinstanton}

By definition, a Yang-Mills instanton is a solution of the classical Euclidean equations of motion with finite action. To find solutions with finite action, we require that the field strength tends to zero at infinity sufficiently fast. Hence, the gauge field asymptotically approaches a pure gauge. All pure gauge configurations, namely $A=U^{-1}dU$, at infinity are classified by
\beq
\pi_3(\mathcal{G})=\mathbb{Z},
\eeq
which is characterized by an integer number, the instanton number. First consider gauge configurations on $\mathbb{R}^4$, which become pure gauge at asymptotic infinity. Given a group $\mathcal{G}$, the instanton number of any such gauge configuration on $\mathbb{R}^4$ is an integer multiple of a minimal positive number. This minimal instanton corresponds to the generator of $\pi_3(\mathcal{G})=\mathbb{Z}$. It is customary in the literature to normalize this minimal instanton so that this has instanton number 1. If $\mathcal{G}$ has a discrete $\mathbb{Z}_2$ subgroup, since $\pi_2(\mathbb{Z}_2)=\pi_3(\mathbb{Z}_2)=\mathbb{Z}_1$, we have
\beq
\pi_3(\mathcal{G}/\mathbb{Z}_2)=\pi_3(\mathcal{G}),
\eeq
which indicates that $\mathcal{G}/\mathbb{Z}_2$ and $\mathcal{G}$ share the same generator for instantons. For any non-abelian group $\mathcal{G}$, an instanton of minimal charge can be obtained by embedding a minimal instanton of $SU(2)$ through an appropriate isomorphism $SU(2) \rightarrow \mathcal{G}$, which is obtained by picking a sub-$SU(2)$ algebra  generated by a long root in the Lie algebra of $\mathcal{G}$. For continuous group $\mathcal{G}$, the instanton number can be calculated from an integral of a local density,
\beq
l_{\mathcal{G}}=c_R\int_{Y_4}\text{Tr}_{R}\left(\frac{F}{2\pi}\wedge\frac{F}{2\pi}\right),
\label{instanton}
\eeq
where $R$ denotes a specific representation we can freely choose, and the coefficient $c_R$ is chosen to make sure that $l_{\mathcal{G}}=1$ for the minimal instanton configuration. Particularly, $c_R$ can be determined by embedding the minimal $SU(2)$ instanton to $\mathcal{G}$ and evaluating the expression above. If we use adjoint representation in Eq. (\ref{instanton}), the normalization coefficient $c_R$ will only depend on the Lie algebra of $\mathcal{G}$ but not the global structure of the group.\footnote{We thank private communication with Yuji Tachikawa for clarifying this point.} Therefore, the formula gives the same result for $\mathcal{G}$ and $\mathcal{G}/\mathbb{Z}_2$, namely $l_{\mathcal{G}}=l_{\mathcal{G}/\mathbb{Z}_2}$. All the instanton numbers we used in the main text are normalized in this way. 

Now let us talk about the relation between the Pontryagin classes and the instanton numbers of $SU(N)$, $SO(N)$ and $Sp(N)$ groups. The first Pontryagin class of a group $\mathcal{G}$ is defined with its fundamental representation as the following,  
\beq
p_1(\mathcal{G})=\frac{1}{2}\int_{Y_4} \text{tr}_f \left(\frac{F}{2\pi}\wedge\frac{F}{2\pi}\right).
\label{p1}
\eeq
For $SU(2)$ group, we get exactly $1$ from Eq. (\ref{p1}) if we plug in the minimal instanton configuration. This indicates that the first Pontryagin class is equal to the instanton number for $SU(2)$ group, namely $p_1(SU(2))=l_{SU(2)}$. This is starting point. Now consider the $SU(N)$ and $Sp(N)$ groups.The minimal instanton number is achieved by embedding the minimal $SU(2)$ instanton configuration in the upper left corner in the gauge configuration as following.
\beq
A_\mu=\begin{bmatrix}
    A^{SU(2)}_\mu &0 \\
    0&0\\
\end{bmatrix}
\eeq
It is obvious that we will get 1 if plug this into the Eq.(\ref{p1}). Therefore, for $SU(N)$ and $Sp(N)$, the first Pontryagin class is equal to their instanton number. 
\beqn
p_1(SU(N))&=&l_{SU(N)}\\
p_1(Sp(N))&=&l_{Sp(N)}
\eeqn

The case for Pontryagin class of $SO(N)$ group is a little complicated. They are defined as the same form in Eq. (\ref{p1}) normalized with the vector representation of $SO(N)$. For $SO(3)$, we can only embed the $SU(2)$ instanton configuration into the $SO(3)$ gauge configurations using the $SU(2)$ adjoint representation. Because of this embedding, for a minimal $SU(2)$ instanton configuration, $p_1(SO(3))$ actually is equal to 4. Hence, $p_1(SO(3))$ is equal to four times of the instanton number. 
\beq
p_1(SO(3))=\frac{1}{2}\int_{Y_4} \text{tr}_{SO(3)} \left(\frac{F}{2\pi}\wedge\frac{F}{2\pi}\right)=4\frac{1}{2}\int_{Y_4} \text{tr}_{SU(2)} \left(\frac{F}{2\pi}\wedge\frac{F}{2\pi}\right)=4l_{SU(2)}=4l_{SO(3)}
\eeq

The embedding for $SO(N)$ with $N>3$ is different. We make use of the fact $SO(N)\supset SO(4)=SU(2)\times SU(2)/\mathbb{Z}_2$, and embed the $SU(2)$ instanton configuration into one of the $SU(2)$ subgroup of $SO(4)$. With this embedding, it is easy to verify that $p_1(SO(N))$ is equal to 2 if we put in a minimal instanton configuration. Therefore, 
\beq
p_1(SO(N>3))=2l_{SO(N)}.
\eeq

The $\theta$-angle of $3+1$-D gauge theories are usually defined so that a configuration of instanton number 1 contributes to the Euclidean action by the phase $\exp(i\theta)$. 

\section{A $2+1$D example of unnecessary continuous phase transition}

In the same spirit as the $3d$ examples, let us give an example in $2d$. We consider the trivial to topological phase transition of the $p\pm ip$ superconductor system with $Z_2\times Z_2^T$ symmetry. The low energy field theory near the phase transition is the following.
\beq
H_{\times 1}=\int d^2x\ \chi^T(i\partial_x \sigma^{10}+i\partial_y \sigma^{30}+m\sigma^{23})\chi, 
\label{2dtsc}
\eeq
where the $Z_2$ symmetry, $Z_2: \chi\rightarrow \sigma^{03}\chi$, is the relative fermion parity symmetry of the two layer. Time reversal symmetry, $\mathcal{T}: \chi\rightarrow i\sigma^{21}\chi$, exchanges the $\pm$ layers. The two symmetries together only admit the mass term in Eq. (\ref{2dtsc}), which guarantees that there is a generic phase transition described by free majorana fermions in the bulk. The edge of the system consists of helical majorana modes described by the following equation.
\beq
H_{edge}=\int dx\ \chi^T(i\partial_x \sigma^3)\chi
\eeq
The $Z_2$ and time reversal transformations are 
\beq
Z_2: \chi\rightarrow \sigma^3\chi,\ \ Z_2^T: \chi\rightarrow i\sigma^2\chi.
\eeq
We can introduce a mass term on the boundary $m_b\chi^T\sigma^2\chi$ which breaks both $Z_2$ and $Z_2^T$ symmetries but preserves a different time reversal symmetry $\tilde{Z_2^T}: \chi\rightarrow -\sigma^1\chi$ (which is the original $Z_2^T$ transformation followed by the $Z_2$ transformation). The domain wall of the $Z_2$ breaking mass term traps a majorana zero mode, labeled by $\gamma$. The $\tilde{Z_2^T}$ symmetry will not change the domain wall background and it just acts trivially on the zero modes, namely $\tilde{Z_2^T}:\gamma\rightarrow \gamma$. 

Now let us consider 8 copies of the same system and impose an $SO(7)$ symmetry which rotates these 8 copies in the spinor representation. This symmetry only allows a uniform mass term. The low energy theory near the phase transition is the following. 
\beq
H_{\times 8}=\int d^2x\ \sum_{i=1}^8\chi^T_i(i\partial_x \sigma^{10}+i\partial_y \sigma^{30}+m\sigma^{23})\chi_i, 
\label{2dtsc}
\eeq
When $m$ is tuned from negative to positive, the system goes through a continuous phase transition described by bulk free majorana fermions. This transition is stable against small interactions. Our goal now is to show that $m<0$ and $m>0$ phases are in fact the same phase. We can always regularize the system such that $m<0$ phase is trivial. In the $m>0$ phase, the natural edge state has 8 copies of helical majorana modes with $Z_2\times Z_2^T\times SO(7)$ symmetry. We will argue that the boundary modes can be gapped out while preserving all the symmetries, which indicates the $m>0$ phase is actually topologically trivial. 

To that end we first break the $Z_2$ and $Z_2^T$ symmetry on the edge by adding $m_b\sum_{i=1}^8\chi_i^T\sigma^2\chi_i$. Then we proliferate the topological defects of this order parameter, namely the domain walls, to restore a symmetric gapped edge. Since there are zero modes residing at the domain wall of the order parameter, we have to be careful about their condensation. The domain wall must have single gapped ground state and it has to be symmetric under the combined $\tilde{Z_2^T}$ symmetry. This can be precisely achieved by the $SO(7)$ invariant Fidkowski-Kitaev interaction. Therefore, with this interaction, we can safely condense the domain wall to get a symmetric gapped edge state. Thus, the $m>0$ phase is topologically trivial. The phase diagram of the system is similar to the previous cases as shown in Fig. (\ref{upt}).

\section{Fermion zero modes and time reversal transformations}\label{zeromodes}
In this appendix, we consider a $2+1$-D Dirac fermion in a $2\pi$ flux background and solve the zero mode wavefunction. Then we will consider the time reversal transformation on the zero mode. 

Let us first write down the Hamiltonian for the $2+1$-D Dirac fermion on a flat 2-dimensional plane with a background gauge field 
\beq
H=\psi^\dagger((i\partial_x-A_x)\sigma^x+(i\partial_y-A_y)\sigma^z)\psi=\psi^\dagger\mathcal{D}\psi, 
\eeq
where we take the Landau gauge $A_x=0, A_y=Bx$. Notice this is equivalent to the spherical geometry, since the flat plane can be viewed as the infinite radius limit of the sphere. The time reversal transformation for the fermion fields is
\beq
Z_2^T: \psi\rightarrow i\sigma^y \psi^\dagger.
\eeq
In component form, the time reversal action is
\beq
Z_2^T:\ \ \psi_1\rightarrow \psi_2^\dagger, \ \ \ \psi_2\rightarrow -\psi_1^\dagger. 
\eeq
This time reversal transformation will flip the electric charge of the Dirac fermions but keep the magnetic flux background invariant. Therefore, it is meaningful to discuss the time reversal transformation of the zero modes trapped in the flux background. 

Consider the Dirac equation
\beq
\mathcal{D}\psi=(i\partial_x\sigma^x+(i\partial_y-Bx)\sigma^z)\psi=\varepsilon \psi.
\eeq
The usual trick to solve the Dirac equation is to square the Dirac operator to get 
 \beq
(p_x^2+(p_y-Bx)^2+B\sigma^y)\psi=\varepsilon^2\psi.
\eeq
The spectrum for $\varepsilon^2$ is (in the unit with $\hbar=c=1$)
\beq
\varepsilon^2=|B|(2n+1)-|B|,\ \ \ n\in\mathbb{Z}. 
\eeq
Notice that the zero mode wavefunction depends on the sign of magnetic field $B$. Consequentially the time reversal transformation on the zero modes are different for $\pm B$. For $B>0$, the zero mode operator is 
\beq
f_+=(\psi_1-i\psi_2)\phi_0(p_y,x).
\eeq
While for $B<0$, the zero mode operator is 
\beq
f_-=(\psi_1+i\psi_2)\phi_0(p_y,x),
\eeq
where $\phi_0(p_y,x)$ is the wavefunction for the ground state of a harmonic oscillator.

Thus we find that the time reversal transformations for the zero modes are   
\beqn
Z_2^T:&& f_+\rightarrow (\psi_2^\dagger-i\psi_1^\dagger)\phi_0^*(p_y,x)=-if_+^\dagger\\
&&f_-\rightarrow (\psi_2^\dagger+i\psi_1^\dagger)\phi_0^*(p_y,x)=if_-^\dagger
\eeqn

\section{Band-theory-forbidden quantum criticality between two band insulators: a simple example}
\label{bbtqcp2d}
A system of free fermions can be in a gapped ground state. Distinct such states in any given system are labeled by topological invariants. Within free fermion theory there are rules on what kinds of continuous phase transitions can occur between these distinct phases. Roughly speaking the topological invariant jumps by the smallest possible amount for a direct second order transition to be possible. The best known example is in a system of fermions with charge-$1$ under a global $U(1)$ symmetry in two space dimensions. Gapped free fermion ground states show an integer quantum Hall effect characterized by a quantized electrical  Hall conductivity $\sigma_{xy} = n \in \mathbb{Z}$, and a quantized thermal Hall conductivity $\kappa_{xy} = \bar{c}\frac{\pi^2 k_B^2 T}{3}$ with $\bar{c} = n$.  These distinct phases can be realized within band theory by fully filling a band with Chern number $n$. In the absence of any other symmetries\footnote{Lattice translation may or may not be present and makes no difference to this discussion.}, within free fermion theory, continuous phase transitions between these distinct phases are possible if and only if $n$ jumps by $1$. 

Can such rules be changed in the presence of interactions?  The classification of the gapped phases can itself of course be changed by interactions but here we are interested in the phenomenon of band-theory-forbidden quantum criticality between band allowed phases. For the standard integer quantum Hall system discussed above it is not known to us if the rule $\Delta n = 1$ survives the inclusion of interactions. However a closely related system provides us with a simple  example where interactions modify an analogous band theory rule.

Consider a system with two species of fermions - denoted $\psi$ and $\chi$ - in two space dimensions. We will assume that there is a global $U(1)$ symmetry under which $\psi$ has charge-$1$ and $\chi$ is neutral. Within free fermion theory gapped ground states of this system  are now characterized by a pair of integers $(n, m)$. The electrical Hall conductivity is $\sigma_{xy} = n$ while the thermal Hall conductivity 
$\kappa_{xy} = \bar{c}\frac{\pi^2 k_B^2 T}{3}$ with $\bar{c} = \frac{m}{2}$. Compared to the standard integer quantum Hall system the presemce of the additional neutral fermion means that $\bar{c}$ can take any multiple of half-integer value and is not tied to $\sigma_{xy}$. Within free fermion theory, a generic continuous transition between these phases satisfies the following rules: (i) $\Delta n =1, \Delta m = 2$, or (ii) $\Delta n = 0$, $\Delta m = 1$.  The former can be understood as a quantum Hall transition of the $\psi$ fermion and the latter as a transition of the $\chi$ fermion. 

Now  we will show that this rule can be violated in the presence of short ranged interactions that preserve the global $U(1)$ symmetry.  Imagine an interaction such that the charged fermion forms a $3$-body bound state  (a ``cluston"\cite{cluston}) $\psi_3 \sim \psi \psi \psi$.  A cluston integer quantum Hall state\cite{cluston} is clearly then possible with $\sigma_{xy} = 9k$, and $\bar{c} = k$ with $k \in \mathbb{Z}$. In this system where both charged and neutral fermions are present, such a cluston integer quantum Hall state can also be accessed within free fermion theory: it corresponds to 
$n = 9k, m = 2k$. Now consider a cluston integer quantum Hall transition which can be second order so long as  $\Delta k=1$. 
This corresponds to $\Delta n = 9, \Delta m = 2$ which violates the band theory rules discussed in the previous paragraph even though both phases are band-allowed.  The critical theory has gapless clustons but the $\psi, \chi$ particles are gapped.

\section{A $2+1$-D bosonic Mott insulator to $Z_2$ topological order transition}\label{2dZ2SL}

Here we provide another example of continuous phase transition in which modifying the properties of a gapped spectator field changes the nearby phase however not the universality class of the transition. We consider a transition from a $2+1$-D bosonic Mott insulator to a $Z_2$ topological order. 

Consider a bosonic system in a Mott insulating phase. The physical bosons $b$ are gapped. We assume the system has a time reversal symmetry $\mathcal{T}$ and the physical bosons are Kramer's singlets. Now consider partons of the physical boson. We decompose the physical boson into two bosonic partons which we call the $e$ particles. This fractionalization introduces a $Z_2$ gauge field and the $e$ particles carry $Z_2$ gauge charge 1. The $Z_2$ gauge field also has $\pi$ flux excitations which we label as $m$ particle. The $e$ and $m$ particles have mutual Berry's phase $\pi$. The Mott insulating phase is the confined phase of the $Z_2$ gauge field. The $Z_2$ confined phase can be viewed as a condensed phase of the $m$ particles.  

Let us imagine by tuning some parameter we can drive the system through a deconfinment transition to a $Z_2$ topological order. We can view the deconfinement transition as the proliferation of the vortices of the $m$ particle condensation. The transition is in the Ising universality class. After the transition, the $m$ particle is gapped and the $Z_2$ gauge field is deconfined. The resultant phase has a $Z_2$ topological order. Throughout the transition, the $e$ particle remains gapped and does not participate in the low energy theory. We can view them as the massive spectator fields in our system. Since the system has time reversal symmetry, there are actually different classes of $Z_2$ topological orders distinguished by their time reversal properties. These are called Symmetry Enriched Topological (SET) orders. In our case, the time reversal properties of the spectator $e$ particle precisely determine which SET state we get for the deconfined phase. There are two choices. One is $e$ particle is a Kramer's singlet. In this case, the resultant deconfined phase is a vanilla $Z_2$ topological order which we can label as $e0m0$ meaning that both $e$ and $m$ are Kramer's singlet. The other choice is that the $e$ particle actually carries a Kramer's doublet.\footnote{We can fractionalize the boson as $b\sim (e_1\partial_x e_2-e_2\partial_x e_1)$. The time reversal transformation on the $e$'s is $\mathcal{T}: e\rightarrow i\sigma^y e$.} In this case, we get a non-trivial symmetry enriched $Z_2$ topological order labeled by $eTm0$. $eTm0$ and $e0m0$ are distinct phases if the system preserves the time reversal symmetry. However, since the $e$ particle remains gapped during the transition, it cannot change the universality class of the transition. 

\section{Fractionalizing loops}
\label{fractionalloops}
Consider a quantum system with a Hilbert space of unoriented loops in $3+1$-D.   In this Appendix we briefly describe phases of such a system where the loops have `fractionalized'. 
Loop fractionalization played an important role in the topological order discussed in Sec. \ref{4ddrcdual}. 

We can think of the system of unoriented loops  as pure $Z_2$ gauge theory, {\em i.e} without any matter.  Formally such a system has a global $Z_2$ 1-form symmetry (denoted $(Z_2)_1$) associated with the unbreakable unoriented loops. There are some obvious phases of this loop system. First there is a phase in which the loops have line tension and there are no other excitations. This is the confined phase of the pure $Z_2$ gauge theory. The 1-form $(Z_2)_1$ symmetry remains unbroken in this phase. Second there is a distinct phase where the loops have zero line tension, and the pure $Z_2$ gauge theory is in its deconfined phase. Then $(Z_2)_1$ is spontaneously broken. 

Here we are interested in a different kind of phase where the microscopic loops (denoted strength-$1$) are tension-full but have `fractionalized' into other loops. In other words the 1-form global symmetry has been fractionalized. 
We  will describe a simple example of such a loop fractionalized phase where there are two kinds of excitations: 
\begin{enumerate}
\item
 A strength-$1/2$  loop with  line tension. Two strength-$1/2$ loops fuse into a single microscopic strength-$1$ loop.
  
 \item
 A point particle excitation that braids with $\pi$ phase with the fractionalized strength-$1/2$ loop. 
 \end{enumerate}
 
 This excitation structure is that of an emergent deconfined $Z_2$ gauge theory (not to be confused with the microscopic pure $Z_2$ gauge theory).  Let us explicitly construct this phase. 
 To that end we begin  by considering first  a $U(1)$ gauge theory with a $(Z_2)_1$ symmetry. This theory has a gapless photon, gapped electric charges $E$, gapped magnetic charges $M$, and their bound states. Now assume that all particles with odd magnetic charge are thrown out of the $U(1)$ gauge theory. Then odd strength magnetic loops cannot end and there is an exact $(Z_2)_1$ symmetry. This symmetry is broken spontaneously in the $U(1)$ gauge theory (the odd strength magnetic loops are tension-less).  Consider now a Higgs transition obtained by condensing the basic $E$ particle. All magneic flux loops will then have line tension, and we will get the ``trivial" phase of loops with unbroken $(Z_2)_1$. If instead we consider a Higgs transition obtained by condensing $E^2$ without condensing $E$, we will get a $Z_2$ gauge theory where $E$ survives as the $Z_2$ gauge charge. We also get strength-$1/2$ magnetic flux loops with line tension which braid with $\pi$ phase with the $Z_2$ gauge charge. Of course strength-$1$ magnetic loops also have line tension, and cannot break. We identify them with the microscopic loops.  This state preserves $(Z_2)_1$ and is exactly the loop fractionalized phase described above\footnote{An alternate construction of the same phase is to start with a standard deconfined $Z_4$ gauge theory, and throw out all particles with odd $Z_4$ charge. This builds in a $(Z_2)_1$ symmetry into the theory associated with the $Z_4$ flux loop with  even flux. This loop does not braid non-trivially with any other excitation, and has line tension. However it is fractionalized into two odd flux loops which themselves braid with phase $\pi$ with the particle with even $Z_4$ charge.} 
 
 An effective field theory for this loop fractionalized phase is readily written down. Consider the Lagrangian
 \begin{equation}
 {\cal L} = \frac{1}{\pi} \beta \wedge d\alpha + \frac{1}{2\pi} B \wedge d\alpha
 \end{equation}
 where $\alpha$ is a $1$-form dynamical gauge field, and $\beta$ is a $2$-form dynamical gauge field. $B$ is a $2$-form background gauge field that couples to the global $(Z_2)_1$ symmetry. The first term is the standard ``BF" theory description of $Z_2$ gauge theory. It dictates that the strings that are charged under $\beta$ are seen as $\pi$ flux of the $\alpha$. These strings are the tension-full loops of the $Z_2$ gauge theory. The `microscopic' loops that couple to $B$ however have $2\pi$ flux of $\alpha$. Thus this action correctly captures the loop fractionalized phase described above.

\section{The $SU(2)+N_f^A$ theory with $N_f^A\in 2\mathbb{Z}$}
\label{nfeven}

In this appendix, we provide generalizations of the previous fermionic deconfined quantum critical points. We extend the $SU(2)+N_f^A$ theories to even $N_f^A$ cases. 

Let us consider an $SU(2)$ gauge theory coupled to $N_f^A=2$ flavors of adjoint Dirac fermions. The $3+1$-D Lagrangian of this theory is 
\beq
\mathcal{L}^{SU(2)}_{Adj2}=\sum_{i=1}^2i\bar{\psi}_i\gamma_\mu(\partial_\mu-a_{\mu}^\alpha T^\alpha)\psi_i-m\bar{\psi}_i\psi_i+...,
\label{su2adj2}
\eeq
Analytically, this theory is expected to be inside the conformal window\cite{BZfund,BZadj2}. Numerically, it is found that the infrared limit for the $m=0$ theory is consistent with a conformal field theory\cite{BZadj1}. We want to understand what phase transition this theory describes. To be more precise we will content ourselves with determining the topological distinction between the phases with the two signs of $m$ assuming large $|m|$.  We will not attempt to answer the question of whether there is are other intermediate phases at small $|m|$. Accordingly whenever we talk about the massive theory below we implicitly mean the theory at large $|m|$.  If we tune $m$ to be non-zero, the fermions are gapped. As usual, we can regularize the theory such that for $m<0$ integrating out the massive fermion generates zero $\Theta$-angle for the $SU(2)$ gauge theory, in which case the theory will enter a confined phase in the low energy. For $m>0$, the massive fermions contribute an $8\pi$ $\Theta$-angle for the $SU(2)$ gauge fields. Since the $\Theta$ term is $2\pi$ periodic, the $SU(2)$ gauge theory will again confine in the infrared limit. The question is what is the nature of the gapped phases for $m<0$ and $m>0$. The two states can only differ in  their topological aspects. They can be different SPT states of certain global symmetry. 

For general mass $m$, the global symmetry of the theory is $\mathcal{G}=SO(4)\times Z_2^T$. The time reversal symmetry transformation is as usual
\beq
Z_2^T: \psi\rightarrow \gamma_0\gamma_5\psi^\dagger,
\eeq
where we suppressed the flavor and gauge indices. To see the $SO(4)$ symmetry, we decompose the 2 flavors of Dirac fermions into 4 flavors of Majorana fermions. The $SO(4)$ symmetry is then a flavor rotation between the 4 Majoranas. Since the $SU(2)$ adjoint representation is a real representation, the $SO(4)\times Z_2^T$ symmetry commutes with the gauge group. This is not anomalous and is an exact symmetry for any $m$.

Let us first discuss the classification of interacting fermion SPT with $SO(4)\times Z_2^T$ symmetry. In the free fermion limit, the $3+1$-D fermion SPT classification is $\mathbb{Z}$. The root state for this class is 4 copies of topological superconductors with $Z_2^T$ symmetry (DIII class), where the $SO(4)$ rotates among the 4 copies. The typical surface theory of such root state is 4 copies of gapless Majorana fermions. In the free fermion limit, since DIII class is $\mathbb{Z}$ classified, the classification of $SO(4)\times Z_2^T$ SPT is $\mathbb{Z}$ as well. With interaction, the classification becomes $\mathbb{Z}_4\times \mathbb{Z}_2$. The $\mathbb{Z}_2$ part corresponds to the pure $Z_2^T$ SPT state labeled by its anomalous surface $Z_2$ topological order $efmf$, which only appears in interacting system and has no free fermion correspondence. The free fermion $\mathbb{Z}$ classification is reduced to $\mathbb{Z}_4$ by interaction. The reason is the following. The pure time reversal anomaly on the $2+1$-D surface is $\mathbb{Z}_{16}$ classified, which means that multiples of 16 copies of $2+1$-D Majorana fermions are time reversal anomaly free. Therefore, we at least need 4 copies of the root state to cancel the time reversal anomaly on the surface. Next, we need to consider the mixed anomaly between $SO(4)$ and $Z_2^T$. This is related to the generalized parity anomaly. According to \cite{EWnonorientable}, by considering the system on general unorientable manifold, the surface theory of 4 copies of the root states will be free from the mixed anomaly between $SO(4)$ and $Z_2^T$ symmetry. Physically, it means that the $SO(4)$ monopole in the $3+1$-D bulk carries trivial time reversal quantum number. Combining the two constraints, we conclude that the interaction classification reduced from the free fermion states is $\mathbb{Z}_4$. We can also see this from a surface argument. Let us take 4 copies of the root states. The boundary theory consists of 16 copies of Majorana fermions. The 16 Majorana forms 4 copies of vector representations under $SO(4)$. The question is whether we can gap them out while preserving the $SO(4)\times Z_2^T$ symmetry. We can group the 16 Majorana fermions into 8 Dirac fermions and assume there is an extra $U(1)_e$ symmetry for the Dirac fermions which we will eventually explicitly break. The Dirac fermion has two indices, an $SO(4)$ vector index $v=1,2,3,4$ and another flavor index $i=1,2$. 
\beq
H_{\times 4}=\sum_{i=1}^{2}\sum_{v=1}^4\psi_{i,v}^\dagger(i\partial_x\sigma^x+i\partial_y\sigma^z)\psi_{i,v}
\eeq
The symmetry transformations are
\beqn
U(1)_e&:& \psi_{i,v}\rightarrow e^{i\theta}\psi_{i,v} \\
Z_2^T&:& \psi_{i,v}\rightarrow i\sigma^y\psi_{i,v}^\dagger \\
SO(4)&:& \psi_{i,v}\rightarrow O_{v,w}\psi_{i,w}, \ \ O\in SO(4)
\eeqn 
Now we introduce a superconducting order parameter just as in Eq. (\ref{pairing}). This breaks both $U(1)_e$ and $Z_2^T$ but preserves a combination of $Z_2^T$ and $U(\pi/2)$ rotation. Consider the $\pi$ vortex of the superconductor order parameter. It carries 8 Majorana zero modes labeled by $\chi_{i,v}$. We can combine them into 4 complex zero modes, $f_{v}=\chi_{1,v}+i\chi_{2,v}$. We can write down an $SO(4)$ invariant four fermion interaction of the form $H_{int}=-V(f_1^\dagger f_2^\dagger f_3^\dagger f_4^\dagger+h.c.) $\cite{smg2017}. This interaction leads to an $SO(4)$ symmetric ground state $|\psi_v\rangle=(|0\rangle+f_1^\dagger f_2^\dagger f_3^\dagger f_4^\dagger|0\rangle)/\sqrt{2}$ and a gapped spectrum for the vortex core. Now we can condense the $\pi$ vortices and restore the $U(1)_e$ and $Z_2^T$ symmetry. The resultant surface state is a trivial gapped symmetric state under $U(1)_e\times SO(4)\times Z_2^T$ symmetry. We can then turn on a small explicit $U(1)_e$ breaking term. Since the surface is now trivially gapped, it is stable against any small perturbation. Thus, we proved that the surface of 4 copies of the root states can be trivially gapped while preserving the $SO(4)\times Z_2^T$ symmetry, which is equivalent to saying that the bulk state is topologically trivial.  

Next we want to determine which SPT state the $m>0$ phase falls into. We can always regularize the system such that $m<0$ phase is the trivial class of the SPT states under this global symmetry. To detect the topological properties of the $m>0$ phase, we can derive the topological response for the background $SO(4)$ gauge field on an orientable manifold. 
\beqn
\mathcal{S}_{topo}&=&i\frac{\pi}{2}\left(p_1(A^{SO(12)})-\frac{12}{8}\sigma\right)\\ \nonumber
&=&i\frac{\pi}{2}\left(3p_1(A^{SO(4)})+4p_1(a^{SO(3)})-\frac{6}{4}\sigma\right)\\ \nonumber
&=&i3\pi\left(\frac{1}{2}p_1(A^{SO(4)})-\frac{1}{4}\sigma\right)+i2\pi p_1(a^{SO(3)})\\
&=&i3\pi\left(S_\Theta^{SO(4)}-\frac{1}{4}\sigma\right).
\eeqn
This non-trivial response theory tells us that the $m>0$ state is indeed a non-trivial SPT protected by the $SO(4)\times Z_2^T$ symmetry. 

As before, to understand the theory we need to introduce the spin-$\frac{1}{2}$ spectator field. Let us take the simplest case where the spectator is a scalar under $SO(4)$ and a singlet under $Z_2^T$ as in Eq. (\ref{spectator1}). In this case, we can do similar surface analysis as in previous sections to understand the $m>0$ phase. The natural surface state of the $m>0$ system is $SU(2)$ $QCD_3$ with 2 flavors of adjoint massless Dirac fermions. We can condense the trivial spectator boson to Higgs out the $SU(2)$ gauge field. The surface state results in 6 physical Dirac fermions or 12 physical Majorana fermions with identical time reversal symmetry transformation. This state corresponds to $n=3\sim -1$ in the $\mathbb{Z}_4$ classification. 

Now if the spectator is a scalar under $SO(4)$, and a Kramers doublet under $Z_2^T$ transforming as in Eq. (\ref{TRforz}),  we can run a similar argument as in previous section.  Condensation of the spectator field will not break the {\em physical} time reversal symmetry. The system will be invariant under a gauge equivalent time reversal transformation $\tilde{Z}_2^T$ as in Eq. (\ref{newTRz}). Since the gauge transformations also change the time reversal transformation on the adjoint fermions as in Eq. (\ref{newTRf}), the resulting state is now $n=-2+1=-1\sim3$ in the $\mathbb{Z}_4$ classification.  

Notice in this case the topological index of the $m>0$ phase actually does \emph{not} depend on the two choices of the spectator fields. This is indeed consistent with the bulk analysis. We will show that in this case, the neutral $SO(3)$ monopole in the bulk can be a Kramers singlet boson. Therefore, the two choices of the spectator fields do not have different surface time reversal anomaly. To consider the zero modes in the $SO(3)$ monopole, we consider the system with a sphere geometry and set the background $SO(3)$ gauge field such that there is $2\pi$ magnetic flux coming out of the sphere along $z$ direction in the flavor space. For the $m>0$ phase, the surface theory hosts gapless Dirac fermions which contribute zero modes for the monopole configuration. Let us write down the surface state. 
\beq
\label{surfaceQED2}
\mathcal{L}^{N_f^A=2}_{surf}=\sum_{i=1}^2i\bar{\psi}_{+,i}\gamma_\mu(\partial_\mu-ia_\mu^z)\psi_{+,i}+\sum_{i=1}^2i\bar{\psi}_{-,i}\gamma_\mu(\partial_\mu+ia_\mu^z)\psi_{-,i}+\sum_{i=1}^2i\bar{\psi}_{0,i}\gamma_\mu\partial_\mu\psi_{0,i}
\eeq
Here we see there are three classes of Dirac fermions which carry $\pm1$ and $0$ gauge charge under $a_\mu^z$ respectively. Each class has two flavors. For each class, we can decompose the 2 Dirac fermions into 4 Majorana fermions and they form a vector representation of the global $SO(4)$ symmetry. Now let us consider the zero modes in the $2\pi$ flux. There are in total 4 complex zero modes labeled by $f_{+,1},f_{+,2},f_{-,1},f_{-,2}$. The $\pm$ denote the gauge charge they carry. The Hilbert space spanned by these zero modes has 16 states. It is very easy to spot which states are gauge invariant but a little difficult to construct an $SO(4)$ scalar. To start let us consider the 4 states constructed from $f_{+,1}$ and $f_{+,2}$. They can be group into two classes: $\{|0\rangle, f_{+,1}^\dagger f_{+,2}^\dagger|0\rangle\}$ and $\{ f_{+,1}^\dagger|0\rangle, f_{+,2}^\dagger|0\rangle\}$. These two sets of states  form the left and right handed spinor representation of the $SO(4)$ group.\footnote{In general, for $2n$ Majorana zero modes, they form a vector representation of an $SO(2n)$ group and they host $2^n$ dimensional Hilbert space. This Hilbert space can always be decomposed into left and right handed spinor representations of the $SO(2n)$ symmetry.} We know that for $SO(4)$ two left or two right spinors can be combined into an $SO(4)$ scalar.\footnote{This is actually true for all $SO(4\mathbb{Z})$ group.} Therefore, combining two left or right handed spinors from $f_{+}$ sector and $f_{-}$ sector, we can form such a gauge neutral and $SO(4)$ singlet state, for example $(f_{+,1}^\dagger f_{-,2}^\dagger-f_{+,2}^\dagger f_{-,1}^\dagger)|0\rangle$. Under time reversal transformation, $Z_2^T: |0\rangle\rightarrow f_{+,1}^\dagger f_{+,2}^\dagger f_{-,1}^\dagger f_{-,2}^\dagger|0\rangle, f_{\pm,i}\rightarrow \mp if_{\pm,i}^\dagger$, this state goes back to itself. Therefore, the gauge and global neutral $SO(3)$ monopole is a Kramers singlet boson. This corresponds to a trivial $m$ particle for the surface $Z_2$ topological order which indicates that it is not anomalous. Hence, the two spectator choices make no difference on the topological index of the $m>0$ phase. 

The above analysis also suggests a possible duality between the $3+1$-D $SU(2)+N_f^A=2$ theory and two free Dirac fermions with $SO(4)\times Z_2^T$ symmetry as they both describe the continuous phase transition between the $n=0$ and $n=-1$ SPT states in this symmetry class. However we will leave  to future study an analysis of the emergent symmetries and anomalies of the gauge theory.

For even $N_f^A>2$, the global symmetry of the system is $SO(2N_f^A)\times Z_2^T$ and the interacting fermionic SPT classification is the same as the $N_f^A=2$ case. The $SU(2)+N_f^A$ theory is also a theory of quantum phase transition between $n=0$ and $n=-1$ SPT states in this symmetry class. However, in this case the gauge theory is free in the infrared limit. Therefore, we can tell with confidence that it is distinct from the phase transition theory in the free fermion setting. Thus this provides other examples of multiple universality classes for the same phase transition. 

\section{The $SU(2)+N_f^A$ theory with $N_f^A\in\mathbb{Z}+\frac{1}{2}$}
\label{nfhalf}
The meaning of $N_f^A$ being half integer  is  that we consider, instead of Dirac fermions, Majorana fermions. Since the adjoint representation of $SU(2)$ is a real representation, we can easily generalize the theory to Majorana fermions. We thus consider $2N_f^A=2k+1$ ($k\in\mathbb{Z}$) flavors of $SU(2)$ adjoint Majorana fermions whose $3+1$-D action can be written down as
\beq
\mathcal{L}^{SU(2)}_{AdjMaj}=\sum_{i=1}^{2k+1}i\chi^T_i\gamma_\mu(\partial_\mu-a^\alpha_\mu T^\alpha)\chi_i-m\bar{\chi}_i\chi_i+...
\eeq
(We still assume massive spin-$\frac{1}{2}$ spectator fields in the spectrum of our system.)The massless theory with $N_f^A=\frac{3}{2}$ or $k=1$ is inside the conformal window of adjoint $SU(2)$ gauge theory. For $N_f^A>2$ or $k>2$, the massless theory flows to the free fixed point in the infrared. 

Let us first discuss the dynamical properties of the massive phase. As before, the $m<0$ phase can be regularized to have a trivial $\Theta$-angle for the $SU(2)$ gauge theory and it enters a confined phase at low energy. In the $m>0$ side, the $\Theta$-angle for $SU(2)$ is $4k\pi+2\pi$, which is also trivial because it is a multiple of $2\pi$. Therefore, the $m>0$ side also enters a confined phase. As in all the other examples before, the two phases are not distinguished by their dynamical properties but their topological properties. 

The global symmetry in this system is $SO(2k+1)\times Z_2^T$. The fermion SPT classification for this symmetry is $\mathbb{Z}_{16}\times \mathbb{Z}_2$. $\mathbb{Z}_2$ part is the $efmf$ state protected by $Z_2^T$ only. The $\mathbb{Z}_{16}$ part is descendent from the free fermion classification. The root state is $2k+1$ copies of topological superconductor in DIII class. The $2k+1$ copies form a vector representation of $SO(2k+1)$. Since the time reversal anomaly for DIII class is $\mathbb{Z}_{16}$ fold and $2k+1$ is coprime with 16, at least we need 16 root states to cancel the time reversal anomaly on the surface. For 16 copies of the root states, there is also no mixed anomaly between $SO(2k+1)$ and $Z_2^T$. (Using the argument in the previous section, the mixed anomaly is 4-fold periodic.) Therefore 16 copies of root states is the minimal number for an anomaly free surface. Hence, the interaction reduced classification is $\mathbb{Z}_{16}$. 

Let us now discuss the nature of the $m>0$ phase. We can derive the topological response theory for the background $SO(2k+1)$ gauge field on the $m>0$ side on an orientable manifold. 
\beqn
\mathcal{S}_{topo}&=&i\frac{\pi}{2}\left(p_1(A^{SO(3(2k+1))})-\frac{3(2k+1)}{8}\sigma\right)\\ \nonumber
&=&i\frac{\pi}{2}\left(3p_1(A^{SO(2k+1)})+(2k+1)p_1(a^{SO(3)})-\frac{3(2k+1)}{8}\sigma\right)\\ \nonumber
&=&i3\pi\left(\frac{1}{2}p_1(A^{SO(2k+1)})-\frac{2k+1}{16}\sigma\right)+i\pi(4k+2)p_1(a^{SU(2)})\\
&=&i3\pi\left(S_\Theta^{SO(2k+1)}-\frac{2k+1}{16}\sigma\right).
\eeqn
Here we use the fact that for both choices of spectator field on an orientable manifold the gauge bundle must satisfy $w_2(SO(3))=0$ mod $2$, which means the gauge bundle is a pure $SU(2)$ bundle. This response theory, while not revealing all the information about the $m>0$ phase, does tell us  that the $m>0$ phase is topologically non-trivial. We still need to determine which SPT the $m>0$ phase is in. 

We find that the nature of the $m>0$ phase depends on the properties of the spectator field. Assuming a spectator boson which is a $SO(2k+1)$ scalar and time reversal singlet as in Eq. (\ref{spectator1}), the topological index for the $m>0$ phase is $n=3$ state in the $\mathbb{Z}_{16}$ classification. For the other case of a time reversal doublet spectator as in Eq. (\ref{TRforz}), the topological index is $n=-1$. The arguments for these results are straightforward generalization of surface arguments in Section IV. We note that the difference between the two cases is $n=4$ state which is not the $eTmT$ state in this situation. ($eTmT$ state would correspond to $n=8$ state in the $\mathbb{Z}_{16}$ classification.)

The time reversal singlet spectator case gives us another example of band-theory-forbidden continuous transition between band-theory-allowed insulating states. For the time reversal doublet spectator case, with $k=1$ or $N_f^A=\frac{3}{2}$, the massless $SU(2)+N_f^A=\frac{3}{2}$ theory is a strongly coupled conformal field theory in the gauge theory description.  For $k>1$ or $N_f^A>2$, the massless $SU(2)+N_f^A$ theory is free in the infrared. This theory is clearly different from $2k+1$ free massless Majorana fermions. However, both theories describe the same $n=0$ to $n=-1$ transition. Therefore, this provides more examples for multiversality classes. 

\begin{table}
\begin{center}
\label{moreNf}
\begin{tabular}{| c |c |c |c|}
\hline
$N_f^A$ & $\mathbb{Z}+\frac{1}{2}$ & $2\mathbb{Z}+1$ & $2\mathbb{Z}$ \\ 
\hline
 Symmetry $\mathcal{G}$ & $SO(2\mathbb{Z}+1)\times Z_2^T$ & $SO(4\mathbb{Z}+2)\times Z_2^T$ & $SO(4\mathbb{Z})\times Z_2^T$ \\  
\hline
 fSPT classification & $\mathbb{Z}_{16}\times \mathbb{Z}_2$ & $\mathbb{Z}_{8}\times \mathbb{Z}_2$ & $\mathbb{Z}_{4}\times \mathbb{Z}_2$ \\  
\hline
 $\mathcal{T}^2=1$ spectator & $n=3$ & $n=3$ & $n=3\sim -1$\\ \hline
 $\mathcal{T}^2=-1$ spectator & $n=-1$ & $n=-1$ & $n=-1\sim 3$ \\ \hline
\end{tabular}
\end{center}
\caption{A summary for $SU(2)+N_f^A$ theory with general $N_f^A$. The global symmetry associated with the system is $SO(2N_f^A)\times Z_2^T$. The SPT classification depends on the $N_f^A$. The last two rows show the topological index for the $m>0$ phase for both choices of the spectator field.} 
\end{table}

A summary of all the results from different $N_f^A$ series is tabulated in Table \ref{moreNf}.
 
\section{An $SU(4)$ generalization}
\label{su4}
In this appendix, we explore a generalization with a different gauge group. To make things simple let us first restrict our attention to only 1 flavor of adjoint Dirac fermion. 
Let us consider $3+1$-D $SU(4)$ gauge theory coupled to one flavor of adjoint fermion. The adjoint Dirac fermion has 15 components. The Lagrangian is written as the following,
\beq
\mathcal{L}^{SU(4)}_{Adj}=i\bar{\psi}\gamma_\mu(\partial_\mu-a_{\mu}^aT^a)\psi-m\bar{\psi}\psi+...,
\label{su4adj}
\eeq
where $T^a$'s, the generators of $SU(4)$ group, are $15\times 15$ matrices. The infrared limit of the massless theory is still unclear. Let us assume it is inside the conformal window for the moment. 

We first consider the dynamical properties of the massive phases. For $m<0$, the fermions are massive and we can integrate them out. We will choose a regularization such that the $\Theta$-angle for the $SU(4)$ gauge theory is 0. The $SU(4)$ gauge theory enters a confined phase at low energy. With this regularization, we can calculate the $\Theta$-angle of the $SU(4)$ gauge theory for the $m>0$ phase as the following,\footnote{We have use the fact that, for $SU(N)$ group, Tr$_AT^aT^b=N\delta^{ab}$ and Tr$_fT^aT^b=\frac{1}{2}\delta^{ab}$. }
\beq
\mathcal{L}^{SU(4)}_{\Theta}=\frac{\pi}{2}\text{Tr}_{A}\frac{F^{A}}{2\pi}\wedge\frac{F^A}{2\pi}=8\pi \left(\frac{1}{2}\text{Tr}_f\frac{F^f}{2\pi}\wedge \frac{F^f}{2\pi}\right).
\eeq
The $\Theta$-angle is $8\pi$ which is equivalent to trivial because of the $2\pi$ periodicity. Therefore the $SU(4)$ gauge theory on the $m>0$ side is also confined. Next we will discuss the topological difference between the two massive phases. 

First let us identify the symmetries. The 0-form global symmetry of the theory is $U(1)\times Z_2^T$.\footnote{We can check this in an explicit way. The 15 components of Dirac fermion by decomposing into Majorana fermions can have at most $SO(30)$ flavor symmetries. We can explicitly check that there is only one generator in $SO(30)$ that commutes with all the $SU(4)$ generators in the adjoint representation. This generates an $SO(2)$ or $U(1)$ global symmetry.} The time reversal and $U(1)$ transformation are the same as the AIII class in Eq. (\ref{AIIIT}) and (\ref{AIIIU1}).  The global symmetry commutes with the $SU(4)$ gauge group. We also assume a massive bosonic spectator $z$ that carries $SU(4)$ fundamental representation. This breaks the 1-form $Z_4$ center symmetry in the system. There are clearly gauge invariant fermions in the system such as $(z^\dagger T^az)\psi^a$. Therefore the massless theory describes a critical point in a fermionic system.

Let us consider the case where the spectator is neutral under global $U(1)$ and a singlet under time reversal transformation.
\beq
U(1): z\rightarrow z; \ \ Z_2^T: z\rightarrow z^*.
\eeq
We note that the $\mathcal{T}^2=\pm1$ is meaningless in this case for the spectator boson. We can redefine the time reversal transformation to be $\tilde{Z}_2^T: z\rightarrow e^{i\pi/2}z^*$, where the phase rotation is an element of the center of the $SU(4)$ gauge group. This gauge equivalent time reversal has $\mathcal{T}^2=-1$ for the spectator boson. We also notice that the adjoint fermion has identical time reversal transformation for $\tilde{Z}_2^T$ and $Z_2^T$.  

Let us regularize the $m<0$ phase such that it is in the topologically trivial state. Then consider the $m>0$ phase. We again consider the surface state of the system to determine the topological properties of the system. The natural surface state of the system is $2+1$-D QCD of $SU(4)$ gauge theory coupled to one adjoint fermion. On the surface, we can condense the spectator field, which Higgs the $SU(4)$ gauge field completely while preserving the $U(1)\times Z_2^T$ symmetry. The 15 Dirac fermions in the $SU(4)$ adjoint fermion become physical fermions with identical $U(1)\times Z_2^T$ transformations. Therefore, this state has topological index in AIII class $n=15\sim -1$. Thus in the large mass limit we either get a trivial insulator or the simplest topological superconductor. Study of the small mass limit within this framework may reveal interesting possible evolutions between these two familiar phases. However, we will leave this to future work. 

\section{Gauge invariant operators}
\label{operators}
We organize the gauge invariant operators according to their quantum numbers under Lorentz group and the emergent global symmetry group. We will only list Lorentz scalars and spinors composed from the adjoint fermions $\psi^a$ and gluon fields $F_{\mu\nu}^a$ (up to product of three operators). 

The time reversal transformations in our system ($\mathcal{CT}$ to be more precise) on the Weyl fermions and the gluon fields are as the following. 
\beq
\mathcal{CT}: \xi_1\rightarrow \epsilon\xi_2,\ \ \xi_2\rightarrow \epsilon\xi_1\ \ \text{with} \ \ \epsilon=i\sigma^y.
\eeq
\beq
\mathcal{CT}: F_{\mu\nu}^a\rightarrow s_\mu s_\nu F_{\mu\nu}^a\ \ \ \text{where} \ \ s_\mu=(+,-,-,-).
\eeq
\begin{table}
\begin{center}
\label{operators}
\begin{tabular}{| c |c |c |c |c |}
\hline
Operator $\mathcal{O}$ & Weyl Rep & $U(1)_c$ & $U(1)_A$ & $Z_2^T (\text{or}\ \mathcal{CT})$ \\ 
\hline
 $\bar{\psi}^a\psi^a$ & $\xi_1^{a\dagger}\epsilon\ \xi_2^{a*}-\xi_2^{aT}\epsilon\ \xi_1^{a}$ & 0 & - & $\mathcal{O}\rightarrow \mathcal{O}$ \\ \hline
 $\bar{\psi}^ai\gamma_5\psi^a$ & $i\xi_1^{a\dagger}\epsilon\ \xi_2^{a*}+i\xi_2^{aT}\epsilon\ \xi_1^{a}$ & 0 & - & $\mathcal{O}\rightarrow -\mathcal{O}$ \\ \hline
 $i\bar{\psi}^a\psi^a+\bar{\psi}^ai\gamma_5\psi^a$ & $2i\xi_1^{a\dagger}\epsilon\ \xi_2^{a*}$ & 0 & -2 & $\mathcal{O}\rightarrow -\mathcal{O}$ \\ \hline
 $-i\bar{\psi}^a\psi^a+\bar{\psi}^ai\gamma_5\psi^a$ & $2i\xi_2^{aT}\epsilon\ \xi_1^{a}$ & 0 & +2 & $\mathcal{O}\rightarrow -\mathcal{O}$ \\ \hline 
 $\psi^{aT}C\psi^a$ & $-\xi_1^{aT}\epsilon\ \xi_1^{a}+\xi_2^{a\dagger}\epsilon\ \xi_2^{a*}$ & 2 & - & $\mathcal{O}\rightarrow \mathcal{O}^\dagger$ \\ \hline
$\psi^{aT}C\gamma_5\psi^a$ & $-\xi_1^{aT}\epsilon\ \xi_1^{a}-\xi_2^{a\dagger}\epsilon\ \xi_2^{a*}$ & 2 & - & $\mathcal{O}\rightarrow -\mathcal{O}^\dagger$ \\ \hline
$\mathcal{O}_{+2}\sim\psi^{aT}C\psi^a+\psi^{aT}C\gamma_5\psi^a$ & $-2\xi_1^{aT}\epsilon\ \xi_1^{a}$ & 2 & 2 & $\mathcal{O}_{+2}\rightarrow \mathcal{O}_{-2}$ \\ \hline
 $\mathcal{O}_{-2}\sim(\psi^{aT}C\psi^{a})^\dagger-(\psi^{aT}C\gamma_5\psi^a)^\dagger$ & $-2\xi_2^{aT}\epsilon\ \xi_2^{a}$ & -2 & 2 & $\mathcal{O}_{-2}\rightarrow \mathcal{O}_{+2}$ \\ \hline
$\epsilon_{\mu\nu\lambda\rho}F^a_{\mu\nu}F^a_{\lambda\rho}$ & - & 0 & 0 & $\mathcal{O}\rightarrow -\mathcal{O}$ \\ \hline
$\epsilon_{abc}F^a_{\mu\nu}F^b_{\nu\rho}F^c_{\rho\mu}$ & - & 0 & 0 & $\mathcal{O}\rightarrow \mathcal{O}$ \\ \hline 
\end{tabular}
\end{center}
\caption{A summary of Lorentz scalars} 
\end{table}

\begin{table}
\begin{center}
\label{operators}
\begin{tabular}{| c |c |c|c |c |}
\hline
Operator & Weyl Rep & $U(1)_c$ & $U(1)_A$ & $Z_2^T (\text{or}\ \mathcal{CT})$ \\ 
\hline
$f_{\gamma_0}\sim\epsilon_{abc}(\bar{\psi}^a\psi^b)\psi^c$ &-& 1 & - & $f_{\gamma_0}\rightarrow \gamma_0\gamma_5f_{\gamma_0}^\dagger$ \\ \hline 
$f_{\gamma_5}\sim\epsilon_{abc}(\bar{\psi}^ai\gamma_5\psi^b)\psi^c$ &-& 1 & - & $f_{\gamma_5}\rightarrow -\gamma_0\gamma_5f_{\gamma_5}^\dagger$ \\ \hline 
$f_{+3}\sim\epsilon_{abc}((i\bar{\psi}^a\psi^b)\gamma_5+(\bar{\psi}^ai\gamma_5\psi^b))\psi^c$ & $\eta_\lambda \sim (-1)^\lambda\epsilon_{abc}i(\xi^a_1\epsilon \xi^b_2) \xi^c_\lambda$& 1 & 3 & $f_{+3}\rightarrow -\gamma_0\gamma_5f_{+3}^\dagger$ \\ \hline
$f_{-1}\sim\epsilon_{abc}(-(i\bar{\psi}^a\psi^b)\gamma_5+(\bar{\psi}^ai\gamma_5\psi^b))\psi^c$ & $\eta_\lambda \sim (-1)^\lambda\epsilon_{abc}i(\xi^{a\dagger}_2\epsilon \xi^{b*}_1) \xi^c_\lambda$& 1 & -1 & $f_{-1}\rightarrow -\gamma_0\gamma_5f_{-1}^\dagger$ \\ \hline
$f_g\sim\sigma_{\mu\nu}F^a_{\mu\nu}\psi^a$ & - & 1 & 1 & $f_g\rightarrow\gamma_0\gamma_5f_g^\dagger$ \\ \hline
$f_g'\sim\epsilon_{\mu\nu\lambda\rho}\sigma_{\mu\nu}F^a_{\lambda\rho}\psi^a$ & - & 1 & 1 & $f_g'\rightarrow -\gamma_0\gamma_5f_g'^{\dagger}$ \\ \hline
$f_{gg}\sim\epsilon_{abc}F^a_{\mu\nu}F^b_{\mu\nu}\psi^c$ & - & 1 & 1 & $f_{gg}\rightarrow \gamma_0\gamma_5f_{gg}^\dagger$ \\ \hline
$f_{gg}'\sim\epsilon_{abc}\epsilon_{\mu\nu\lambda\rho}F^a_{\mu\nu}F^b_{\lambda\rho}\psi^c$ & - & 1 & 1 & $f_{gg}'\rightarrow -\gamma_0\gamma_5f_{gg}'^\dagger$ \\ \hline
\end{tabular}
\end{center}
\caption{A summary of Lorentz spinor} 
\end{table}

\bibliography{3dDQCP}

\end{document}